\documentclass[12pt]{article}
\usepackage[utf8]{inputenc}
\usepackage{amsmath,amsfonts,amssymb}
\usepackage{hyperref}
\usepackage{epsfig}
\usepackage{cite}
\def\bra#1{\langle #1 |\, }
\def\ket#1{\, | #1 \rangle}
\newcommand{\ba}{\begin{array}}
\newcommand{\ea}{\end{array}}
\newcommand{\no}{\nonumber}
\newcommand{\be}{\begin{equation}}
\newcommand{\ee}{\end{equation}}
\newcommand{\beqn}{\begin{eqnarray}}
\newcommand{\eeqn}{\end{eqnarray}}
\newcommand{\eqn}[1]{(\ref{#1})}
\newcommand{\cO}{{\cal O}}
\newcommand{\bel}[1]{\be\label{#1}}

\newcommand{\smtau}{\frac{s}{m_{\tau}^{2}}}
\newcommand{\ImPi}{\operatorname{Im}\Pi}
\def\lett#1{\fontsize{#1}{1.2#1}\selectfont}

\usepackage{color}
\definecolor{orange}{RGB}{255,127,0}
\definecolor{brown}{RGB}{102,51,0}
\definecolor{myred}{RGB}{192,0,0}
\begin{document}
\bibliographystyle{unsrt}
\thispagestyle{empty}

\begin{flushright}
IFIC/16-13\\  FTUV/16-0522
\end{flushright}

\vspace{1.5cm}

\begin{center}
{\fontsize{20}{24}\selectfont\bf
Determination of the QCD Coupling\\[10pt] from  ALEPH $\tau$ Decay Data}

\vspace{1.5cm}

Antonio~Pich and Antonio Rodr\'{\i}guez-S\'anchez

\vspace*{.7cm}

\vspace*{0.1cm}

Departament de F\'\i sica Te\`orica, IFIC, Universitat de Val\`encia -- CSIC,\\
Apt. Correus 22085, E-46071 Val\`encia, Spain,

\vspace*{0.7cm}
%\today

\end{center}

\vspace*{1cm}

\begin{abstract}
We present a comprehensive study of the determination of the strong coupling from $\tau$ decay, using the most recent release of the experimental ALEPH data. We critically review all theoretical strategies used in previous works and put forward various novel approaches which allow to study complementary aspects of the problem. We investigate the advantages and disadvantages of the different methods, trying to uncover their potential hidden weaknesses and test the stability of the obtained results under slight variations of the assumed inputs. We perform several determinations, using different methodologies, and find a very consistent set of results. All determinations are in excellent agreement, and allow us to extract a very reliable value for $\alpha_s(m_\tau^2)$. The main uncertainty originates in the pure perturbative error from unknown higher orders. Taking into account the systematic differences between the results obtained with the CIPT and FOPT prescriptions, we find
$\alpha_{s}^{(n_f=3)}(m_\tau^2)  = 0.328 \pm 0.013$ which implies
$\alpha_{s}^{(n_f=5)}(M_Z^{2}) = 0.1197\pm 0.0015$.

\end{abstract}

%\end{frontmatter}

\clearpage

\section{Introduction}

The hadronic decay width of the $\tau$ lepton provides one of the most precise determinations of the strong coupling \cite{Pich:2013lsa,Pich:2015ivv,d'Enterria:2015toz,Agashe:2014kda,Pich:2013sqa,Deur:2016tte}. The inclusive ratio
\be\label{rtau}
R_{\tau}\; =\;\frac{\Gamma[\tau^{-}\rightarrow \nu_{\tau} \mathrm{hadrons}]}{\Gamma[\tau^{-}\rightarrow \nu_{\tau} e^{-}\overline{\nu}_{e}]}\, ,
\ee
can be rigorously calculated within QCD with high precision \cite{Narison:1988ni,Braaten:1988hc,Braaten:1988ea,Braaten:1991qm} and turns out to be very sensitive to the input value of $\alpha_s(m_\tau^2)$. Non-perturbative corrections are very suppressed for well-understood theoretical reasons \cite{Braaten:1991qm}; moreover, their quantitative size can be directly extracted from the measured distribution of the final hadrons in $\tau$ decays \cite{LeDiberder:1992zhd}. The predicted value of $R_\tau$ is completely dominated by the perturbative contribution, which is already known to $\cO(\alpha_s^4)$ \cite{Baikov:2008jh}, and includes renormalization-group resummations of higher-order logarithm-induced corrections \cite{LeDiberder:1992jjr,Pivovarov:1991rh}.

Owing to the low value of the $\tau$ mass scale, $\alpha_s(m_\tau^2)$ is sizeable with a numerical value around $0.33$ \cite{Pich:2013lsa}. This makes $R_\tau$ more sensitive to the strong coupling than higher-energy observables, even if some of them can be predicted more accurately. Although $\alpha_s(m_\tau^2)$ has been only determined with a 4\% accuracy, evolving it up in energy with the QCD renormalization-group equations, it implies a 1\% precision on $\alpha_s(M_Z^2)$ \cite{Pich:2013lsa,Pich:2015ivv}, which is a factor of two more accurate than the direct measurement of the strong coupling at the $Z$ peak
\cite{d'Enterria:2015toz,Agashe:2014kda,Pich:2013sqa,Deur:2016tte}. The excellent agreement between these two determinations of $\alpha_s$, at very different mass scales, constitutes at present the most precise quantitative test of asymptotic freedom \cite{Gross:1973id,Politzer:1973fx,Coleman:1973sx}.

Since the strong coupling is not small at $\mu = m_\tau$, the predicted value of $R_\tau$ is quite sensitive to higher-order perturbative corrections. The induced perturbative uncertainties dominate in fact the final error on $R_\tau$ and are, at present, the main limitation on the potentially achievable accuracy \cite{Pich:2013lsa,Pich:2015ivv,Pich:2011bb}. Nevertheless, at the current level of $\cO(\alpha_s^4)$ precision, it is also necessary to analyze carefully the numerical role of the small non-perturbative contributions.

The most precise experimental analysis, performed with the ALEPH $\tau$ decay data \cite{Schael:2005am}, bounds non-perturbative effects to be safely below 1\% \cite{Davier:2013sfa,Davier:2008sk,Davier:2005xq}, in agreement with theoretical expectations \cite{Braaten:1991qm} and previous experimental studies \cite{Schael:2005am,Barate:1998uf,Buskulic:1993sv,Ackerstaff:1998yj,Coan:1995nk} which confirmed the predicted suppression of this type of contributions.
However, the ALEPH results have been strongly criticized in recent years in a series of papers
\cite{Boito:2011qt,Boito:2012cr,Boito:2014sta},
advocating to pursue a slightly different type of analysis \cite{Cata:2008ye,Cata:2008ru},
focused on observables which maximize the role of non-perturbative effects in order to better study them. Unfortunately, trying to stress the advantages of their approach, these papers adopt an overly conservative/pessimistic attitude when judging previous work on the subject, while the uncertainties of their own analyses appear to be largely underestimated. Legitimate
criticisms are mixed up with some not fully-correct or even slightly misleading statements. Studying observables which are more sensitive to some types of non-perturbative contributions is interesting per-se and can help us to better understand QCD in the strong-coupling regime, but it is not necessarily the best strategy to perform a clean and accurate measurement of $\alpha_s$.

In this paper we attempt a fresh numerical analysis of the ALEPH data, trying to assess the advantages and disadvantages of different possible approaches. Ideally, all sound theoretical methods should finally give similar results,
complementing each other so that a combination of them would allow to maximize the amount and quality of the extracted information. However, current $\tau$
data suffer from strong correlations and large uncertainties, specially in the highest energy range, which severely limits the potential scope of a realistic statistical analysis and
the maximum number of parameters to be fitted.

We present first in section~\ref{sec:theory} a short overview of the theoretical ingredients underlying all QCD analyses of the inclusive $\tau$ decay width, so that the paper is self-contained. Some
technical details on the data handling are briefly given in section~\ref{sec:data}. The standard analysis of the data \cite{Braaten:1991qm,LeDiberder:1992zhd,LeDiberder:1992jjr}, adopted by the ALEPH
Orsay group \cite{Davier:2005xq,Davier:2008sk,Davier:2013sfa}, is revised in section~\ref{sec:ALEPH}, which performs a complete numerical study and comments on the quality and potential weaknesses of
the final results. Sections~\ref{sec:optimal} and \ref{sec:improvements} discuss some possible improvements and their limitations, and analyze the stability of the results, compared with the ones
previously obtained in section~\ref{sec:ALEPH}.

The approach followed in Refs.~\cite{Boito:2011qt,Boito:2012cr,Boito:2014sta}, aimed to uncover duality violation effects, is critically studied in section~\ref{sec:DV}. While we are able to reproduce
most of the numerical results presented in those references, they are based on an ad-hoc assumption on the functional form of the spectral function whose validity is unknown.
Small modifications of this assumption translate into sizeable changes in the fitted value of $\alpha_s(m_\tau^2)$ which turns out to be model dependent. Although compatible with the more solid determinations
performed in previous sections, the values of $\alpha_s$ extracted from this approach are not precise enough to be competitive once the real uncertainties are properly estimated.

In section~\ref{sec:Borel} we follow an alternative strategy, based on the Borel transform of the spectral distribution, in order to change the weights of different contributions/effects. While having its own
weaknesses, this approach provides an additional handle to judge the reliability of the results extracted from current data. The numerical determinations of $\alpha_s(m_\tau^2)$ obtained with all approches turn
out to be consistent, within their estimated errors. We compile all of them in section~\ref{sec:summary}, and conclude giving our final value for the determination of the strong coupling from $\tau$ decay.

\section{Theoretical framework}
\label{sec:theory}

The ratio $R_{\tau}$ can be calculated from the spectral identity \cite{Narison:1988ni,Braaten:1988hc,Braaten:1988ea,Braaten:1991qm}
\bel{eq:RtauSpectral}
R_{\tau} \; =\; 12 \pi\, S_{\mathrm{EW}}\, \int^{m_ {\tau}^{2}}_{0}\frac{ds}{m_{\tau}^{2}}\left(1-\smtau\right)^2
\left[\left(1+2\smtau\right)\ImPi^{(1)}(s)+\ImPi^{(0)}(s)\right]\, ,
\ee
with
\be
\Pi^{(J)}(s)\;\equiv\; \sum_{q=d,s}|V_{uq}|^{2}\left(\Pi^{(J)}_{uq, V}(s)+ \Pi^{(J)}_{uq, A}(s)  \right) \, ,
\ee
where $\Pi^{(J)}_{uq, V/A}(s)$ are the two-point correlation functions for the vector
$V_{ij}^{\mu}=\overline{q}_{j}\gamma^{\mu} q_{i}$ and axial-vector
$A_{ij}^{\mu}=\overline{q}_{j} \gamma^{\mu}\gamma_5 q_{i}$ colour-singlet quark currents ($i,j=u,d,s; \mathcal{J}=V,A$):
\be
i \int d^{4}x\; e^{iqx} \;
\bra{0}T[\mathcal{J}_{ij}^{\mu}(x)\mathcal{J}_{ij}^{\nu \dagger}(0)]\ket{0}
\; =\;
(-g^{\mu\nu}q^{2}+q^{\mu}q^{\nu})\; \Pi^{(1)}_{ij,\mathcal{J}}(q^{2})
+ q^{\mu}q^{\nu}\;\Pi^{(0)}_{ij,\mathcal{J}}(q^{2}) \, .
\ee
The factor $S_{\mathrm{EW}} = 1.0201 \pm 0.0003$ contains the renormalization-group-improved electroweak correction, including a next-to-leading order resummation of large logarithms
\cite{Marciano:1988vm,Braaten:1990ef,Erler:2002mv}.

Since non-strange hadronic $\tau$ decay data can be separated into the $V$ and $A$ channels, identifying the invariant-mass and spin of the final hadronic system, we have experimental access to the different spectral functions
$\rho_{ud,\mathcal{J}}^{(J)}(s)\equiv\frac{1}{\pi}\ImPi^{(J)}_{ud,\mathcal{J}}(s)$, with $J=0, 1$. From now on, we will use the updated ALEPH non-strange spectral functions $\rho_{ud,V/A}^{(0+1)}(s)$ \cite{Davier:2013sfa}.

On the other side, the theoretical correlators\footnote{The $ud$ subscript and the $(1+0)$ superscript will be omitted from now on.}
$\Pi_{V/A}(s)\equiv \Pi_{ud,V/A}^{(1+0)}(s)$
are predicted by QCD for large Euclidean momenta through their Operator Product Expansion (OPE) \cite{Shifman:1978bx}:
\be\label{eq:ope}
\Pi^{\mathrm{OPE}}_{V/A}(s=-Q^{2})\; =\; \sum_{D}\frac{1}{(Q^2)^{D/2}}\sum_{\mathrm{dim} \, \mathcal{O}=D} C_{D, V/A}(Q^{2},\mu)\;\langle\mathcal{O}(\mu)\rangle
\;\equiv\; \sum_{D}\;\dfrac{\mathcal{O}_{D,\, V/A}}{(Q^2)^{D/2}}\, .
\ee
Although the OPE of the correlators is not valid at low Minkowskian momenta, the region where
we have experimental data, their known analytic structure can be used to relate both regions. Since they are analytic functions in all the complex plane, except for a cut in the positive real axis, the integral along the circuit shown in Figure~\ref{fig:circuit} must be zero, so that \cite{Braaten:1991qm,LeDiberder:1992jjr,LeDiberder:1992zhd}:
%%%%%%%%%%%%%%%%%%%%%%%%%%%%% FIGURE %%%%%%%%%%%%%%%%%%%%%%%%%%%%%
\begin{figure}[tb]\centering
\includegraphics[width=0.4\textwidth]{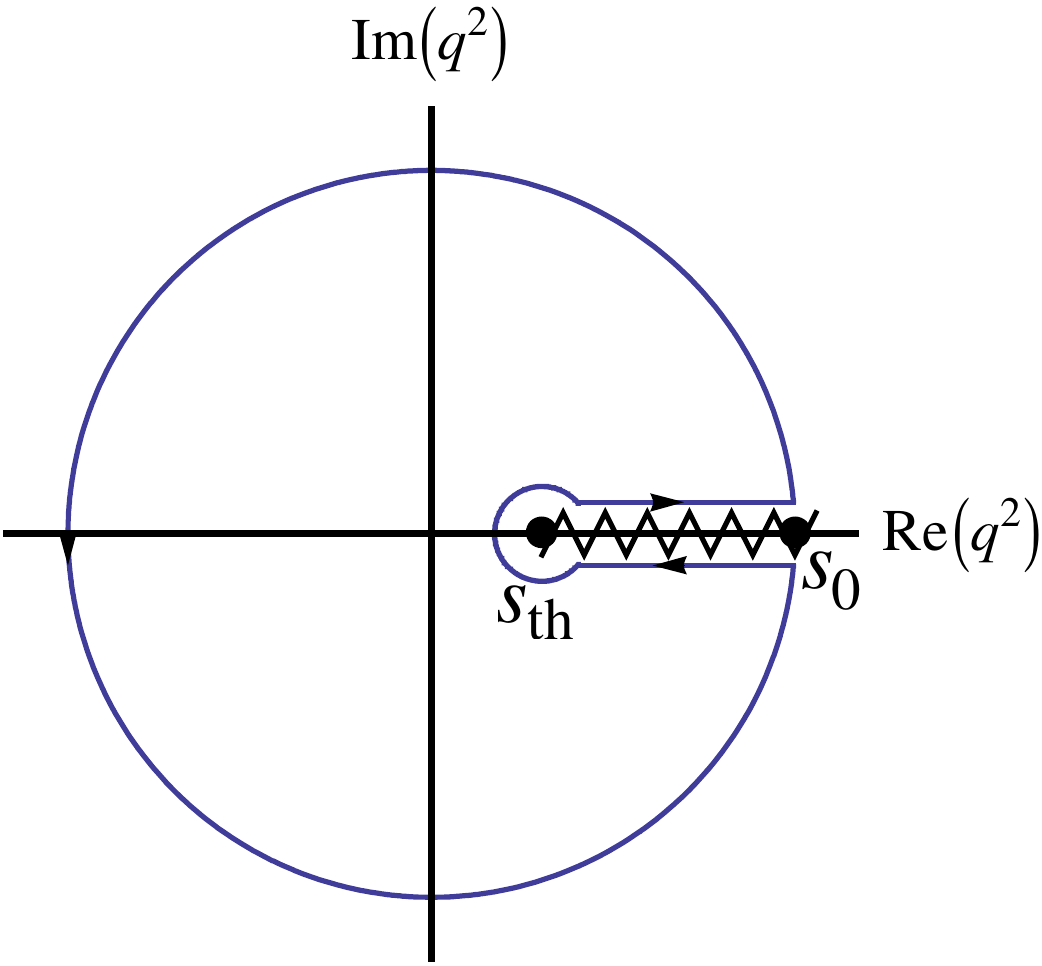}
\caption{Complex integration contour, used to derive Eq.~\eqn{aomega}.}
\label{fig:circuit}
\end{figure}
%%%%%%%%%%%%%%%%%%%%%%%%%%%%%%%%%%%%%%%%%%%%%%%%%%%%%%%%%%%%%%%%%%
\beqn\label{aomega}
A^{\omega}_{V/A}(s_{0})\;\equiv\; \int^{s_{0}}_{s_{\mathrm{th}}} \frac{ds}{s_{0}}\;\omega(s)\, \ImPi_{V/A}(s)\; =\; \frac{i}{2}\;\oint_{|s|=s_{0}}
\frac{ds}{s_{0}}\;\omega(s)\, \Pi_{V/A}(s)\, ,
\eeqn
where $\omega(s)$ is any weight function analytic inside the circuit, $s_{\mathrm{th}}$ is the hadronic mass-squared threshold, and the complex integral in the right-hand side runs counter-clockwise around the circle $|s|=s_{0}$.

While experimental data can be used in the left-hand side of Eq. (\ref{aomega}), in the right-hand side we can use the analytic continuation of the OPE of the correlator, which is valid for values of $s_{0}$ large enough except, as said above, in the positive real axis. Differences between the physical values of the $A^{\omega}_{V/A}(s_{0})$ integrals and their OPE approximations
are known as quark-hadron duality violations \cite{Chibisov:1996wf,Shifman:2000jv,Cirigliano:2002jy,Cirigliano:2003kc,Cata:2005zj,GonzalezAlonso:2010xf,Dominguez:2016jsq,Rodriguez-Sanchez:2016jvw}.
They are minimized by taking ``pinched'' weight functions  \cite{Braaten:1991qm,LeDiberder:1992zhd} which vanish at $s=s_0$, suppressing in this way the contributions from the region near the real axis to the integral in the right-hand side of Eq.~\eqn{aomega}.

With the weight functions appearing in Eq.~\eqn{eq:RtauSpectral}, the purely perturbative contribution dominates the theoretical prediction for $R_\tau$ \cite{Braaten:1991qm}, making possible to perform a rather clean measurement of $\alpha_{s}$. However, the quantitative estimate of the small non-perturbative corrections is far from trivial and necessarily involves the analysis of additional weighted integrals, more sensitive to power corrections. The choice of an optimal set of weight functions, allowing for a reliable estimate of the non-perturbative OPE corrections to $R_\tau$ without introducing background noise from unwanted effects such as duality violations, may become challenging.

\subsection{Perturbative contribution}

The main contribution to $A_{V/A}^{\omega}(s_{0})$ comes from the perturbative part $A^{\omega, P}(s_{0})$, which for massless quarks is identical for the vector and the axial-vector correlators, due to chiral symmetry.
It can be extracted from the renormalization-scale-invariant Adler function
\cite{Adler:1974gd}:
\beqn\label{adler}
D(s)\;\equiv\; -s\,\frac{d\,\Pi^{P}(s)}{ds}\; =\;\frac{1}{4\pi^{2}}\;\sum_{n=0}\tilde K_{n}(\xi)\;
a_{s}^{n}(-\xi^2 s) \, ,
\eeqn
where $a_{s}(s)\equiv \alpha_{s}(s)/\pi$  satisfies the renormalization-group equation:
\beqn\label{running}
2\,\frac{s}{a_s}\,\frac{d\, a_{s}(s)}{d s}\; =\; \sum_{n=1} \beta_{n}\, a_{s}^{n}(s) \, .
\eeqn
The perturbative coefficients $K_n\equiv \tilde K_{n}(\xi=1)$ are  known up to $n\le 4$. For $N_f=3$ flavours, one has:
$K_0 = K_1 = 1$, $K_2 = 1.63982$, $K_3^{\overline{\mathrm{MS}}} = 6.37101$ and
$K_4^{\overline{\mathrm{MS}}} = 49.07570$
\cite{Baikov:2008jh,Gorishnii:1990vf,Surguladze:1990tg,Chetyrkin:1979bj,Dine:1979qh,Celmaster:1979xr}. The homogeneous renormalization-group equation satisfied by the Adler function determines the corresponding scale-dependent parameters $\tilde K_{n}(\xi)$ \cite{LeDiberder:1992jjr,Pich:1999hc}.
Although the dependence on the renormalization scale cancels exactly in the infinite sum, the truncation to a finite perturbative order leads to a scale dependence from the missing higher-order terms, which must be taken into account when estimating perturbative uncertainties.

Integrating by parts Eq. (\ref{aomega}), we can rewrite $A^{\omega, P}(s_{0})$ in terms of the Adler function:
\beqn\label{intbyparts}
A^{\omega, P}(s_{0})\; =\; \frac{i}{2s_{0}}\;\oint_{|s|=s_{0}}\frac{ds}{s}\;\left[ W(s)-W(s_{0})\right]\; D(s)\, ,
\eeqn
with $W(s)\equiv \int^{s}_{0}ds'\,\omega(s')$. Introducing Eq. (\ref{adler}) in Eq. (\ref{intbyparts}) and parametrizing the circumference as $s=-s_{0}\, e^{i\varphi}$, one gets:
\beqn
A^{\omega, P}(s_{0})\; =\; -\frac{1}{8\pi^{2}s_{0}}\;\sum_{n=0}\; \tilde K_{n}(\xi)\;\int^{\pi}_{-\pi} d\varphi\;
\left[ W(-s_{0}\, e^{i\varphi})-W(s_{0})\right]\; a_{s}^{n}(\xi^2 s_{0} \, e^{i\varphi}) \, . \label{pertu}
\eeqn
The contour integral on the right-hand side only depends on $a_{s}(\xi^2 s_{0})$.
The integration can be performed, either truncating the integrand to a fixed perturbative order in $\alpha_s(\xi^2 s_{0})$ (fixed-order perturbation theory, FOPT) \cite{Braaten:1991qm}, or solving exactly the differential $\beta$-function equation in the $\beta_{n>n_{\mathrm{max}}}=0$ approximation (contour-improved perturbation theory, CIPT) \cite{LeDiberder:1992jjr,Pivovarov:1991rh}.
This second procedure should be preferred, as it sums big corrections arising for large values of $|\varphi|$, due to the long running of $a_{s}^{n}(\xi^2 s_{0} \, e^{i\varphi})$ along the contour integration \cite{LeDiberder:1992jjr}. Taking $n_{\mathrm{max}}= 1,2,3,4$, one easily checks that CIPT leads to a fast perturbative convergence for the integrals and the numerical results are stable under changes of the renormalization scale \cite{LeDiberder:1992jjr,Pich:2010xb}. On the other side, the slow convergence of the FOPT series leads to a much larger renormalization-scale dependence.

Since perturbation theory is known to be at best an asymptotic series, it has been argued  that, in the asymptotic large-$n$ regime, the expected renormalonic behaviour of the $K_n$ coefficients could induce cancellations with the running corrections, which would be missed by CIPT. This happens actually in the large-$\beta_1$ limit, which however does not approximate well the known $K_n$ coefficients (it predicts an alternating series) \cite{Ball:1995ni,Neubert:1995gd,Altarelli:1994vz}.
Models of higher-order corrections with this behaviour have been advocated \cite{Beneke:2008ad,Beneke:2012vb}, but the results are model dependent \cite{DescotesGenon:2010cr,Jamin:2005ip}. The implications of a renormalonic behaviour have been also studied using an optimal conformal mapping in the Borel plane and properly implementing the CIPT procedure within the Borel transform. Assuming that the known fourth-order Adler series is already dominated by the lowest ultraviolet ($u=-1$) and infrared ($u=2,3$) renormalons, the conformal mapping generates a full series of higher-order coefficients which result, after Borel summation, in a perturbative correction which is numerically close to the naive FOPT result \cite{Caprini:2009vf,Caprini:2011ya,Abbas:2012fi,Abbas:2012py,Abbas:2013usa}.

For a fixed value of $\alpha_s(m_\tau^2)$, FOPT predicts a slightly larger perturbative contribution to $R_\tau$ than CIPT. Therefore, it leads to a smaller fitted value of $\alpha_s(m_\tau^2)$. In the absence of a better understanding of higher-order perturbative corrections, we will perform all our analyses with both procedures.
Within a given perturbative approach, either CIPT or FOPT, we will estimate the perturbative uncertainty varying the renormalization scale in the interval $\xi^2 \in (0.5\, ,\, 2)$. Additionally, we will truncate the perturbative series at $n=5$, taking $K_{5}=275 \pm 400$ \cite{Pich:2011bb} as an educated guess of the maximal range of variation of the unknown fifth-order contribution. These two sources of theoretical uncertainty will be combined quadratically. 

In order to give a combined determination for the strong coupling, we will finally average the CIPT and FOPT results. Since the previously estimated perturbative uncertainties do not fully account for the difference between these two prescriptions, we will conservatively assess the final error adding in quadrature half the difference between the CIPT and FOPT values to
the smallest of the CIPT and FOPT errors. We want to emphasize that the perturbative errors are at present the largest source of uncertainty in the determination of the strong coupling from $\tau$ decays.

\subsection{Non-perturbative contribution}

Since OPE corrections are going to be small, we can safely neglect the logarithmic dependence on $s$ of the Wilson coefficients $C_{D, V/A}$, appearing in Eq.~(\ref{eq:ope}), so that
$\mathcal{O}_{D,\, V/A}$ is an effective $s$-independent vacuum condensate of dimension $D$. To simplify notation, together with the genuine non-perturbative contributions which have $D\ge 4$, we include also in the sum the inverse-power corrections of pure perturbative origin, induced by the finite quark masses, which give tiny contributions to $R_\tau$ smaller than $10^{-4}$  \cite{Braaten:1991qm,Pich:1998yn,Pich:1999hc}.

The lowest-dimensional vacuum condensate contributions are \cite{Braaten:1991qm}:
\bel{eq:O4cont}
\cO_{4,V/A}\; =\; \frac{1}{12}\left[1-\frac{11}{18}\, a_{s}  \right]\,\langle a_{s}GG\rangle \, +\,\left[ 1+\frac{\pm36-23}{27}\, a_{s}\right]\,\langle(m_{u}+m_{d})\,\bar{q}q\rangle \, .
\ee
The size of the quark condensate is determined by chiral symmetry to be \cite{GellMann:1968rz,Pich:1995bw,Ecker:1994gg}
\be
\langle(m_{u}+m_{d})\,\bar{q}q\rangle\; =\; -m_\pi^2 f_\pi^2\; \approx\; -1.6\cdot 10^{-4}\;\mathrm{GeV}^4
\; \approx\; -1.6\cdot 10^{-5}\; m_\tau^4
\ee
and, therefore, is not going to be very relevant in our numerical analyses. The gluon condensate has been analyzed in many works \cite{NarisonBook}, since its first phenomenological estimate in Ref.~\cite{Shifman:1978bx}, but unfortunately its numerical size is still quite uncertain. As a conservative estimate, one can quote the range \cite{Braaten:1991qm}
\bel{eq:GluonCond}
\langle a_{s}GG\rangle \; \approx\; (0.02\pm 0.01)\;\mathrm{GeV}^4
\; \approx\; (1.7\pm 0.8)\cdot 10^{-4}\,\times\, (12\, m_\tau^4)\, ,
\ee
where in the last expression we have included the factor $1/12$ in Eq.~\eqn{eq:O4cont} to better appreciate its possible numerical impact in the $\tau$ hadronic width. As we are going to see next, $R_\tau$ is insensitive to the $D=4$ OPE contribution \cite{Narison:2009vy} and, given the small numerical size of \eqn{eq:GluonCond}, the invariant-mass distribution in $\tau$ decays does not help much in pinning down the gluon condensate.

Inserting Eq. (\ref{eq:ope}) in Eq. (\ref{aomega}), one finally gets the non-perturbative contribution to $A_{V/A}^{\omega}$:
\be\label{nonpertu}
A_{V/A}^{\omega, NP}(s_{0})\; =\; \frac{i}{2}\;\sum_{D}\,\mathcal{O}_{D,\, V/A}\;\oint_{|s|=s_{0}}\frac{ds}{s_{0}}\;\dfrac{\omega(s)}{(-s)^{D/2}}
\; =\;\pi\;\sum_{D} a_{-1,\, D} \; \dfrac{\mathcal{O}_{D,\, V/A}}{s_{0}^{D/2}} \, ,
\ee
where $a_{-1,\, D}$ is the $-1$ coefficient of the Laurent expansion of $\omega(s=-s_{0}x)/x^{D/2}$:
\beqn
\omega(-s_{0}x)\; =\; \sum_{n}\, a_{n,\, D}\; x^{n+D/2}\, .
\eeqn

With the phase-space and spin-1 factors appearing in Eq.~\eqn{eq:RtauSpectral},\footnote{There is in addition a small correction from the $s\,\Pi^{(0)}(s)$ term, which vanishes for massless quarks
because the vector and axial-vector currents are conserved in the chiral limit.}
the measured invariant-mass distribution in hadronic $\tau$ decays weights the $\Pi_{V/A}(s)$ correlators with the function $\omega(x) = (1-x)^2 (1+2x) = 1-3x^2+2x^3$. This implies that the inclusive hadronic width is only sensitive to OPE corrections with $D=6$ and 8, which are strongly suppressed by the corresponding powers of the $\tau$ mass. Moreover, owing to its different chirality, the $D=6$ contributions to the vector and axial-vector correlators are expected to have opposite signs leading to a partial cancellation in the $V+A$ case \cite{Braaten:1991qm}. Therefore, non-perturbative corrections to the total hadronic decay width $R_{\tau}$ are very suppressed, while perturbative contributions are of $\cO(20\%)$. This is what makes $R_{\tau}$ such a clean observable to measure the strong coupling. Note also that the factor $(1-x)^2$ in Eq.~\eqn{eq:RtauSpectral} naturally suppresses the contributions to the contour integral from the region near the real axis ($x\sim 1$), minimizing any possible uncertainties from duality-violation effects.

In the next sections we will try to take advantage of the measured invariant-mass distribution of the final hadrons in order to increase the sensitivity to non-perturbative effects. One can either use weight functions with the appropriate powers of $x$ to pin down a given dimension-$D$ term in the OPE, or lower the value of $s_0$, the upper end of the integration range, to get a smaller suppression from the $s_0^{-D/2}$ factor. Using non-pinched weight functions one can also investigate violations of quark-hadron duality, but paying the price of a weaker theoretical control since one becomes sensitive
to regions where the use of the OPE is not justified.

\section{Data handling}\label{sec:data}

In this work we use the updated ALEPH invariant mass-squared distributions \cite{Davier:2013sfa}, which incorporate an improved unfolding of the measured mass spectra from detector effects and
correct some problems \cite{Boito:2010fb} in the correlations between unfolded mass bins. The improved
unfolding brings an increased statistical uncertainty near the edges of phase space. It has
also reduced the number of bins in the spectral distribution, as a larger bin size has been
adopted.

From these distributions we can get the spectral functions, using Eq. (\ref{eq:RtauSpectral}):
\begin{align}\nonumber
\frac{1}{N}\,
\frac{\Delta N^{(1)}_{V/A}(s_{i})}{\Delta s_{i}}\; &\approx\; \frac{1}{N}\,\frac{dN^{(1)}_{V/A}}{ds} \; =\; B_{e}\;\dfrac{d R_{\tau,V/A}^{(1)}}{ds}(s)
\\[5pt]
 &=\;\dfrac{12\pi}{m_{\tau}^{2}}\;  B_{e}\, S_{\mathrm{EW}}\, |V_{ud}|^{2}\;
\left(1-\frac{s}{m_{\tau}^{2}}\right)^{2}\left(1+ \frac{2 s}{m_{\tau}^{2}}  \right)\,
\operatorname{Im} \Pi_{V/A}^{(1)}(s)\, ,
\label{invariant1}
\end{align}
\begin{align}\nonumber
\frac{1}{N}\,
\frac{\Delta N^{(0)}_{V/A}(s_{i})}{\Delta s_{i}}\; &\approx\; \frac{1}{N}\,\frac{dN^{(0)}_{V/A}}{ds} \; =\; B_{e}\;\dfrac{d R_{V/A}^{(0)}}{ds}(s)
\\[5pt]
 &=\;\dfrac{12\pi}{m_{\tau}^{2}}\; B_{e}\, S_{\mathrm{EW}}\, |V_{ud}|^{2}\;
\left(1-\frac{s}{m_{\tau}^{2}}\right)^{2}\,
\operatorname{Im} \Pi_{V/A}^{(0)}(s) \, ,
\label{invariant0}
\end{align}
where $B_e$ is the $\tau^-\to e^-\bar\nu_e\nu_\tau$ branching ratio,
$\Delta N^{(0,1)}_{V/A}(s_{i})$ the number of $V/A$ events with $J=0,1$ in the bin centered at $s_i$,
$\Delta s_{i}$ the corresponding bin size and
$N$ the total number of events in the ALEPH data sample.

In the massless quark limit, the vector and axial-vector currents are conserved which implies
$dN_V^{(0)}(s) = 0$ and $dN^{(0)}_{A}(s)/N = B_{\pi}\, \delta(s-m_{\pi}^2)\, ds$, with
$B_\pi =\mathrm{Br}(\tau^-\to\pi^-\nu_\tau)$. The non-zero contribution to the longitudinal axial distribution originates in the Goldstone nature of the pion ($m_\pi = 0$, for massless quarks), associated with the chiral symmetry breaking of QCD.

Inserting Eqs. (\ref{invariant1}) and (\ref{invariant0}) into Eq. (\ref{aomega}), we get the experimental values of the moments $A_{V/A}^{\omega}(s_{0})$:
\begin{align}\label{exp1}
A^{\omega}_{V}(s_{0})\; =\; & F\;\sum_{s_{i}}^{s_{0}-\frac{\Delta s_{0}}{2}}\dfrac{\Delta N_{V}(s_{i})}{N}\;\omega_{i}(s_{i},s_{0})\,H(s_{0},s_{i}) \, ,
\\ \nonumber
A^{\omega}_{A}(s_{0})\; =\; & F\;\sum_{s_{i}}^{s_{0}-\frac{\Delta s_{0}}{2}}\dfrac{\Delta N_{A}(s_{i})}{N}\;\omega_{i}(s_{i},s_{0})\,H(s_{0},s_{i})\, ,
\\&\,
+ F \;\frac{m_{\tau}^{2}}{s_{0}}\left(1-\frac{m_{\pi}^{2}}{m_{\tau}^{2}}\right)^{-2}B_{\pi}\;
\omega_{i}(m_{\pi}^{2},s_{0}) \, , \label{exp2}
\end{align}
where
\begin{equation}
F\; =\;\left[12\pi\, S_{\mathrm{EW}}\, |V_{ud}|^{2} B_{e}\right]^{-1}
\end{equation}
collects all normalization factors,
\begin{equation}
H(s_{0},s_{i})\; =\;\frac{m_{\tau}^{2}}{s_{0}}\,\left(1-\frac{s_{i}}{m_{\tau}^{2}}\right)^{-2}\left(1+ \frac{2 s_{i}}{m_{\tau}^{2}}\right)^{-1}
\end{equation}
and $\Delta s_{0}$ is the bin width of the bin centered at $s_{0}-\frac{\Delta s_{0}}{2}$.

%%%%%%%%%%%%%%%%%%%%%%%%%%%%% ALEPH Figures %%%%%%%%%%%%%%%%%%%%%%%%%%%%%%%%%%
\begin{figure}[tb]
\centerline{
\includegraphics[width=0.36\textwidth]{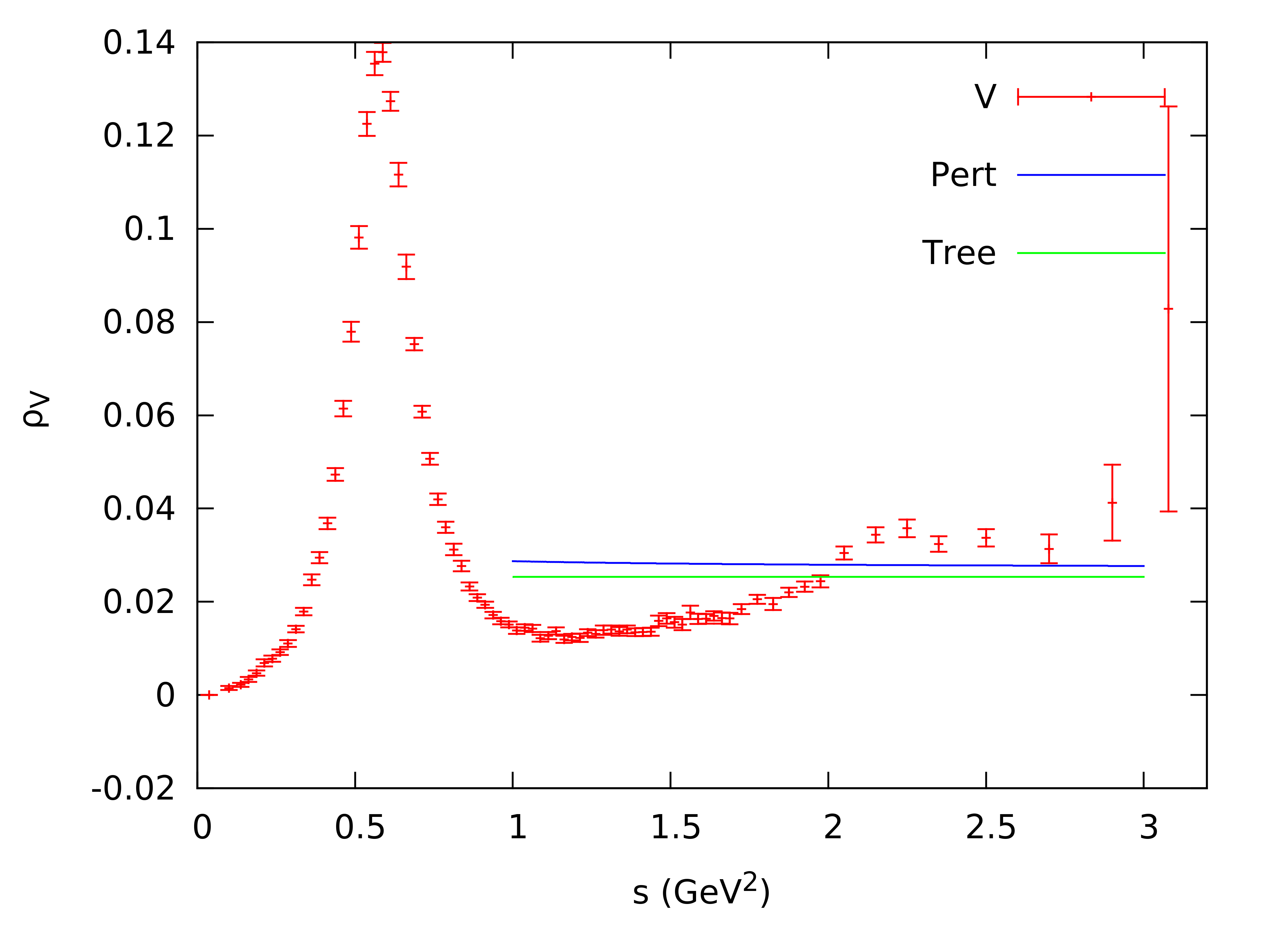}\hskip -.15cm
\includegraphics[width=0.36\textwidth]{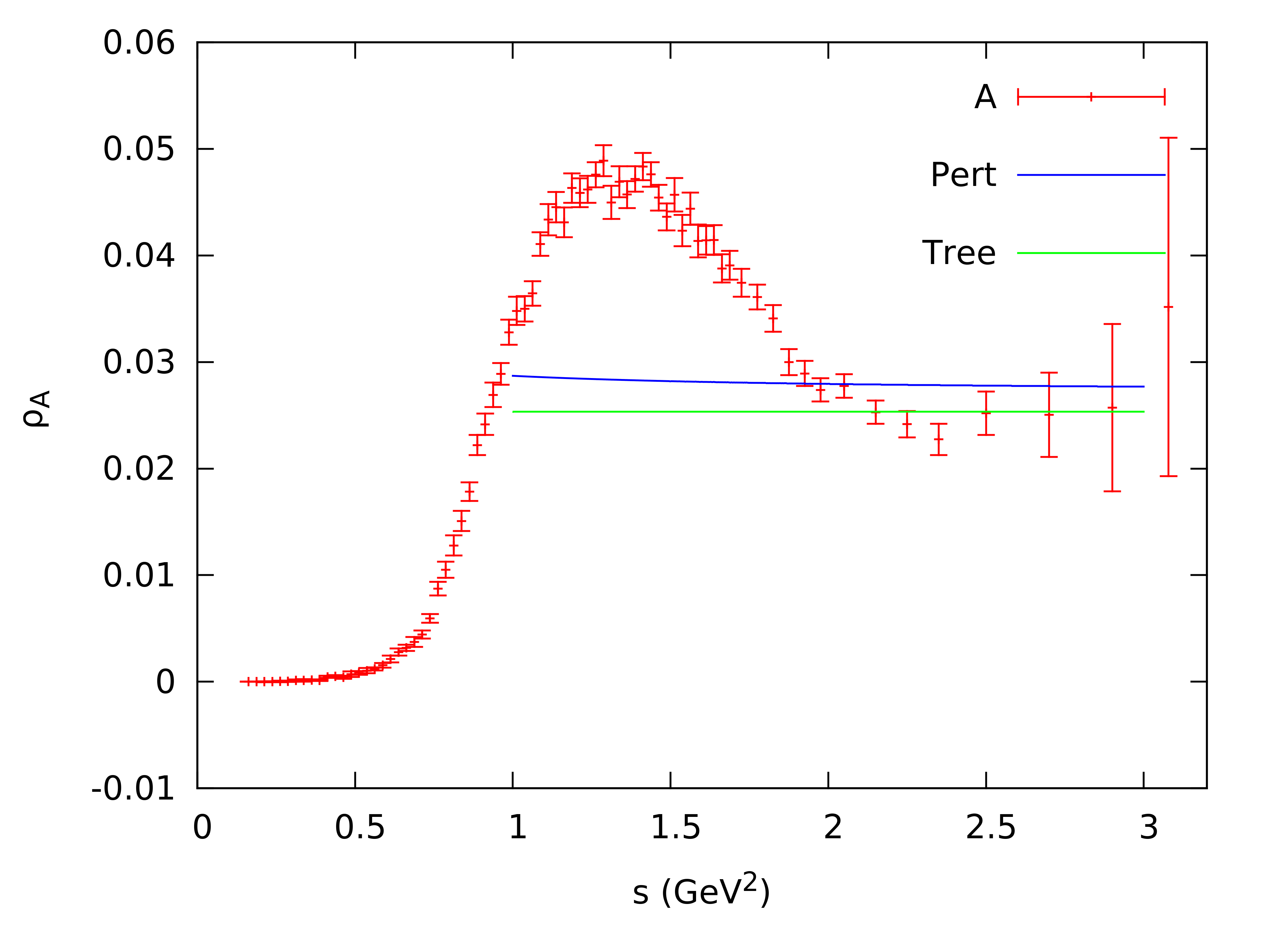}\hskip -.15cm
\includegraphics[width=0.36\textwidth]{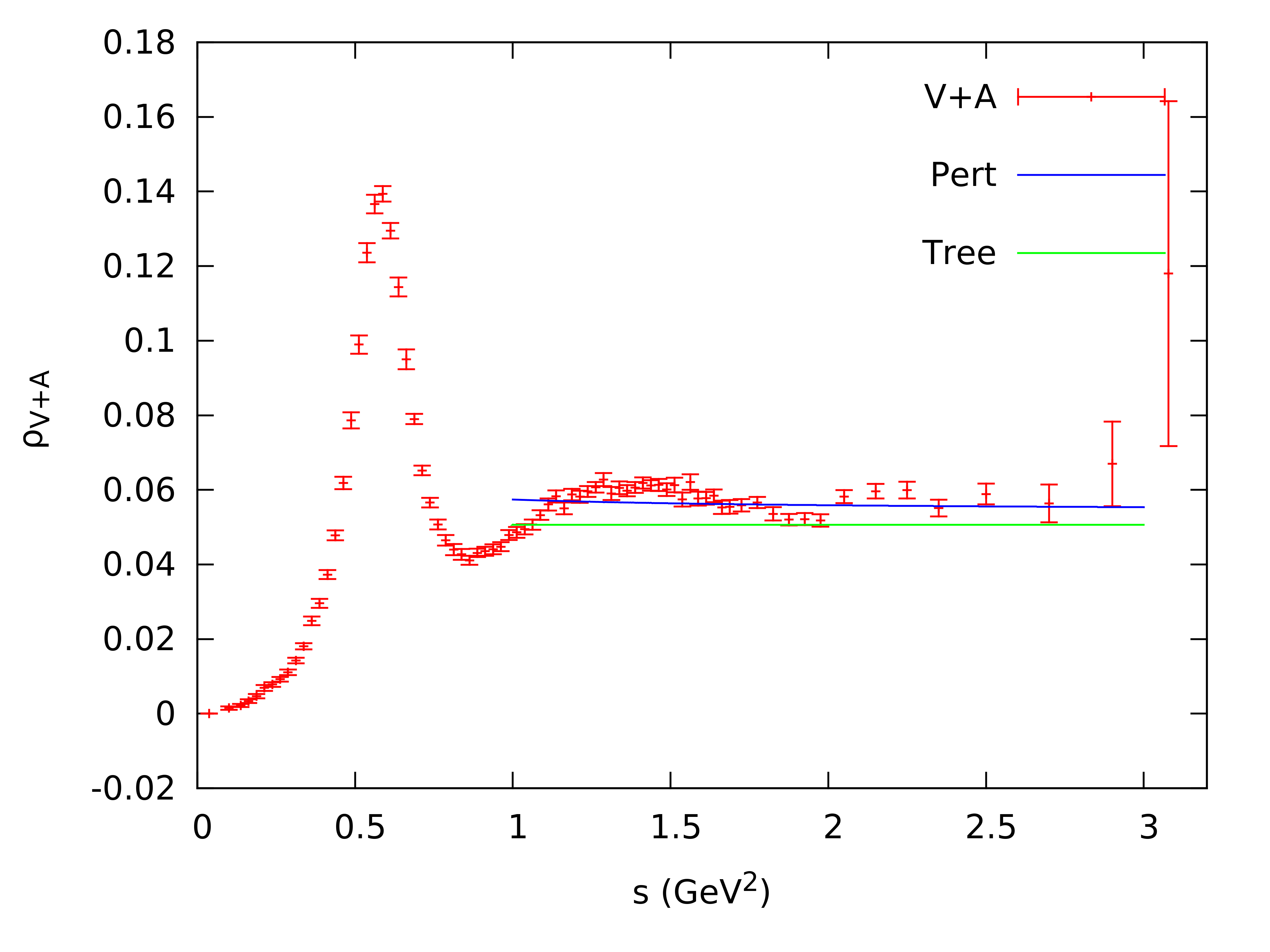}
}
\caption{\label{fig:ALEPHsf}
ALEPH spectral functions for the $V$, $A$ and $V+A$ channels \cite{Davier:2013sfa}.}
\end{figure}
%%%%%%%%%%%%%%%%%%%%%%%%%%%%%%%%%%%%%%%%%%%%%%%%%%%%%%%%%%%%%%%%%%%%%%%%%%%%%%%%%%%%

In Figure~\ref{fig:ALEPHsf} we show the updated spectral functions measured by the ALEPH collaboration \cite{Davier:2013sfa}. Together with the experimental data points,
the figure shows the naive parton-model expectations (horizontal green lines) and the massless perturbative QCD predictions, using $\alpha_s(m_{\tau}^2) = 0.329$ (blue lines).
This comparison shows beautifully, how the data approach the QCD predictions at the highest available energy bins, without any obvious need for non-perturbative corrections
at $s=m_\tau^2$. Resonance structures are clearly visible at lower values of the hadronic invariant mass, specially the prominent $\rho (2\pi)$ and $a_1 (3\pi)$ peaks, but
as $s$ increases the opening of higher-multiplicity hadronic thresholds results in much smoother inclusive distributions, as expected from quark-hadron duality considerations
\cite{Poggio:1975af}. The flattening of the spectral distribution is specially good in the most inclusive channel, $V+A$, where perturbative QCD seems to work even
at $s\sim 1.2\;\mathrm{GeV}^2$, a surprisingly low value. The onset of the asymptotic perturbative QCD behaviour appears obviously later in the semi-inclusive $V$
and $A$ distributions. In the vector case perturbative QCD seems to work well above $s\sim 2\;\mathrm{GeV}^2$, while the axial distribution, which is affected by the
tail of the nearby $a_1$ resonance, only flattens at higher values of $s$.

Unfortunately, the experimental uncertainties on the last two (three in the axial distribution) experimental bins are very large, precisely in the highest energy domain where the short-distance QCD methods become more precise. Additionally, the error correlations among the different bins are quite large, which limits the amount of information that can be extracted from current data.

\section{ALEPH determination of $\boldsymbol{\alpha_s(m_\tau^2)}$}
\label{sec:ALEPH}

The determination of Ref. \cite{Davier:2013sfa} takes $s_{0}=m_{\tau}^{2}$, the maximum energy for which we have data from $\tau$ decays, where the
OPE is supposed to be a better approximation. The weight functions chosen in this analysis have the functional form:
\begin{equation}\label{kl}
\omega_{kl}(s)\; =\;\left(1-\frac{s}{m_{\tau}^{2}}\right)^{2+k}\left( \frac{s}{m^{2}_{\tau}} \right)^{l}\left( 1+\frac{2s}{m_{\tau}^{2}}   \right) \, .
\end{equation}
with $(k,l)=\{(0,0), (1,0), (1,1), (1,2), (1,3)\}$. All these weights have at least a double zero at $s=s_0=m_\tau^2$, to numerically suppress the contributions to the contour integral from the region near the positive real axis, so that duality-violation effects are minimized.
The corresponding moments are normalized with the moment $(k,l)=(0,0)$, in order to reduce experimental correlations and to incorporate in the $V+A$ fit the more precise determination of $R_{\tau,V+A}$
with a universality-improved leptonic branching ratio,
subtracting the small contribution of final states with non-zero strangeness.

From Eq.~(\ref{nonpertu}), we see that the moments $A_{kl,V/A}^{^{\mathrm{ALEPH}}}\equiv A^{\omega_{kl}}_{V/A}(m_{\tau}^{2})$ depend on the following free parameters:
\beqn
A_{00,V/A}^{^{\mathrm{ALEPH}}} &=& A_{00,V/A}^{^{\mathrm{ALEPH}}}(a_{s},\mathcal{O}_{6\, V/A},\mathcal{O}_{8\, V/A}) \, ,
\no\\
A_{10,V/A}^{^{\mathrm{ALEPH}}} &=& A_{10,V/A}^{^{\mathrm{ALEPH}}}(a_{s},\langle a_{s}GG\rangle,\mathcal{O}_{6\, V/A},\mathcal{O}_{8\, V/A},\mathcal{O}_{10\, V/A}) \, ,
\no\\
A_{11,V/A}^{^{\mathrm{ALEPH}}} &=& A_{11,V/A}^{^{\mathrm{ALEPH}}}(a_{s},\langle a_{s}GG\rangle,\mathcal{O}_{6\, V/A},\mathcal{O}_{8\, V/A},\mathcal{O}_{10\, V/A},\mathcal{O}_{12\, V/A}) \, ,
\no\\
A_{12,V/A}^{^{\mathrm{ALEPH}}} &=& A_{12,V/A}^{^{\mathrm{ALEPH}}} (a_{s},\mathcal{O}_{6\, V/A},\mathcal{O}_{8\, V/A},\mathcal{O}_{10\, V/A},\mathcal{O}_{12\, V/A},\mathcal{O}_{14\, V/A}) \, ,
\no\\
A_{13,V/A}^{^{\mathrm{ALEPH}}} &=& A_{13,V/A}^{^{\mathrm{ALEPH}}} (a_{s},\mathcal{O}_{8\, V/A},\mathcal{O}_{10\, V/A},\mathcal{O}_{12\, V/A},\mathcal{O}_{14\, V/A},\mathcal{O}_{16\, V/A}) \, .
\label{eq:ALEPHmom}
\eeqn
Since every new moment adds at least one additional unknown parameter to the previous ones, it seems that no new information is introduced by adding them. This would not be the case if, as it is assumed in Ref.~\cite{Davier:2013sfa}, the contribution of the condensates of dimension $D>8$, $A_{kl,V/A}^{^{\mathrm{ALEPH}}}\bigr|_D\sim \pi \,\mathcal{O}_{D, V/A}/m_{\tau}^{D}$, is negligible. Assuming that, the fit becomes possible and we obtain the results shown in
Table \ref{tablaant}, in good agreement with the ones obtained in Ref.~\cite{Davier:2013sfa}.

%%%%%%%%%%%%%%%%%%%%%%%%%%%%%%%%%%% Table 1 %%%%%%%%%%%%%%%%%%%%%%%%%%%%%%%%%%%%%%%%%
\begin{table}[tb]
\renewcommand\arraystretch{1.2}
\centering
\begin{tabular}{|c|c|c|c|c|} % c|
\hline
Channel  & $\alpha_{s}(m_{\tau}^{2})$ & $<a_{s}GG> $ & $\mathcal{O}_{6}$ & $\mathcal{O}_{8}$
% & \tp{$\chi^2/\mathrm{d.o.f.}$}
\\
&& {\lett{10}{($10^{-3}\;\mathrm{GeV}^4$)}} & {\lett{10}{($10^{-3}\;\mathrm{GeV}^6$)}}  &
{\lett{10}{($ 10^{-3}\;\mathrm{GeV}^8$)}}
% &
\\ \hline

V (FOPT)    & $0.328\,{}^{+0.013}_{-0.007}$ &$8\,{}^{+7}_{-14}$    &$-3.2\,{}^{+0.8}_{-0.5}$ &$5.0\,{}^{+0.4}_{-0.7}$
% & \tp{$0.4$}
\\ \hline
V (CIPT)    & $0.352\,{}^{+0.013}_{-0.011}$ &$-8\,{}^{+7}_{-7}$  & $-3.5\,{}^{+0.3}_{-0.3}$ & $4.9\,{}^{+0.4}_{-0.5}$
% & \tp{$0.8$}
\\ \hline

A (FOPT)   & $0.304\,{}^{+0.010}_{-0.007}$ &$-15\,{}^{+5}_{-8}$ &$4.4\,{}^{+0.5}_{-0.4}$ & $-5.8\,{}^{+0.3}_{-0.4}$
% & \tp{$4.7$}
\\ \hline

A (CIPT)   & $0.320\,{}^{+0.011}_{-0.010}$ &$-25\,{}^{+5}_{-5}$ & $4.3\,{}^{+0.2}_{-0.2}$&$-5.8\,{}^{+0.3}_{-0.3}$
% & \tp{$4.3$}
\\ \hline

V+A (FOPT) & $0.319\,{}^{+0.010}_{-0.006}$ &$-3\,{}^{+6}_{-11}$& $1.3\,{}^{+1.4}_{-0.8}$ &$-0.8\,{}^{+0.4}_{-0.7}$
% & \tp{$2.4$}
\\ \hline

V+A (CIPT) & $0.339\,{}^{+0.011}_{-0.009}$ &$-16\,{}^{+5}_{-5}$ & $0.9\,{}^{+0.3}_{-0.4}$  & $-1.0\,{}^{+0.5}_{-0.7}$
%  & \tp{$1.7$}
\\ \hline
\end{tabular}
\caption{\label{tablaant}
Fitted parameters from the $V$, $A$ and $V+A$ spectral functions, using the $\omega_{kl}(s)$
weight functions in Eq.~\eqn{kl} with $(k,l)=\{(0,0), (1,0), (1,1), (1,2), (1,3)\}$.
The results are given for two different treatments of the perturbative contributions, FOPT and CIPT. The quoted uncertainties include experimental and theoretical errors.}
\end{table}
%%%%%%%%%%%%%%%%%%%%%%%%%%%%%%%%%%%%%%%%%%%%%%%%%%%%%%%%%%%%%%%%%%%%%%%%%%%%%%%%%%%%%%%%

Using the five moments in Eq.~\eqn{eq:ALEPHmom}, we have fitted four parameters: $\alpha_s(m_\tau^2)$, the gluon condensate, $\cO_6$ and $\cO_8$. Table~\ref{tablaant} gives the fitted results, separately for the $V$, $A$ and $V+A$ channels. Moreover, all analyses have been done twice, using the two different treatments of the perturbative QCD series, FOPT and CIPT. As expected, the values of $\alpha_s(m_\tau^2)$ obtained with FOPT are systematically lower than the CIPT results.
All fits result in very precise values of the strong coupling, while rather large errors are obtained for the three vacuum condensates. This just reflects the high sensitivity of the moments to $\alpha_s$, and the minor numerical impact of the non-perturbative power corrections at $s_0=m_\tau^2$.

As it was already observed in the pioneering experimental determinations of $\alpha_s(m_\tau^2)$ \cite{Barate:1998uf}, there is some tension among the parameters fitted from different channels,
which may indicate underestimated uncertainties, either in the experimental data or from non-perturbative effects not yet included in the analysis, such as higher-dimensional condensate contributions or unaccounted duality violations.
On pure theoretical grounds \cite{Braaten:1991qm}, one expects the separate $V$ and $A$ correlators to be more sensitive to higher-dimensional OPE corrections than $V+A$. On the other hand, the recent detailed analysis of the $V-A$ two-point function \cite{Rodriguez-Sanchez:2016jvw} suggests that violations of duality are indeed very efficiently suppressed in pinched moments.

The uncertainties quoted in Table~\ref{tablaant} have been estimated as follows. First we do a direct fit to the data, ignoring theoretical uncertainties, but taking into account all experimental errors and correlations. The statistical quality of these fits, as measured by their $\chi^2/\mathrm{d.o.f.}$, is better when the CIPT approach is used. The vector channel gives quite satisfactory fits ($\chi^2/\mathrm{d.o.f.} = 0.4$ and 0.8 for FOPT and CIPT), while the axial one has a bad $\chi^2/\mathrm{d.o.f.}\sim 4$ for both CIPT and FOPT, being worse in the last case. For $V+A$ one gets $\chi^2/\mathrm{d.o.f.} = 2.4$ (FOPT) and $1.7$ (CIPT). While these $\chi^2$ values do not have a real confidence-level meaning (theoretical errors are not yet included), they do give some indication about the relative quality of the different fits and their expected sensitivity to missing contributions. We then repeat all fits varying the renormalization scale and the fifth-order Adler coefficient within their allowed ranges, $\xi^2\in (0.5,2)$  and $K_5 = 275\pm 400$, and use the variation of the results to estimate the theoretical uncertainties. Theoretical and experimental uncertainties are finally combined in quadrature, giving the final errors indicated in the table. One could instead use the results of the first fit to estimate the theoretical covariance matrices and then perform a full $\chi^2$ minimization, including theoretical and experimental errors together. We have checked that both methods give consistent results, but the first one allows us to better assess the non-linear dependence with $\xi^2$.

Since we have five moments in this fit, we have freedom for fitting also the $D=10$ condensate instead of simply neglecting it. Incorporating $\cO_{10}$ in the global fit, we obtain the results shown in Table~\ref{tablaanto10}. Obviously, we can no-longer estimate the fit quality since there are now as many fitted parameters as moments, but we can still evaluate the statistical errors through the $\chi^2$ function.
One observes that introducing a new degree of freedom results in a sizeable increase of the uncertainties of the fitted condensates. This is not a surprise, given the large correlations present in the data which strongly limit the amount of true information that can be extracted. Adding more free parameters, one is just artificially increasing their possible range of variation by allowing correlated cancellations among them. However, this also puts a word of caution on the reliability of the different extracted parameters. Neglecting $\cO_{10}$, has an important effect on the fitted value of $\cO_8$ which is forced to reabsorb the missing higher-dimensional contributions. Assuming a reasonable convergence of the OPE, the induced uncertainty on $\cO_6$ and $\cO_4$ should be much smaller.

%%%%%%%%%%%%%%%%%%%%%%%%%%%%%%%%%%%%% Table 2 %%%%%%%%%%%%%%%%%%%%%%%%%%%%%%%%%%%%%%%%%%%%
\begin{table}[tb]
\renewcommand\arraystretch{1.2}
\centering
\begin{tabular}{|c|c|c|c|c|c|}
\hline
Channel           & $\alpha_{s}(m_{\tau}^{2})$                       & $<a_{s}GG> $                         & $\mathcal{O}_{6}$                      & $\mathcal{O}_{8}$ & $\mathcal{O}_{10}$
\\
&&  {\lett{10}{($10^{-3}\;\mathrm{GeV}^4$)}}   &
{\lett{10}{($10^{-3}\;\mathrm{GeV}^6$)}}    & {\lett{10}{($10^{-3}\;\mathrm{GeV}^8$)}} &
{\lett{10}{($10^{-3}\;\mathrm{GeV}^{10}$)}}  \\ \hline

V (FOPT)    & $0.320\,{}^{+0.016}_{-0.014}$ &$10\,{}^{+9}_{-17}$    & $-4\,{}^{+3}_{-2}$ & $6\,{}^{+2}_{-2}$  & $-2\,{}^{+5}_{-5}$    \\ \hline

V (CIPT)    & $0.337\,{}^{+0.020}_{-0.019}$ &$-1\,{}^{+10}_{-10}$   & $-5\,{}^{+2}_{-2}$ & $6\,{}^{+2}_{-2}$  &  $-4\,{}^{+4}_{-4}$  \\ \hline

A (FOPT)   & $0.347\,{}^{+0.022}_{-0.021}$ &$-31\,{}^{+16}_{-33}$   & $11\,{}^{+5}_{-4}$  & $-12\,{}^{+4}_{-4}$ & $15\,{}^{+9}_{-9}$  \\ \hline

A (CIPT)   & $0.373\,{}^{+0.029}_{-0.029}$ &$-50\,{}^{+18}_{-16}$   & $10\,{}^{+3}_{-3}$  & $-11\,{}^{+3}_{-3}$ & $14\,{}^{+7}_{-7}$  \\ \hline

V+A (FOPT) & $0.333\,{}^{+0.013}_{-0.012}$ &$-8\,{}^{+10}_{-24}$    & $7\,{}^{+7}_{-4}$  & $-5\,{}^{+4}_{-6}$ &  $12\,{}^{+12}_{-9}$  \\ \hline

V+A (CIPT) & $0.355\,{}^{+0.016}_{-0.015}$ &$-23\,{}^{+10}_{-8}$    & $5\,{}^{+3}_{-3}$  & $-5\pm 3$ &  $10\,{}^{+8}_{-8}$  \\ \hline
\end{tabular}
\caption{\label{tablaanto10}
Fitted parameters from the $V$, $A$ and $V+A$ spectral functions, using the $\omega_{kl}(s)$ weight functions in Eq.~\eqn{kl} with $(k,l)=\{(0,0), (1,0), (1,1), (1,2), (1,3)\}$, but including $\mathcal{O}_{10}$ in the fit. The quoted uncertainties include experimental and theoretical errors.}
\end{table}
%%%%%%%%%%%%%%%%%%%%%%%%%%%%%%%%%%%%%%%%%%%%%%%%%%%%%%%%%%%%%%%%%%%%%%%%%%%%%%%%%%%%%%%%

The strong coupling value turns out to be very stable in all fits because it is basically determined by the lowest moment $(k,l)=(0,0)$, getting only small corrections from $\cO_6$ and $\cO_8$ which are very suppressed by the corresponding $m_\tau^{-6}$ and  $m_\tau^{-8}$ factors. The largest variation on the fitted $\alpha_s$ value occurs in the $A$ channel, the one with the worse $\chi^2$, where the strong coupling increases sizeably when allowing for a non-zero $\cO_{10}$ contribution.
The most reliable results are the ones from the more inclusive $V+A$ channel, which has a smaller $\cO_6$ correction because there is a cancellation between the $V$ and $A$ contributions \cite{Braaten:1991qm}, as corroborated by the results shown in Tables~\ref{tablaant} and~\ref{tablaanto10}.\footnote{In the $V+A$ channel the tables give the fitted values of the sum $\cO_{D,V}+\cO_{D,A}$. The relevant correction is however the average
$\frac{1}{2}\, (\cO_{D,V}+\cO_{D,A})$.}
%%%%%%%%%%%%%%%%%%%%
If we take as reference the $V+A$ value of Table~\ref{tablaant} and we add quadratically the difference with the $V+A$ value of Table~\ref{tablaanto10},
as a conservative estimate of uncertainties for having neglected the higher-dimensional condensates, we obtain:
\be
\ba{c}
\alpha_{s}(m_\tau^2)^{\mathrm{CIPT}} \; =\; 0.339 \,{}^{+\, 0.019}_{-\, 0.017}
\\[7pt]
\alpha_{s}(m_\tau^2)^{\mathrm{FOPT}} \; =\; 0.319 \,{}^{+\, 0.017}_{-\, 0.015}
\ea
\qquad\longrightarrow\qquad
\alpha_{s}(m_\tau^2) \; =\; 0.329 \,{}^{+\, 0.020}_{-\, 0.018} \, .
\label{strongdav}\ee
In order to quote a final value, we have averaged the CIPT and FOPT results, keeping conservatively the minimum uncertainty and adding quadratically half their difference as an additional systematic error.

The sensitivity to the $D=4$ OPE correction is very low and, comparing the two tables, one observes a strong correlation with the higher-dimensional corrections. The fitted central values suggest an unphysical negative value for the gluon condensate but the uncertainties are too large to be significant. Applying the same procedure as before, we get the averaged value:
\bel{eq:ALEPH-GluonCond}
\langle \mbox{$\frac{\alpha_{s}}{\pi}$}\, GG\rangle\; =\;\left( -9\, {}^{+\, 10}_{-\, 11}\right)\cdot  10^{-3}\; \mathrm{GeV}^{4}\, ,
\ee
which is consistent with zero and, taking into account the large errors, still compatible with the usually quoted range in Eq.~\eqn{eq:GluonCond}.

To test the stability of these results, we have repeated all fits taking away from the weight functions the factor $(1 + 2 s/m_\tau^2)$ in Eq.~\eqn{kl}.
This eliminates the highest-dimensional condensate contribution to each moment, at the price of making $A_{00,V/A}$ sensitive to the gluon condensate. Although one
also loses the additional experimental information from the $\tau$ lifetime, the new weights are less sensitive to the higher energy range of the experimental distribution
which, as shown in Figure~\ref{fig:ALEPHsf}, is poorly-defined. The fitted results for $\alpha_s$ and the vacuum condensates, obtained in this way, are shown
in Tables~\ref{tablaantbis} (taking $\cO_{10}=0$) and~\ref{tablaanto10bis} (including $\cO_{10}$ in the fit). They are in complete agreement with the results of the previous fits, given
in Tables~\ref{tablaant} and \ref{tablaanto10}. However, all $\chi^2/\mathrm{d.o.f.}$ turn out now to be smaller than one.

%%%%%%%%%%%%%%%%%%%%%%%%%%%%%%%%%%%% Table 1bis %%%%%%%%%%%%%%%%%%%%%%%%%%%%%%%%%%%%%%%%%%
\begin{table}[tb]
\renewcommand\arraystretch{1.2}
\centering
\begin{tabular}{|c|c|c|c|c|}
\hline
Channel  & $\alpha_{s}(m_{\tau}^{2})$ & $<a_{s}GG> $ & $\mathcal{O}_{6}$ & $\mathcal{O}_{8}$
\\
&& {\lett{10}{($10^{-3}\;\mathrm{GeV}^4$)}} & {\lett{10}{($10^{-3}\;\mathrm{GeV}^6$)}}  &
{\lett{10}{($ 10^{-3}\;\mathrm{GeV}^8$)}}
\\ \hline

V (FOPT)    & $0.331\,{}^{+0.012}_{-0.006}$ &$-5\,{}^{+7}_{-14}$    &$-2.4\,{}^{+0.9}_{-0.5}$ &$2.4\,{}^{+0.3}_{-0.5}$
\\ \hline
V (CIPT)    & $0.356\,{}^{+0.012}_{-0.009}$ &$-22\,{}^{+7}_{-8}$  & $-2.8\,{}^{+0.2}_{-0.1}$ & $2.1\,{}^{+0.3}_{-1.1}$
\\ \hline

A (FOPT)   & $0.305\,{}^{+0.009}_{-0.005}$ &$-5\,{}^{+4}_{-8}$ &$3.9\,{}^{+0.5}_{-0.3}$ & $-3.2\,{}^{+0.2}_{-0.3}$
\\ \hline

A (CIPT)  & $0.320\,{}^{+0.010}_{-0.007}$ &$-15\,{}^{+4}_{-4}$ & $3.8\,{}^{+0.1}_{-0.1}$&$-3.3\,{}^{+0.1}_{-0.2}$
\\ \hline

V+A (FOPT) & $0.319\,{}^{+0.010}_{-0.005}$ &$-3\,{}^{+5}_{-11}$& $1.5\,{}^{+1.3}_{-0.7}$ &$-0.8\,{}^{+0.4}_{-0.8}$
\\ \hline

V+A (CIPT) & $0.338\,{}^{+0.010}_{-0.008}$ &$-16\,{}^{+5}_{-5}$ & $1.1\,{}^{+0.2}_{-0.2}$  & $-1.0\,{}^{+0.4}_{-1.0}$
\\ \hline
\end{tabular}
\caption{\label{tablaantbis}
Fitted parameters from the $V$, $A$ and $V+A$ spectral functions, using the
same weights as in Table~\ref{tablaant} but taking away the factor $(1 + 2 s/m_\tau^2)$. The quoted uncertainties include experimental and theoretical errors.}
\end{table}

%%%%%%%%%%%%%%%%%%%%%%%%%%%%%%%%%%%%%%%%%%%%%%%%%%%%%%%%%%%%%%%%%%%%%%%%%%%%%%%%%%%%%%%%

%%%%%%%%%%%%%%%%%%%%%%%%%%%%%%%%%%%%% Table 2bis %%%%%%%%%%%%%%%%%%%%%%%%%%%%%%%%%%%%%%%%%

\begin{table}[tb]
\renewcommand\arraystretch{1.2}
\centering
\begin{tabular}{|c|c|c|c|c|c|}
\hline
Channel           & $\alpha_{s}(m_{\tau}^{2})$                       & $<a_{s}GG> $                         & $\mathcal{O}_{6}$                      & $\mathcal{O}_{8}$ & $\mathcal{O}_{10}$
\\
&&  {\lett{10}{($10^{-3}\;\mathrm{GeV}^4$)}}   &
{\lett{10}{($10^{-3}\;\mathrm{GeV}^6$)}}    & {\lett{10}{($10^{-3}\;\mathrm{GeV}^8$)}} &
{\lett{10}{($10^{-3}\;\mathrm{GeV}^{10}$)}}  \\ \hline
V (FOPT)    & $0.318\,{}^{+0.043}_{-0.042}$ &$10\,{}^{+46}_{-48}$    &$-5\,{}^{+7}_{-7}$ &$6\,{}^{+11}_{-11}$&$-4\,{}^{+12}_{-12}$
\\ \hline
V (CIPT)    & $0.336\,{}^{+0.056}_{-0.055}$ &$-1\,{}^{+54}_{-54}$  & $-5\,{}^{+7}_{-7}$ & $6\,{}^{+11}_{-11}$&$-4\,{}^{+12}_{-11}$
\\ \hline

A (FOPT)   & $0.336\,{}^{+0.052}_{-0.051}$ &$-39\,{}^{+62}_{-67}$ &$9\,{}^{+10}_{-10}$ & $-12\,{}^{+15}_{-15}$&$9\,{}^{+16}_{-16}$
\\ \hline

A (CIPT)  & $0.360\,{}^{+0.064}_{-0.064}$ &$-54\,{}^{+68}_{-68}$ & $8\,{}^{+7}_{-7}$&$-11\,{}^{+13}_{-13}$&$8\,{}^{+13}_{-13}$
\\ \hline

V+A (FOPT) & $0.327\,{}^{+0.030}_{-0.029}$ &$-13\,{}^{+35}_{-39}$& $4\,{}^{+11}_{-11}$ &$-6\,{}^{+17}_{-17}$&$5\,{}^{+18}_{-18}$
\\ \hline

V+A (CIPT) & $0.348\,{}^{+0.040}_{-0.039}$ &$-26\,{}^{+41}_{-41}$ & $3\,{}^{+9}_{-9}$  & $-5\,{}^{+16}_{-16}$&$4\,{}^{+17}_{-17}$
\\ \hline
\end{tabular}
\caption{\label{tablaanto10bis}
Fitted parameters from the $V$, $A$ and $V+A$ spectral functions, using the
same weights as in Table~\ref{tablaantbis}
and including $\mathcal{O}_{10}$ in the fit. The quoted uncertainties include experimental and theoretical errors.}
\end{table}

%%%%%%%%%%%%%%%%%%%%%%%%%%%%%%%%%%%%%%%%%%%%%%%%%%%%%%%%%%%%%%%%%%%%%%%%%%%%%%%%%%%%%%%%

Thus, it appears that the quoted uncertainties take properly into account any possible effects from missing contributions. The central values of the fitted parameters are very stable, specially the strong coupling, and the sensitivity of $\alpha_s$, $\cO_4$ and $\cO_6$ to vacuum condensates with $D>10$ is indeed negligible. From the results in Tables~\ref{tablaantbis} and \ref{tablaanto10bis}, applying the same procedure as before, we get the averages:
\be
\ba{c}
\alpha_{s}(m_\tau^2)^{\mathrm{CIPT}} \; =\; 0.338 \,{}^{+\, 0.014}_{-\, 0.012}
\\[7pt]
\alpha_{s}(m_\tau^2)^{\mathrm{FOPT}} \; =\; 0.319 \,{}^{+\, 0.013}_{-\, 0.010}
\ea
\qquad\longrightarrow\qquad
\alpha_{s}(m_\tau^2) \; =\; 0.329 \,{}^{+\, 0.016}_{-\, 0.014} \, ,\quad
\label{strongdavbis}
\ee
and
\be
\langle \mbox{$\frac{\alpha_{s}}{\pi}$}\, GG\rangle\; =\;\left( -10\, \pm 13\right)\cdot  10^{-3}\; \mathrm{GeV}^{4}\, .\quad
\ee
These numbers are in excellent agreement with the previous determination in Eqs.~\eqn{strongdav} and \eqn{eq:ALEPH-GluonCond}, performed with the ALEPH moments, and
the values obtained for $\alpha_s(m_\tau^2)$ are even more accurate.

\section{Optimal moments}
\label{sec:optimal}

The moments $\omega_{kl}(s)$ used in the previous analyses were suggested in Ref.~\cite{LeDiberder:1992zhd} as a way to minimize the large statistical and systematic uncertainties of the initial LEP data. All of them incorporate the kinematical factor $\omega_{00}(s)$, present in Eq.~\eqn{eq:RtauSpectral}, allowing for a direct use of the measured invariant-mass distribution. This makes unnecessary to reconstruct the spectral functions, dividing the raw data by $\omega_{00}(s)$,  which enhances the systematically- and statistically-limited tail of the $s$ distribution. On the negative side, these moments involve higher-dimensional condensates and the experimental precision deteriorates with increasing values of $k$ and/or $l$. Nowadays, since we have well-determined and quite precise spectral functions,\footnote{Note, however, that the large uncertainties observed in the higher-energy bins of the spectral functions in Figure~\ref{fig:ALEPHsf} originate in the kinematical factor $\omega_{00}(s)$ which suppresses the end-point of the $\tau$ decay distribution.}
based on the full LEP data sample, it is possible to investigate whether there are better moments, more suitable for a precise QCD analysis.

In order to reduce duality violations, we could consider the simplest $n$-pinched weight functions
\bel{eq:omega-n0}
\omega^{(n,0)}(x)\; =\; (1-x)^n\; =\;\sum_{k=0}^n \, (-1)^k\, \left( \ba{c} n \\ k \ea\right)\, x^k\, ,
\ee
with $x=s/s_0$. However, the moments generated by these weights get non-perturbative corrections from all condensates with dimension $D\le 2 (n+1)$. We would like to become sensitive to the lowest-dimensional condensates without too much contamination from higher-order terms in the OPE. It is possible to build a family of weight functions which project on one single condensate contribution, while still having a zero at $x=1$:
\bel{eq:omega-1n}
\omega^{(1,n)}(x)\; =\; 1-x^{n+1}\; =\; (1-x)\;\,\sum_{k=0}^{n}\, x^k\, .
\ee
The corresponding moments are only sensitive to $\cO_{2 (n+2)}$.
As an intermediate case, the following weight functions have a double pinch and generate moments with only two condensate contributions, $\cO_{2 (n+2)}$ and $\cO_{2 (n+3)}$:
\bel{eq:omega-2n}
\omega^{(2,n)}(x)\; =\; (1-x)^2\;\,\sum_{k=0}^{n}\, (k+1)\, x^k \; =\;
1-(n + 2)\, x^{n + 1} + (n + 1)\, x^{n + 2}\, .
\ee
Notice that $\omega^{(2,1)}(x) = \omega_{00}(x)$, the lowest moment used in the ALEPH-like analysis.

Since every moment % $A^{(n,m)}(s_0)\equiv A^{\omega^{(n,m)}}\! (s_0)$ 
\be
A^{(n,m)}_{V/A}(s_0)\;\equiv\; A^{\omega^{(n,m)}}_{V/A} (s_0)\; =\;
\int^{s_{0}}_{s_{\mathrm{th}}} \frac{ds}{s_{0}}\;\omega^{(n,m)}(s)\, \ImPi_{V/A}(s)
\ee
introduces a new condensate correction, it is not possible to perform a fully complete fit of $\alpha_s$ and some power corrections only using a few single-pinched or doubly-pinched moments. Nevertheless, one can still make some approximations and a few consistency tests, which we attempt next.

\subsection{OPE corrections neglected}

Neglecting all OPE corrections, one can directly extract $\alpha_s(m_\tau^2)$ from a single $A^{(n,m)}(s_0)$ moment. Comparing the values extracted with different choices of $(n,m)$, one can then assess the size of the neglected contributions. For instance, $A^{(0,0)}(s_0)$ does not get any OPE correction, but it is not protected against duality-violation effects. On the other extreme, $A^{(2,3)}(s_0)$ is well protected by a double pinch and gets inverse power corrections with $D=10$ and $12$.

%%%%%%%%%%%%%%%%%%%%%%%%%%%%%%%%%%%%% Table 5 %%%%%%%%%%%%%%%%%%%%%%%%%%%%%%%%%%%%%%%%%%%
\begin{table}[t]
\renewcommand\arraystretch{1.2}
\centering
\begin{tabular}{|c|c|c||c|c|c|}
\hline
Moment & \multicolumn{2}{|c||}{$\alpha_s(m_\tau^2)$} &
Moment & \multicolumn{2}{|c|}{$\alpha_s(m_\tau^2)$}
\\ \cline{2-3} \cline{5-6}  $(n,m)$ & FOPT    & CIPT   & $(n,m)$ & FOPT    & CIPT
\\ \hline
(1,0) & $0.315 \, {}^{+0.012}_{-0.007}$ & $0.327 \, {}^{+0.012}_{-0.009}$ &
(2,0) & $0.311 \, {}^{+0.015}_{-0.011}$ & $0.314 \, {}^{+0.013}_{-0.009}$
\\ \hline
(1,1) & $0.319 \, {}^{+0.010}_{-0.006}$ & $0.340 \, {}^{+0.011}_{-0.009}$ &
(2,1) & $0.311 \, {}^{+0.011}_{-0.006}$ & $0.333 \, {}^{+0.009}_{-0.007}$
\\ \hline
(1,2) & $0.322 \, {}^{+0.010}_{-0.008}$ & $0.343 \, {}^{+0.012}_{-0.010}$ &
(2,2) & $0.316 \, {}^{+0.010}_{-0.005}$ & $0.336 \, {}^{+0.011}_{-0.009}$
\\ \hline
(1,3) & $0.324 \, {}^{+0.011}_{-0.010}$ & $0.345 \, {}^{+0.013}_{-0.011}$ &
(2,3) & $0.318 \, {}^{+0.010}_{-0.006}$ & $0.339 \, {}^{+0.011}_{-0.008}$
\\ \hline
(1,4) & $0.326 \, {}^{+0.011}_{-0.011}$ & $0.347 \, {}^{+0.013}_{-0.012}$ &
(2,4) & $0.319 \, {}^{+0.009}_{-0.007}$ & $0.340 \, {}^{+0.011}_{-0.009}$
\\ \hline
(1,5) & $0.327 \, {}^{+0.015}_{-0.013}$ & $0.348 \, {}^{+0.014}_{-0.012}$ &
(2,5) & $0.320 \, {}^{+0.010}_{-0.008}$ & $0.341 \, {}^{+0.011}_{-0.009}$
\\ \hline
\end{tabular}
\caption{\label{noOPE} Values of the strong coupling extracted from a single $A^{(n,m)}(s_0)$ moment of the $V+A$ distribution, at $s_0 = 2.8\;\mathrm{GeV}^2$, neglecting all non-perturbative corrections.}
\end{table}
%%%%%%%%%%%%%%%%%%%%%%%%%%%%%%%%%%%%%%%%%%%%%%%%%%%%%%%%%%%%%%%%%%%%%%%%%%%%%%%%%%%%%%%%

Since we are going to test also some non-pinched weights, we take $s_0 = 2.8\;\mathrm{GeV}^2$ as reference point, so that we avoid the problems associated with the last two experimental bins.
The results of this exercise are shown in Table~\ref{noOPE}, for all $A^{(n,m)}(s_0)$ moments of the $V+A$ distribution with $n=1,2$ and $0\le m\le 5$. In all cases, the fitted values are well within the error ranges of our determinations in Eq.~\eqn{strongdav}. Notice the good stability displayed by the results from the moments $(2,m\ge 2)$, suggesting that condensates with $D>6$ play indeed a very minor role. A similar behaviour is observed in the moments $(1,m)$ which, however, result in slightly larger values of the strong coupling for all values of the parameter $m$.

A clean test of the magnitude of duality violation effects is provided by the $A^{(0,0)}(s_0)$ moment,
where OPE corrections are absent. One finds in this case $\alpha_s(m_\tau^2)^{\mathrm{CIPT}} = 0.352\,{}^{+0.023}_{-0.022}$ and $\alpha_s(m_\tau^2)^{\mathrm{FOPT}} = 0.333\,{}^{+0.024}_{-0.018}$, in agreement with Eq.~\eqn{strongdav} even being a non-protected moment.

\subsection{Combined fit to $A^{(n,0)}(s_0)$ moments}

Using all $A^{(n,0)}(s_0)$ moments with $n\le N$, one can determine $\alpha_s(m_\tau^2)$ and $\cO_{D\le 2N}$, neglecting the $\cO_{2(N+1)}$ contribution to the last moment. The strong coupling is mostly affected
by the non-protected $(0,0)$ moment, although the effects of duality violation get modulated by the higher moments which do have pinching protection. We show in Table~\ref{n0Fit} the results from global fits to
the $(n,0)$ moments with $0\le n\le 3$, taking $\cO_8=0$. Again, the agreement with Eq.~\eqn{strongdav} is remarkable.

%%%%%%%%%%%%%%%%%%%%%%%%%%%%%%%%%%%%% Table 6 %%%%%%%%%%%%%%%%%%%%%%%%%%%%%%%%%%%%%%%%%%%
\begin{table}[ht]
\renewcommand\arraystretch{1.2}
\centering
\begin{tabular}{|c|c|c|c|}
\hline
Channel    & $\alpha_s(m_{\tau}^{2})$    & $<a_{s}GG> $     & $\mathcal{O}_{6}$
\\
&&  {\lett{10}{($10^{-3}\;\mathrm{GeV}^4$)}} & {\lett{10}{($10^{-3}\;\mathrm{GeV}^6$)}}
\\ \hline
V (FOPT) & $0.310 \, {}^{+0.010}_{-0.005}$ & $11 \, {}^{+7}_{-12}$ & $-3.8 \, {}^{+0.9}_{-0.6}$        \\ \hline
V (CIPT)   & $0.328 \, {}^{+0.011}_{-0.007}$ & $2 \, {}^{+7}_{-7}$  & $-4.1 \, {}^{+0.4}_{-0.5}$        \\ \hline
A (FOPT)   & $0.328\, {}^{+0.011}_{-0.007}$  & $-28 \, {}^{+9}_{-20}$  & $5.8 \, {}^{+1.3}_{-0.7}$
\\ \hline
A (CIPT) & $0.352 \, {}^{+0.012}_{-0.008}$ & $-41 \, {}^{+8}_{-7}$ & $5.3 \, {}^{+0.4}_{-0.5}$
\\ \hline
V+A (FOPT)   & $0.319\, {}^{+0.010}_{-0.007}$  & $-7 \, {}^{+7}_{-16}$  & $2.0 \, {}^{+2.0}_{-1.1}$
\\ \hline
V+A (CIPT)   & $0.340\, {}^{+0.011}_{-0.009}$  & $-18 \, {}^{+6}_{-5}$  & $1.2 \, {}^{+0.5}_{-0.8}$
\\ \hline
\end{tabular}
\caption{\label{n0Fit} Global fit to the $A^{(n,0)}(s_0)$ moments with $0\le n\le 3$, taking $\cO_8=0$.}
\end{table}
%%%%%%%%%%%%%%%%%%%%%%%%%%%%%%%%%%%%%%%%%%%%%%%%%%%%%%%%%%%%%%%%%%%%%%%%%%%%%%%%%%%%%%%%

\subsection{Combined fit to $A^{(2,m)}(m_{\tau}^{2})$ moments}

From the moments $A^{(2,m)}(m_{\tau}^{2})$ with $1\le m\le N$, one can determine $\alpha_s(m_\tau^2)$ and $\cO_{D\le 2(N+1)}$, neglecting the $\cO_{2(N+2)}$ and $\cO_{2(N+3)}$ contributions to the last two moments. In this case, the whole set of selected moments is well protected from duality violation effects by a double pinch. The results obtained with $N=5$ are shown in Tables~\ref{2nFit} and \ref{2nFitbis}. The first one is a global fit, assuming $\cO_{12}=\cO_{14}=\cO_{16}=0$, while
Table~\ref{2nFitbis} only assumes $\cO_{14}=\cO_{16}=0$ and has then as many fitted parameters as moments.

%%%%%%%%%%%%%%%%%%%%%%%%%%%%%%%%%%%%% Table 6 %%%%%%%%%%%%%%%%%%%%%%%%%%%%%%%%%%%%%%%%%%%
\begin{table}[tb]
\renewcommand\arraystretch{1.2}
\centering
\begin{tabular}{|c|c|c|c|c|}
\hline
Channel    & $\alpha_s(m_{\tau}^{2})$    & $\mathcal{O}_{6}$     & $\mathcal{O}_{8}$ &
$\mathcal{O}_{10}$
\\
&& {\lett{10}{($10^{-3}\;\mathrm{GeV}^6$)}} & {\lett{10}{($10^{-3}\;\mathrm{GeV}^8$)}} &
{\lett{10}{($10^{-3}\;\mathrm{GeV}^{10}$)}}
\\ \hline
V (FOPT)   & $0.315 \, {}^{+0.011}_{-0.007}$ & $-5.2 \, {}^{+0.8}_{-0.5}$ & $6.7 \, {}^{+0.5}_{-0.7}$  & $-4.5 \, {}^{+0.4}_{-0.3}$
\\ \hline
V (CIPT)   & $0.334\, {}^{+0.012}_{-0.009}$  & $-5.3 \, {}^{+0.3}_{-0.4}$  & $6.7 \, {}^{+0.4}_{-0.4}$  & $-4.6 \, {}^{+0.2}_{-0.3}$
\\ \hline
A (FOPT)   & $0.318\, {}^{+0.011}_{-0.007}$  & $6.5 \, {}^{+0.9}_{-0.5}$  & $-7.6 \, {}^{+0.5}_{-0.4}$ & $4.9 \, {}^{+0.4}_{-0.3}$
\\ \hline
A (CIPT)   & $0.338\, {}^{+0.013}_{-0.012}$  & $6.3 \, {}^{+0.3}_{-0.4}$  & $-7.7 \, {}^{+0.4}_{-0.3}$ & $4.8 \, {}^{+0.2}_{-0.4}$
\\ \hline
V+A (FOPT) & $0.317 \, {}^{+0.010}_{-0.005}$ & $1.3 \, {}^{+1.7}_{-1.0}$  & $-0.9 \, {}^{+0.9}_{-1.4}$ & $0.3 \, {}^{+0.8}_{-0.5}$
\\ \hline
V+A (CIPT) & $0.336 \, {}^{+0.011}_{-0.009}$ & $0.9 \, {}^{+0.4}_{-0.7}$  & $-0.9 \, {}^{+0.5}_{-0.5}$ & $0.1 \, {}^{+0.3}_{-0.7}$
\\ \hline
\end{tabular}
\caption{\label{2nFit} Global fit to the $A^{(2,m)}(s_0)$ moments with $1\le m\le 5$, taking $\cO_{12}=\cO_{14}=\cO_{16}=0$.}
%\end{table}
%%%%%%%%%%%%%%%%%%%%%%%%%%%%%%%%%%%%%%%%%%%%%%%%%%%%%%%%%%%%%%%%%%%%%%%%%%%%%%%%%%%%%%%%
%
\vskip .5cm
%
%%%%%%%%%%%%%%%%%%%%%%%%%%%%%%%%%%%%% Table 7 %%%%%%%%%%%%%%%%%%%%%%%%%%%%%%%%%%%%%%%%%%%
%\begin{table}[tb]
%\renewcommand\arraystretch{1.2}
%\centering
\begin{tabular}{|c|c|c|c|c|c|}
\hline
Channel    & $\alpha_s(m_{\tau}^{2})$    & $\mathcal{O}_{6}$     & $\mathcal{O}_{8}$ &
$\mathcal{O}_{10}$     & $\mathcal{O}_{12}$
\\
&& {\lett{10}{($10^{-3}\;\mathrm{GeV}^6$)}} & {\lett{10}{($10^{-3}\;\mathrm{GeV}^8$)}} &
{\lett{10}{($10^{-3}\;\mathrm{GeV}^{10}$)}} & {\lett{10}{($10^{-3}\;\mathrm{GeV}^{12}$)}}
\\ \hline
V (FOPT)   & $0.318 \, {}^{+0.013}_{-0.012}$ & $-5 \, {}^{+3}_{-2}$ & $6 \, {}^{+3}_{-4}$   & $-3 \, {}^{+5}_{-5}$   & $-2 \, {}^{+5}_{-5}$
\\ \hline
V (CIPT)   & $0.336 \, {}^{+0.017}_{-0.016}$ & $-5\, {}^{+2}_{-2}$  & $6 \, {}^{+3}_{-3}$   & $-4 \, {}^{+4}_{-4}$  & $-1\,  {}^{+4}_{-4}$
\\ \hline
A (FOPT)   & $0.339\, {}^{+0.018}_{-0.017}$  & $11 \, {}^{+4}_{-3}$ & $-15 \, {}^{+5}_{-5}$ & $16 \, {}^{+9}_{-8}$  & $-11 \, {}^{+8}_{-8}$
\\ \hline
A (CIPT)   & $0.364\, {}^{+0.024}_{-0.022}$  & $10 \, {}^{+2}_{-2}$ & $-14 \, {}^{+5}_{-5}$ & $14 \, {}^{+7}_{-7}$  & $-9 \, {}^{+7}_{-7}$
\\ \hline
V+A (FOPT) & $0.329 \, {}^{+0.012}_{-0.011}$ & $6 \, {}^{+6}_{-4}$  & $-9 \, {}^{+7}_{-9}$  & $13 \, {}^{+12}_{-10}$ & $-12 \, {}^{+9}_{-11}$
\\ \hline
V+A (CIPT) & $0.349 \, {}^{+0.016}_{-0.014}$ & $4 \, {}^{+3}_{-3}$  & $-8 \, {}^{+6}_{-6}$  & $10 \, {}^{+8}_{-8}$   & $-10 \, {}^{+8}_{-8}$
\\ \hline
\end{tabular}
\caption{\label{2nFitbis} Global fit to the $A^{(2,m)}(s_0)$ moments with $1\le m\le 5$, taking $\cO_{14}=\cO_{16}=0$.}
\end{table}
%%%%%%%%%%%%%%%%%%%%%%%%%%%%%%%%%%%%%%%%%%%%%%%%%%%%%%%%%%%%%%%%%%%%%%%%%%%%%%%%%%%%%%%%

Since the previous tests suggested that the neglected higher-dimensional condensates do not play any significant role on the fitted value of the strong coupling, the results of these fits should be very reliable, specially in the $V+A$ case. Of course, adding one more parameter to the fit allows for a wider range of variation, increasing the fitted errors, which explains the differences between the two tables. The small sensitivity to the vacuum condensates is reflected in their large statistical uncertainties, specially in Table~\ref{2nFitbis}. Their fitted values agree with the results quoted in Tables~\ref{tablaant} and \ref{tablaanto10}. On the other side, the determinations of the strong coupling are quite precise and in excellent agreement with the fits performed in section~\ref{sec:ALEPH}.
Notice the very good stability displayed in Table~\ref{2nFit}, where similar central values for $\alpha_{s}(m_\tau^2)$ are obtained from the $V$, $A$ and $V+A$ channels.

Taking again as reference the results from the $V+A$ fits in Table~\ref{2nFit}, and adding quadratically the differences between the two tables, as a conservative estimate of the uncertainties from neglected higher-dimensional condensates, we get
\be
\ba{c}
\alpha_{s}(m_\tau^2)^{\mathrm{CIPT}} \; =\; 0.336 \,{}^{+\, 0.018}_{-\, 0.016}
\\[7pt]
\alpha_{s}(m_\tau^2)^{\mathrm{FOPT}} \; =\; 0.317 \,{}^{+\, 0.015}_{-\, 0.013}
\ea
\qquad\longrightarrow\qquad
\alpha_{s}(m_\tau^2) \; =\; 0.326 \,{}^{+\, 0.018}_{-\, 0.016} \, .
\label{strong2mMom}\ee
Once more, we get results in perfect agreement with the values of the strong coupling obtained in Eqs.~\eqn{strongdav} and~\eqn{strongdavbis}. Given the different sensitivity to higher-dimensional condensates
of the moments used in each approach, and the many tests we have performed showing the negligible numerical impact of higher-order power corrections, our determination of $\alpha_s(m_\tau^2)$ appears to be very
solid and much more stable than what one could expect from the quoted uncertainties, indicating that our errors are indeed conservative.

\section{Including information from the $\mathbf{s_0}$ dependence}
\label{sec:improvements}

%%%%%%%%%%%%%%%%%%%%%%%%%%%%%%%%% Figure %%%%%%%%%%%%%%%%%%%%%%%%%%%%%%%%%%%%%%%%%%%
\begin{figure}[tb]\centering
\includegraphics[width=0.46\textwidth]{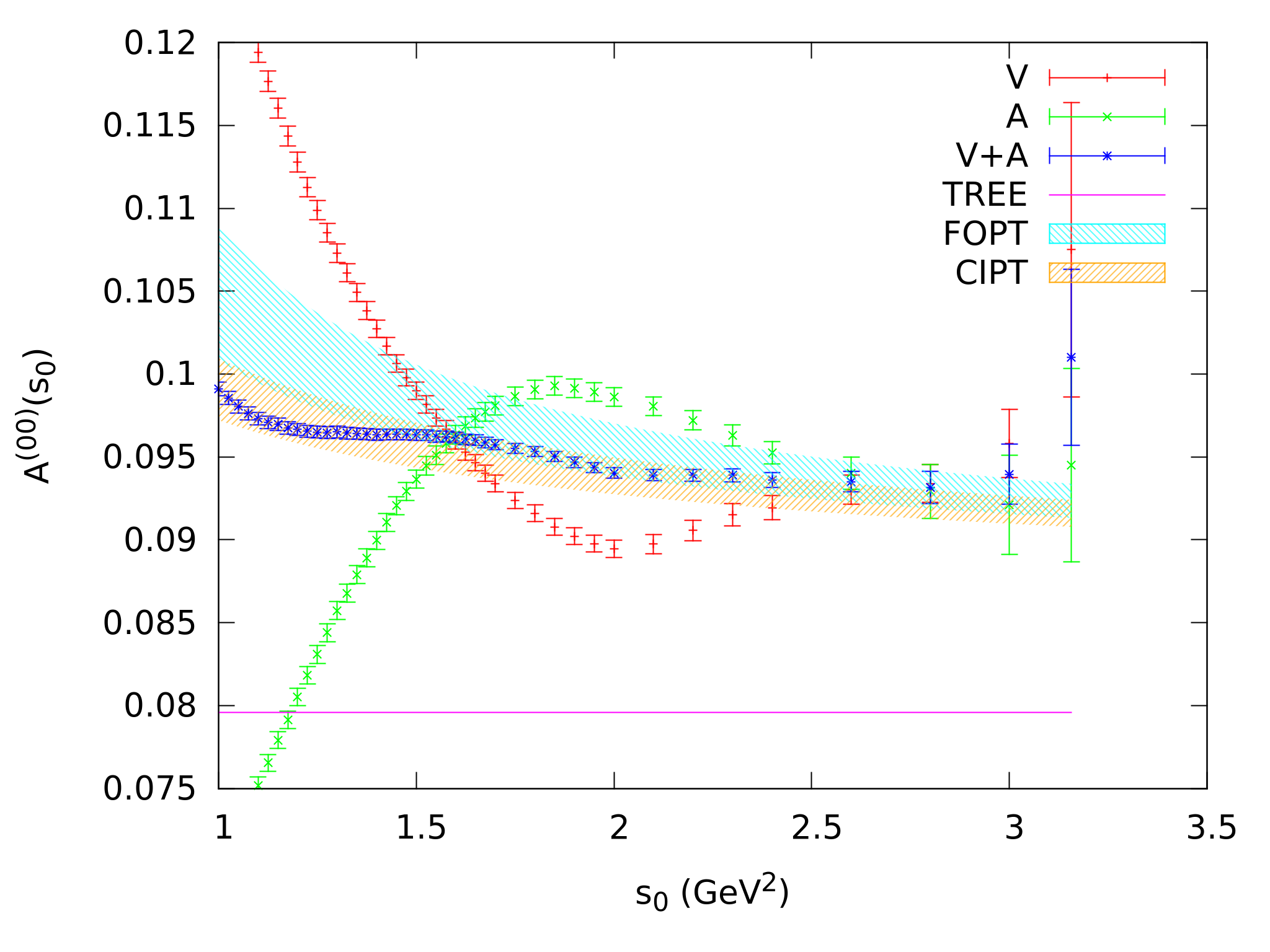}\hskip .5cm
\includegraphics[width=0.46\textwidth]{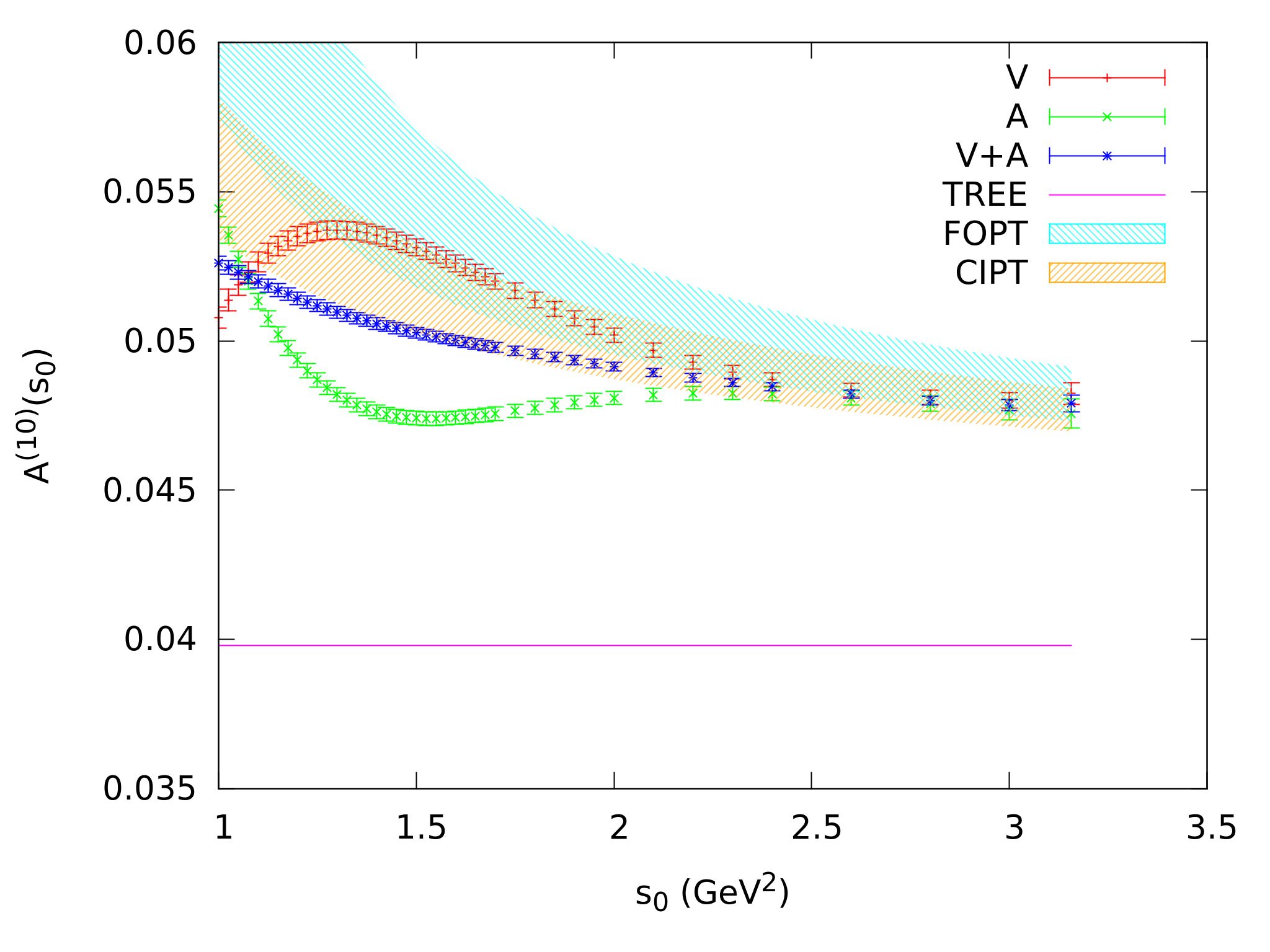}\\
\includegraphics[width=0.46\textwidth]{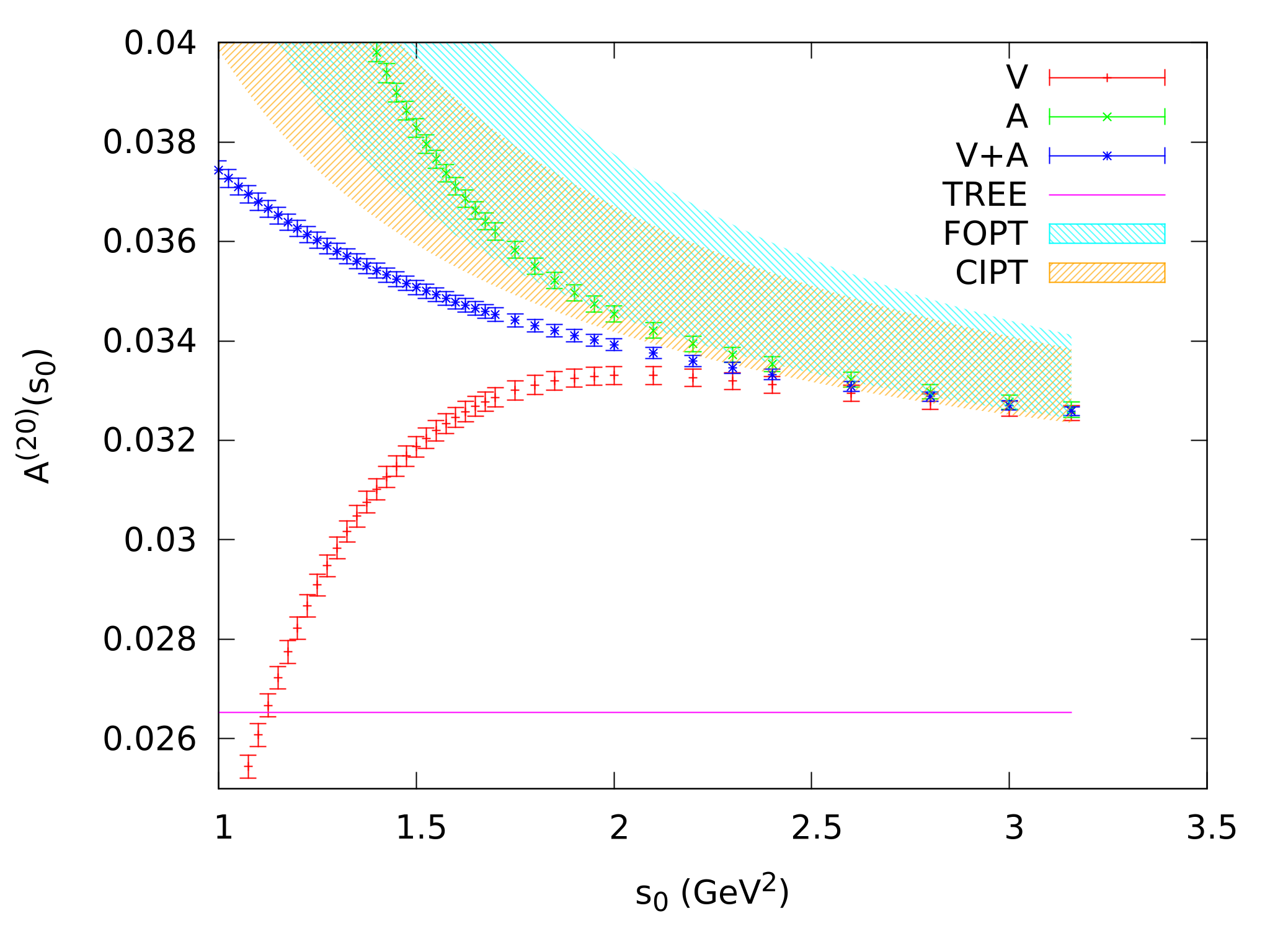}\hskip .5cm
\includegraphics[width=0.46\textwidth]{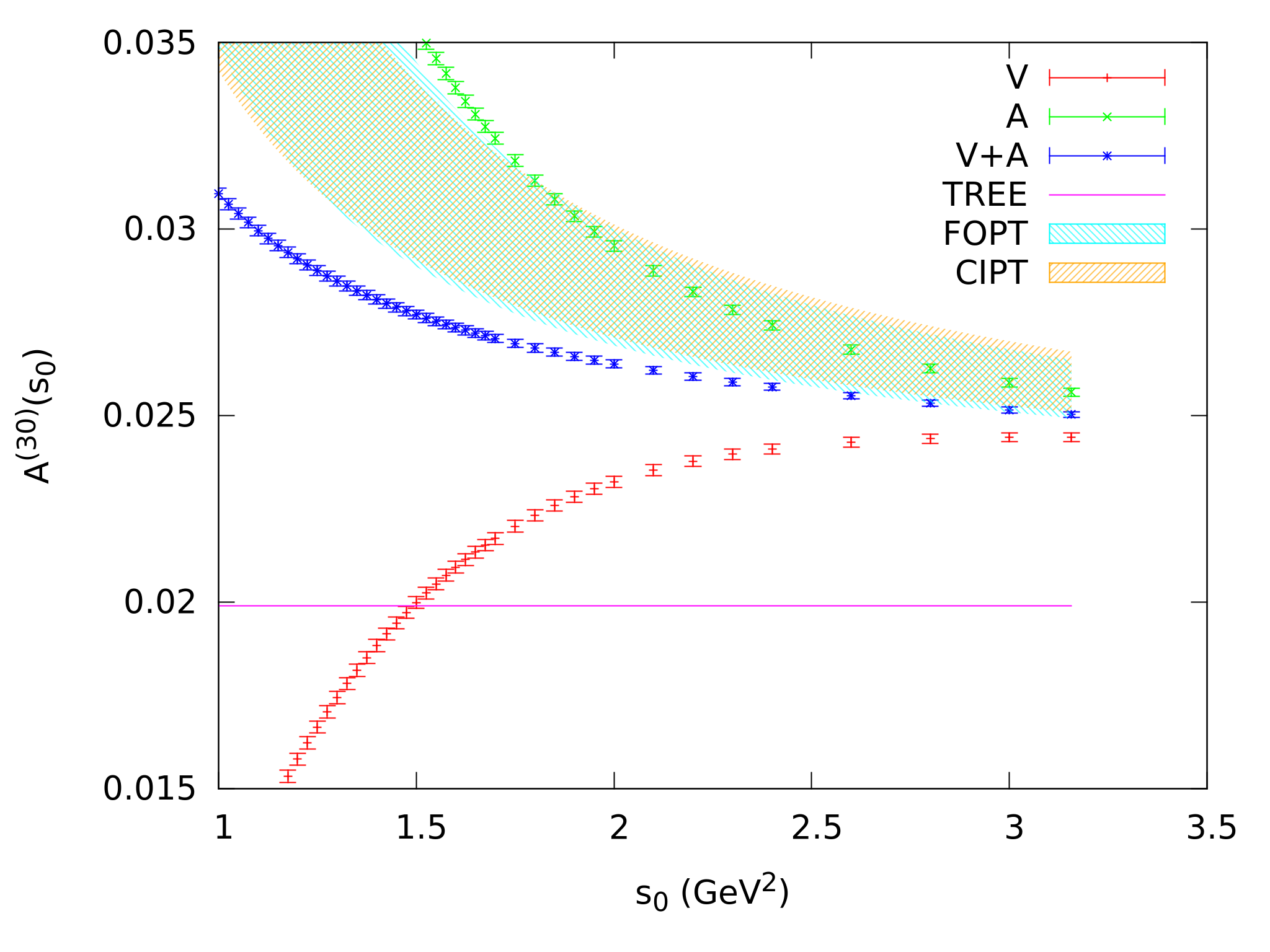}
\caption{\label{pinchs}
Dependence on $s_0$ of the experimental moments $A^{(n,0)} (s_{0})$, associated to the pinched weight functions of Eq.~(\ref{eq:omega-n0}), together with their purely perturbative predictions
calculated using the strong coupling obtained in the r.h.s. of Eq.~(\ref{strongdav}).
Data points are shown for the $V$ (red), $A$ (green) and $\frac{1}{2}\, (V+A)$ (blue)
channels.}
\end{figure}
%%%%%%%%%%%%%%%%%%%%%%%%%%%%%%%%%%%%%%%%%%%%%%%%%%%%%%%%%%%%%%%%%%%%%%%%%%%%%%%%%%%%

Given the large relative uncertainties on the small power-suppressed corrections, one would like to find additional inputs to constrain the range of fitted parameters. One possibility is to look at different values of $s_{0}$.
In Figure~\ref{pinchs} we plot as a function of $s_0$ the experimental moments $A^{(n,0)} (s_{0})$, associated with the simplest $n$-pinched weight functions in Eq.~\ref{eq:omega-n0}, for the $V$, $A$ and $\frac{1}{2}\, (V+A)$
channels, together with the perturbative part of $A^{(n,0)} (s_{0})$ predicted with the value of $\alpha_{s}(m_{\tau}^{2})$ given in Eq.~(\ref{strongdav}).

Perturbation theory appears to reproduce well the data at large values of $s_0\sim m_\tau^2$, without any clear need for sizeable power corrections, except perhaps in $A^{(3,0)}_V (s_{0})$.
Notice the excellent agreement obtained for the $V+A$ channel of $A^{(0,0)} (s_{0})$, the only moment whose non-perturbative OPE contribution is known to be negligible. The agreement with perturbation theory extends to quite
low values of $s_0$, even if this is the moment most exposed to duality violations, suggesting that duality-violation uncertainties are indeed within the quoted errors of $\alpha_{s}(m_{\tau}^{2})$. The moment $A^{(1,0)} (s_{0})$,
which can only get corrections from $\cO_4$, shows above $s_0\sim 2\;\mathrm{GeV}^2$ a surprisingly good agreement with its pure perturbative prediction in all channels ($V$, $A$ and $V+A$). In spite of being only protected by a
single pinch factor, the data points for this moment closely follow the central values predicted by CIPT. In that energy range both, duality violations and $D=4$ power corrections, appear to be too small to become numerically
visible within the much larger perturbative uncertainties covering the shaded areas of the figure. The higher moments $A^{(2,0)} (s_{0})$ and $A^{(3,0)} (s_{0})$ are slightly more sensitive to non-perturbative corrections.
The different curves seem to prefer a power correction with different signs for the $V$ and $A$ distributions, which cancels to a good extent in $V+A$. This fits nicely with the expected $\cO_{6,V/A}$ contribution. However,
in the moment $A^{(2,0)} (s_{0})$, the merging of the $V$, $A$ and $V+A$  curves above $s_0\sim 2.2\;\mathrm{GeV}^2$ suggests a very tiny numerical effect from this source in the high-energy range. Only the moment
$A^{(3,0)} (s_{0})$ appears to have still some sensitivity to power corrections at $s_0\sim m_\tau^2$.

%%%%%%%%%%%%%%%%%%%%%%%%%%%%%%%%%%%%%% Figure %%%%%%%%%%%%%%%%%%%%%%%%%%%%%%%%%%%%%%%%%
\begin{figure}[p]
\includegraphics[width=0.5\textwidth]{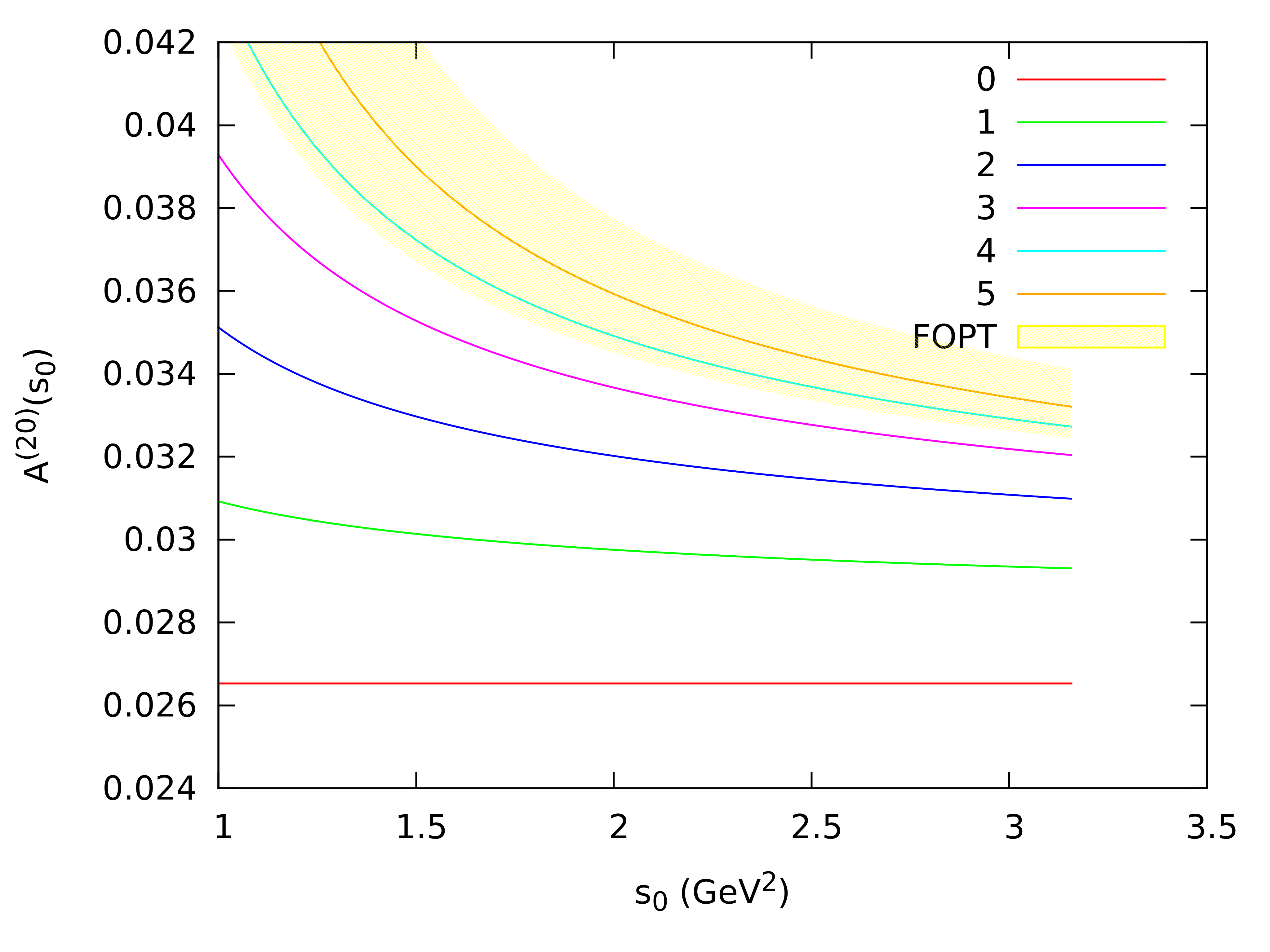}
\includegraphics[width=0.5\textwidth]{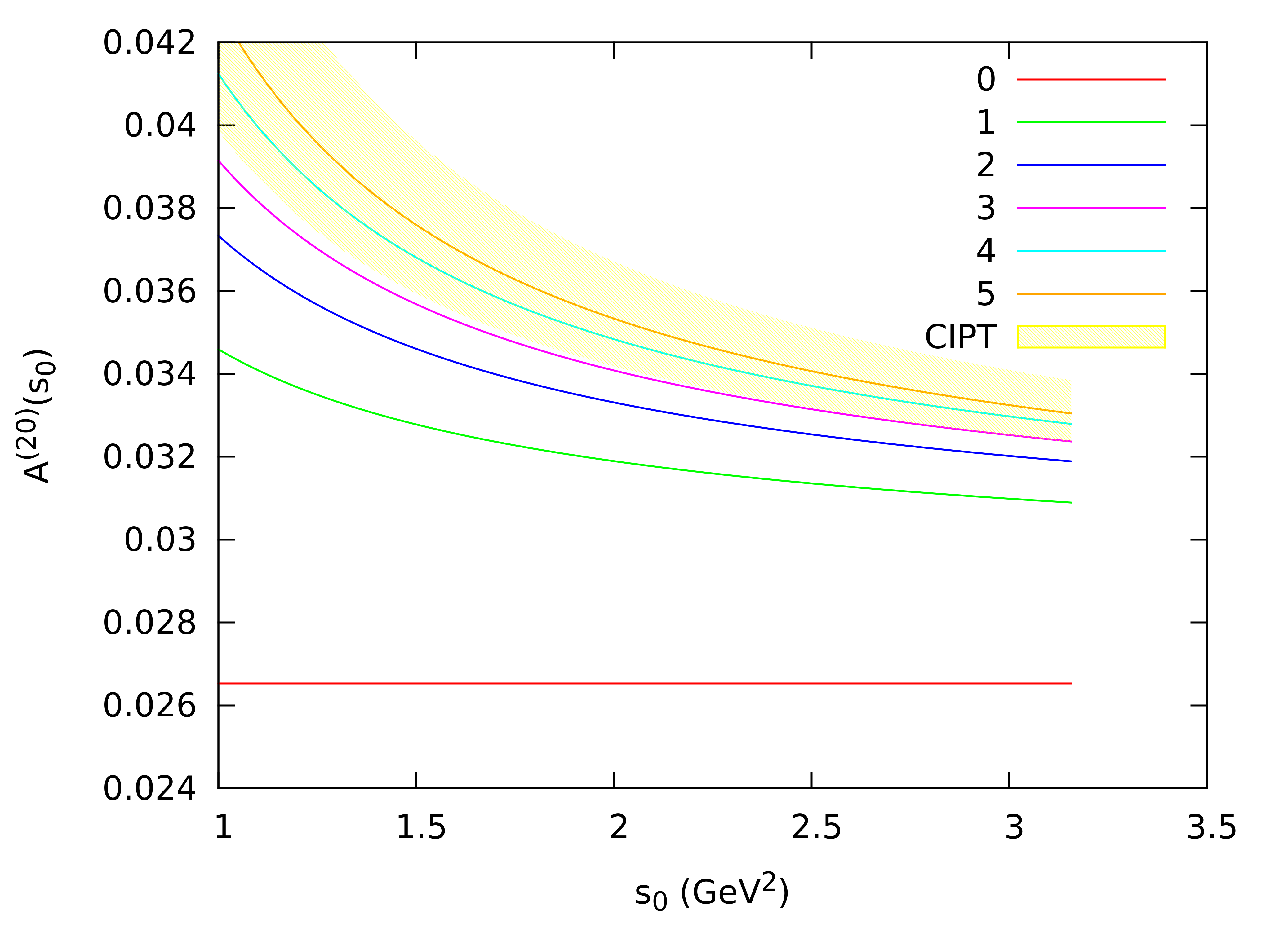}\\
\includegraphics[width=0.5\textwidth]{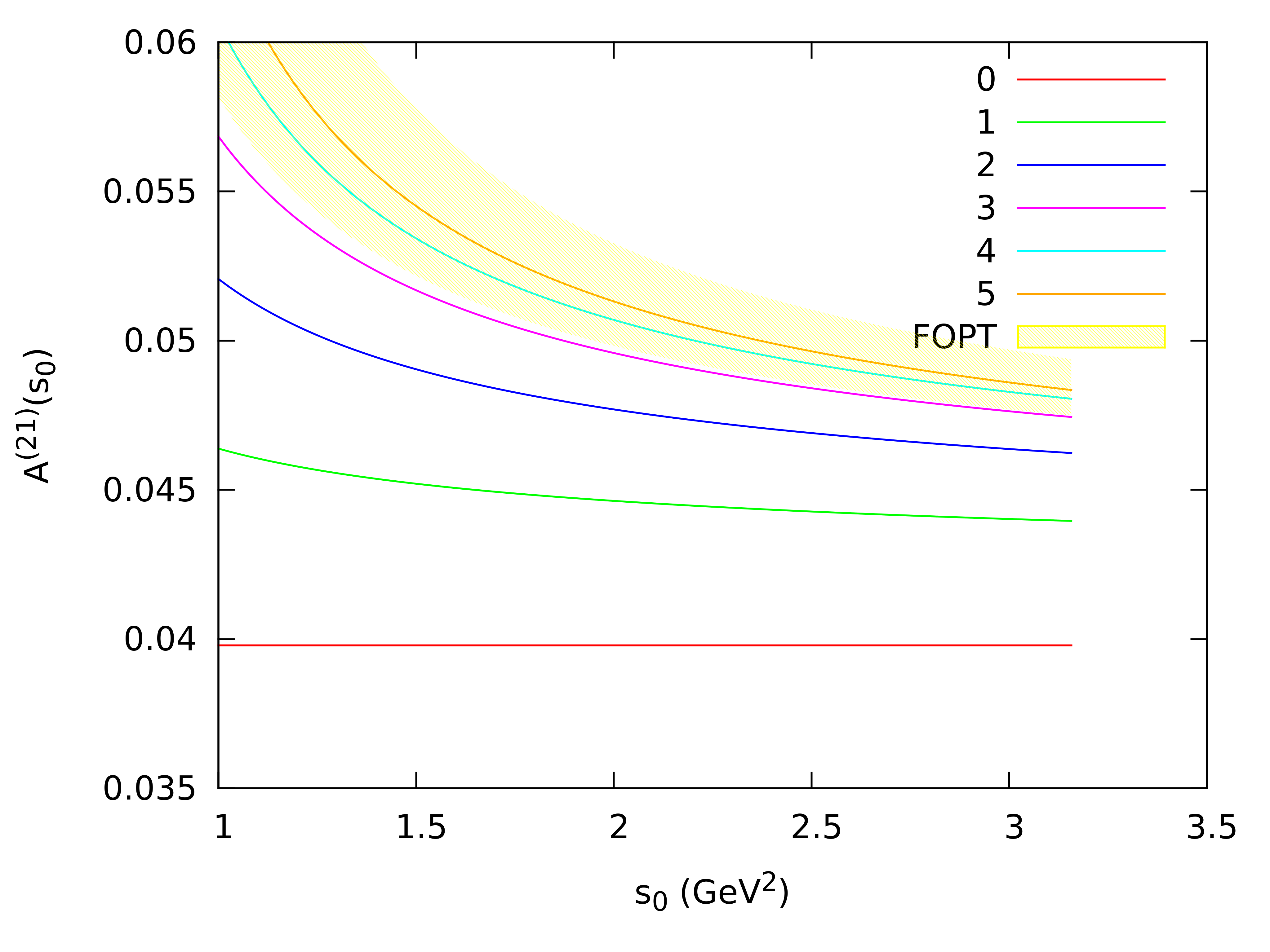}
\includegraphics[width=0.5\textwidth]{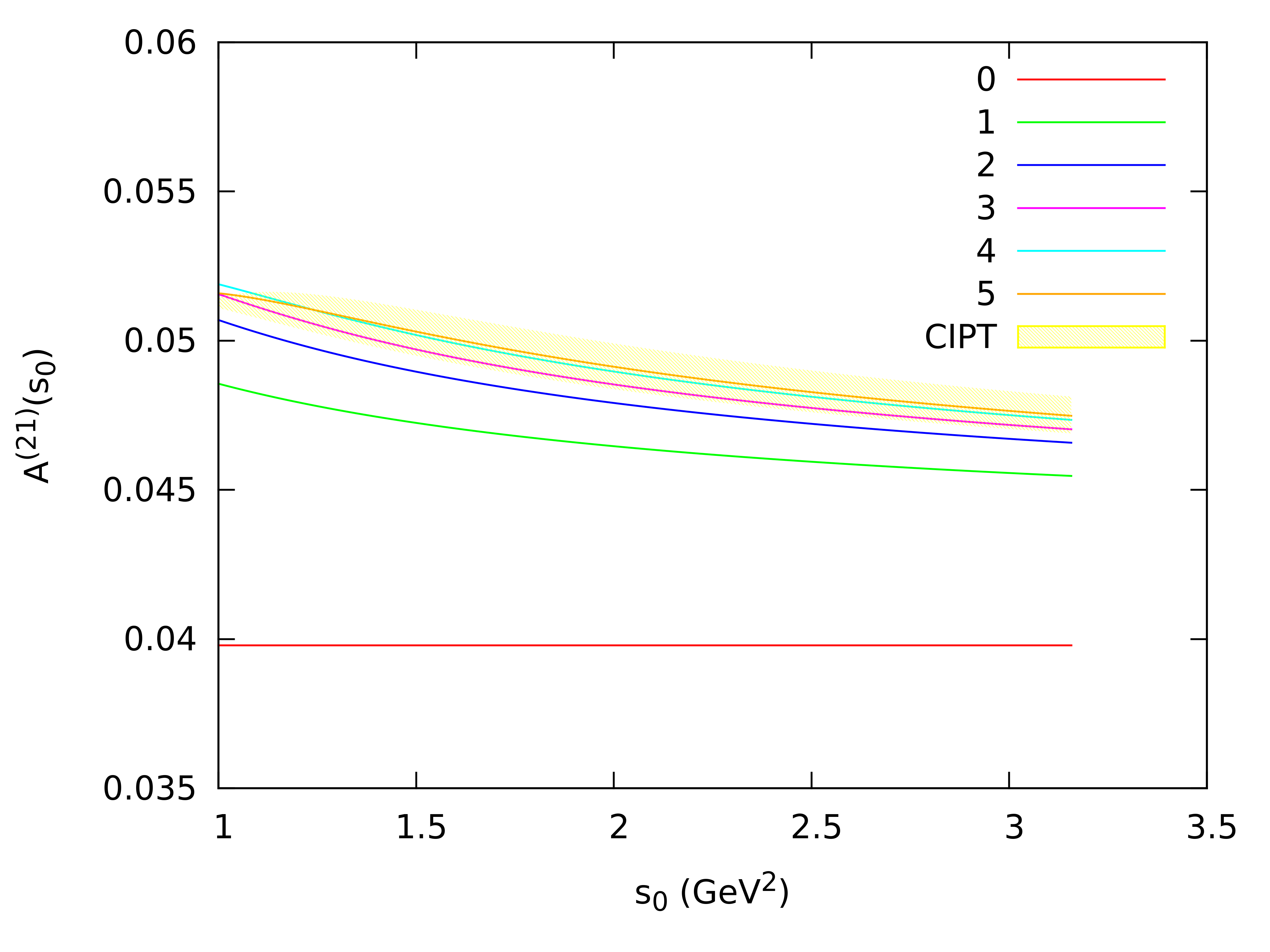}\\
\includegraphics[width=0.5\textwidth]{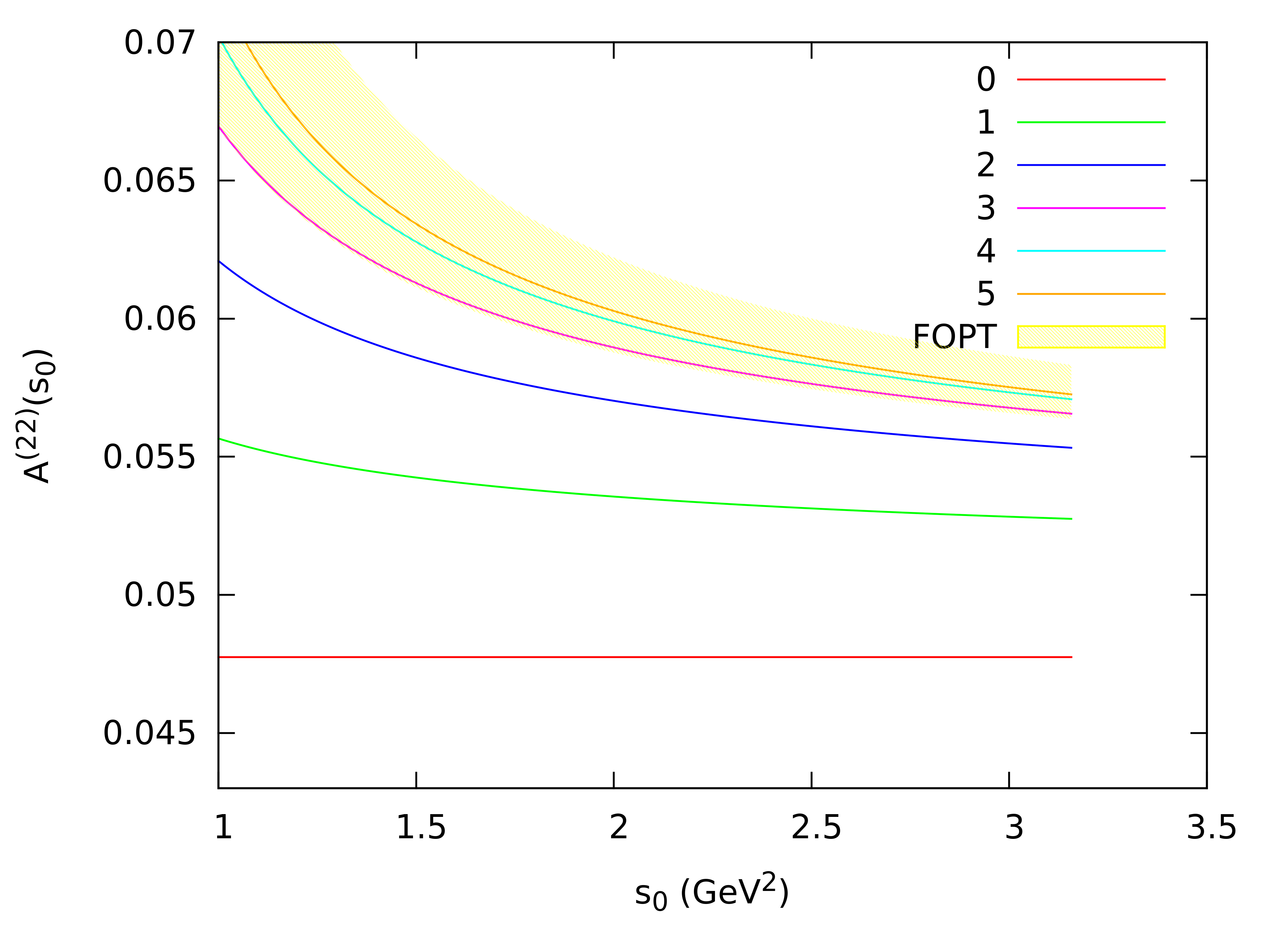}
\includegraphics[width=0.5\textwidth]{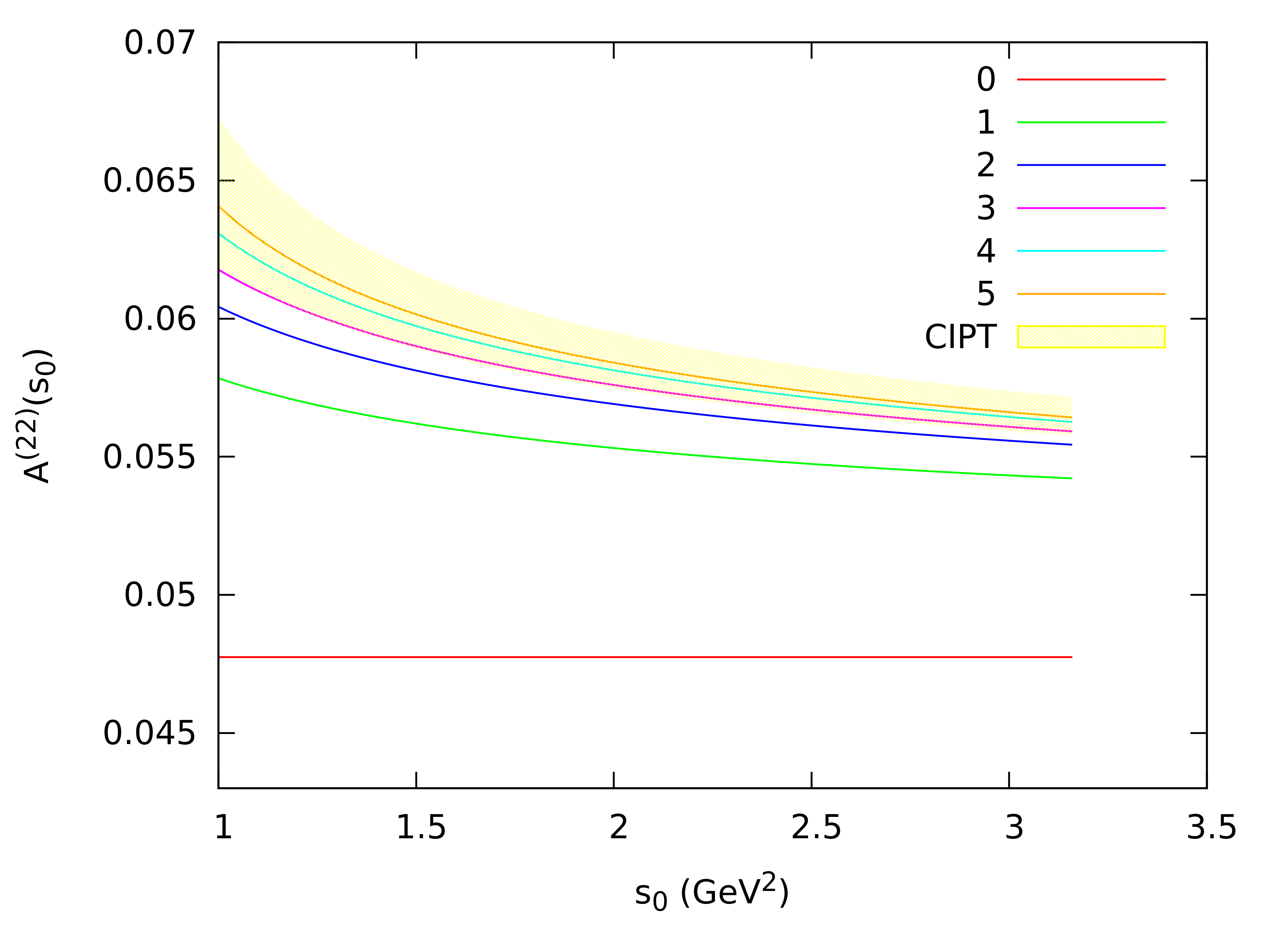}
\caption{\label{pertuseries}
Dependence on $s_0$ of the perturbative contributions to the moments
$A^{(2,0)} (s_{0})$, $A^{(2,1)} (s_{0})$ and $A^{(2,2)} (s_{0})$,
constructed with the doubly-pinched weight functions in Eq. (\ref{eq:omega-2n}),
calculated in FOPT (left panels) and CIPT (right panels) at several loop approximations. The filled areas correspond to $\alpha_{s}(m_{\tau}^{2})=0.329 \,{}^{+\, 0.020}_{-\, 0.018}$.}
\end{figure}
%%%%%%%%%%%%%%%%%%%%%%%%%%%%%%%%%%%%%%%%%%%%%%%%%%%%%%%%%%%%%%%%%%%%%%%%%%%%%%%%%%%%%%%%

In order to better assess the dominant perturbative errors, we present in Figure~\ref{pertuseries}, as a function of $s_0$, the perturbative predictions for the doubly-pinched moments $A^{(2,0)} (s_{0})$, $A^{(2,1)} (s_{0})$ and $A^{(2,2)} (s_{0})$, at different loop approximations within FOPT (left) and CIPT (right), with the same value of $\alpha_{s}(m_{\tau}^{2})$ given above. Note that the $\alpha_s^5$ contribution is just an educated guess estimate, taking for the fifth-order Adler coefficient the value $K_{5}=275$. For the known perturbative orders, CIPT seems to present a better convergence. Additionally, we observe a slightly better perturbative behaviour for the two moments which are independent of the $\mathcal{O}_{4}$ condensate, in agreement with the models considered in Ref.~\cite{Beneke:2012vb}. However, new coefficients of the Adler function would be needed to extract any reliable conclusions. Moreover, a moment with a better perturbative behaviour is not necessarily the best one to determine the strong coupling, since the sensitivity of the moments to $\alpha_{s}(m_{\tau}^{2})$ plays a crucial role too.

Naively, the pinched moments seem suitable for performing phenomenological fits. However, as it was already observed long time ago in Ref.~\cite{LeDiberder:1992zhd}, a fit of the $s_0$ dependence turns out to be nearly equivalent to a direct fit of the spectral function $\rho(s_0) =\frac{1}{\pi}\,\ImPi(s_{0})$, a quantity which cannot be described rigourously with the OPE. This is immediately seen, studying the derivative with respect to the moment end-point $s_0$. For the simpler $n$-pinched moments $A^{(n,0)} (s_{0})$, one finds
\bel{eq:An0-der}
s_0\;\frac{d}{d s_0}\, A^{(n,0)} (s_{0})\; =\; \delta_{n,0}\;\pi\,\rho(s_0) + n\, A^{(n-1, 0)} (s_{0})
- (n+1)\, A^{(n,0)} (s_{0})\, .
\ee
Thus, if we make a fit of consecutive $s_0$ points, we are removing pinchs; {\it i.e.},
the only new experimental information we get adding $A^{(n,0)} (s_{0}+\Delta{s_{0}})$ to a fit with $A^{(n,0)} (s_{0})$ is the same integral with one pinch less. After adding $n$ $s_0$ bins, we are just testing $\rho(s_0)$. A fit with
$m$ $s_0$ points of the moment $A^{(n,0)}(s_{0})$ is going to be equivalent to a fit with:
\be
\left\{A^{(n,0)}(s_{0})\, ,\, A^{(n-1,0)}(s_{0})\, ,\,\cdots\, ,\, A^{(0,0)}(s_{0})\, ,\,\rho(s_{0})\, ,\, \rho(s_{0}+\Delta s_{0})\, ,\, ... , \rho(s_{0}+(m-n-2)\Delta s_{0}) \right\}\, .
\ee
Thus, we are directly using information about the local structure of the spectral function.

Not surprisingly, the functional dependence of the moments with $s_0$ manifests the violations of quark-hadron duality which are obviously present in the physical hadronic spectrum. Once this is properly understood, an analysis of the $s_0$ dependence can nevertheless provide enlightening information on the relevance of duality violation in different energy regimes.
With this caveat in mind, we study next the doubly-pinched moments $A^{(2,0)} (s_{0})$, $A^{(2,1)} (s_{0})$ and $A^{(2,2)} (s_{0})$, making different fits with the 9 available bins above $s_{0}=2\, \mathrm{GeV}^{2}$. In order to avoid too large data correlations, we will restrict every fit to just one moment $A^{(2,m)} (s_{0})$, with three free parameters: $\alpha_s(m_\tau^2)$, $\cO_{2(m+2)}$ and $\cO_{2(m+3)}$.

%%%%%%%%%%%%%%%%%%%%%%%%%%%%%%%%%%% Figure %%%%%%%%%%%%%%%%%%%%%%%%%%%%%%%%%%%%%
\begin{figure}[tb]\centering
\includegraphics[width=0.60\textwidth]{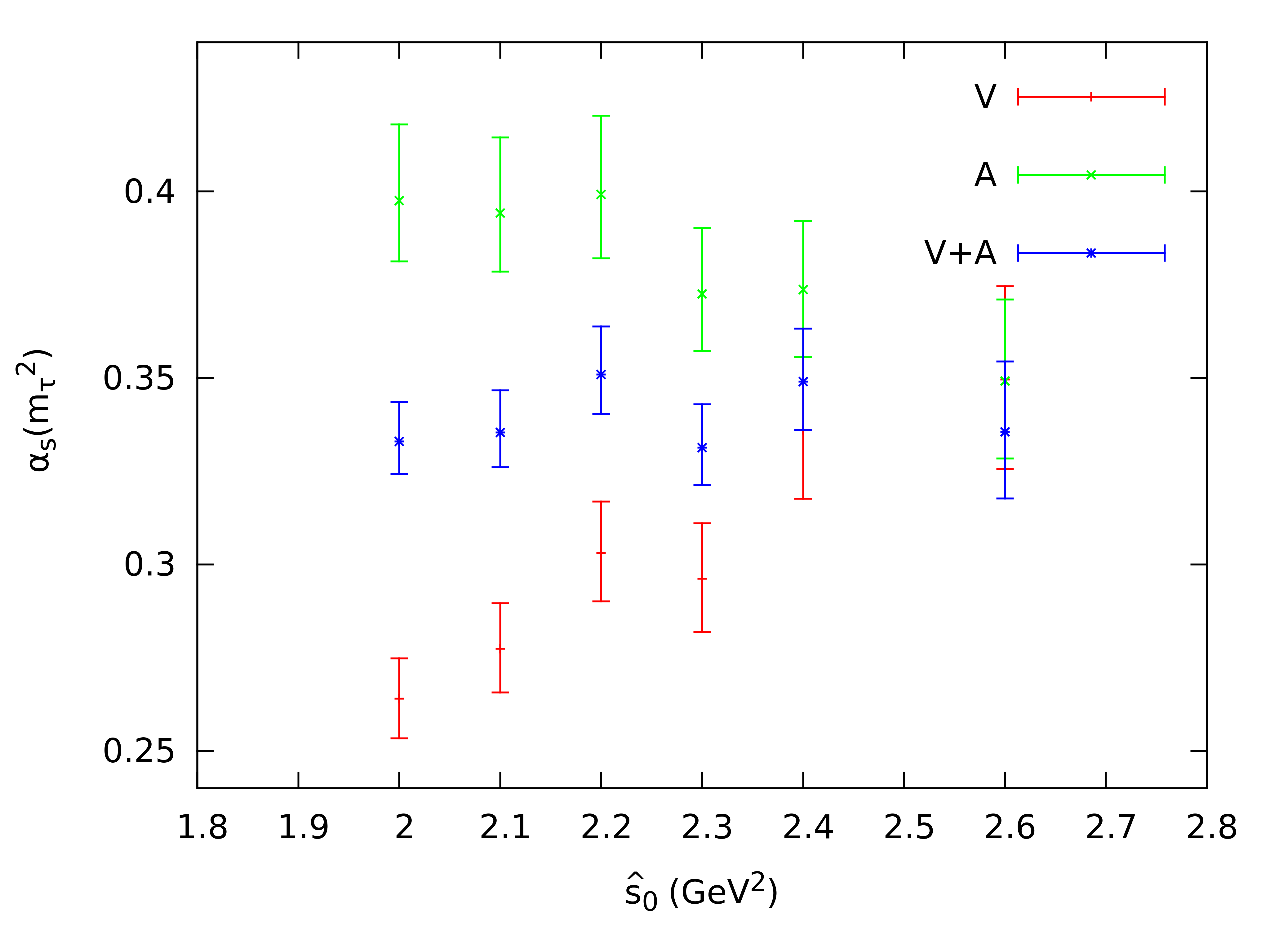}\\
\includegraphics[width=0.49\textwidth]{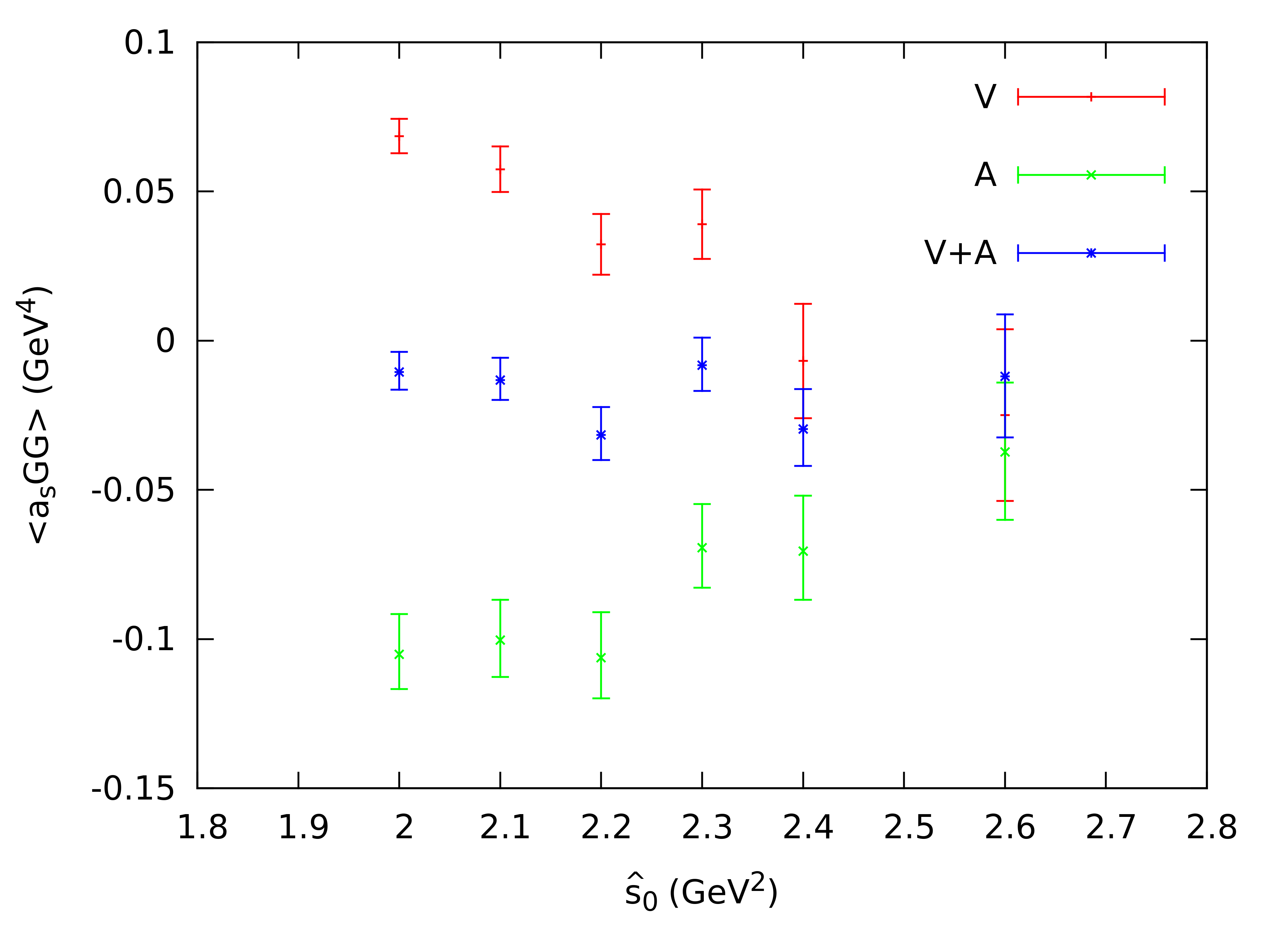}
\includegraphics[width=0.49\textwidth]{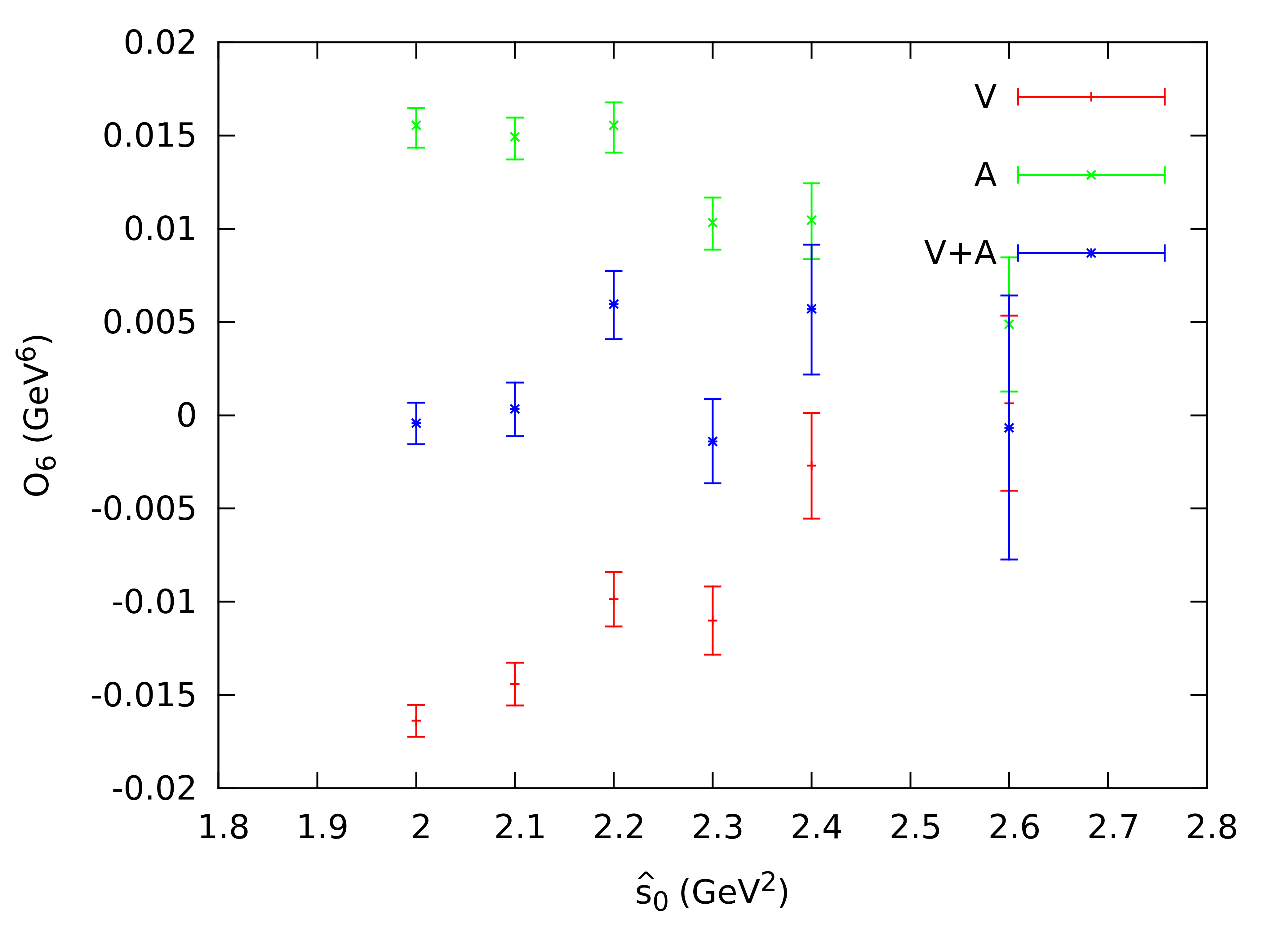}
\caption{\label{dualvio}
Parameters fitted in CIPT with the moment $A^{(2,0)}(s_{0})$, as a function of the starting $s_{0}$ value of the fit, $\hat s_0$.}
\end{figure}
%%%%%%%%%%%%%%%%%%%%%%%%%%%%%%%%%%%%%%%%%%%%%%%%%%%%%%%%%%%%%%%%%%%%%%%%%%%%%%%%%

In Figure~\ref{dualvio} we plot the values of the three parameters fitted in CIPT with the moment $A^{(2,0)}(s_{0})$ (very similar conclusions are obtained with FOPT and for the other moments) as a function of the starting $s_{0}$ value of the fit, $\hat s_{0}$, for the $V$, $A$ and $V+A$ channels. The points with error bars shown at a given value of $\hat s_0$ represent the results of the fit using only the values of the moments at $s_0\ge \hat s_0$, {\it i.e.}, at the bins above $\hat s_0$. Since we need to fit three parameters, four points at least are needed. Thus, the highest value $\hat s_0 = 2.6\;\mathrm {GeV}^2$ gives the results of a fit to the last four $s_0$ bins, while for the lowest value $\hat s_0 = 2.0\;\mathrm {GeV}^2$ the nine bins are included in the fit.

As expected, the figure shows a strong dependence on $\hat s_{0}$ in the $V$ and $A$ channels, as well as completely incompatible values for $\alpha_s(m_\tau^2)$ and $\langle \frac{\alpha_{s}}{\pi}\, GG\rangle$ in the
lower $\hat s_{0}$ range, where local duality has been assumed. However, when we go to  higher values of $\hat s_{0}$, the $V$ and $A$ fitted parameters start to converge towards the much more stable $V+A$ results.
The stability of the $V+A$ fit in the whole range of $\hat s_0$ values analyzed is quite surprising.
Duality-violation effects are present (we are sensitive to the spectral function itself) and clearly manifest in the $V$ and $A$ plotted points, but their size seems to be quite suppressed in the more inclusive $V+A$ distribution. This qualitative behaviour is easily
understood looking at the experimental spectral functions in Fig.~\ref{fig:ALEPHsf} and observing the flattening of the highest energy points in the $V+A$ curve, with a clear compensation of the vector and axial-vector departures from local duality.

Ignoring completely any possible effects from violations of duality, a direct fit of the $V+A$ moments for the 9 available points above $s_{0}=2\, \mathrm{GeV}^{2}$ gives the results shown in Table~\ref{fit9}. 
%
%%%%%%%%%%%%%%%%%%%%%%%%%%%%%%%%%%%%%%% Table %%%%%%%%%%%%%%%%%%%%%%%%%%%%%%%%%%%%%%%
\begin{table}[tb]
\renewcommand\arraystretch{1.2}
\centering
\begin{tabular}{|c|c|c|c|c|}
\hline
Moment & Method     & $\alpha_{s}(m_{\tau}^{2})$  &   Lower-$D$ Condensate     & Higher-$D$ Condensate                  \\
&&& ($10^{-3}\;\mathrm{GeV}^D$) & ($10^{-3}\;\mathrm{GeV}^D$)\\ \hline

$A^{(2,0)}(s_{0})$ & FOPT       & $0.331\,{}^{+\, 0.013}_{-\, 0.018}$ &$-9\,{}^{+\, 12}_{-\, 4}$     & $-4\,{}^{+\, 3}_{-\, 7}$ \\
$A^{(2,0)}(s_{0})$ & CIPT  & $0.333 \,{}^{+\, 0.011}_{-\, 0.009}$ &$-11\,{}^{+\, 7}_{-\, 6}$    & $0 \pm 1$  \\ \hline
$A^{(2,1)}(s_{0})$ & FOPT       & $0.322 \,{}^{+\, 0.010}_{-\, 0.006}$ &$3\,{}^{+\, 1}_{-\, 2}$     &$0 \pm 2$ \\
$A^{(2,1)}(s_{0})$ & CIPT  & $0.334 \,{}^{+\, 0.011}_{-\, 0.009}$ &$0 \pm 1$    & $2 \pm 2$  \\ \hline
$A^{(2,2)}(s_{0})$ & FOPT       & $0.319 \,{}^{+\, 0.009}_{-\, 0.006}$ & $-2 \,{}^{+\, 3}_{-\, 2}$     & $-1 \,{}^{+\, 4}_{-\, 5}$ \\
$A^{(2,2)}(s_{0})$ & CIPT  & $0.334 \,{}^{+\, 0.011}_{-\, 0.009}$ &$2 \pm 2$    & $-5 \pm 4$  \\ \hline
\end{tabular}
\caption{\label{fit9} Fitted results in the $V+A$ channel, using the weight functions of Eq. (\ref{eq:omega-2n}) and ignoring duality-violation effects. The value given for the $D=4$ condensate refers to $\langle \frac{\alpha_{s}}{\pi}\, GG\rangle$.
The quoted uncertainties include experimental and perturbative errors.}
\end{table}
%%%%%%%%%%%%%%%%%%%%%%%%%%%%%%%%%%%%%%%%%%%%%%%%%%%%%%%%%%%%%%%%%%%%%%%%%%%%%%%%%%%%%%
%
Each horizontal line corresponds to the fit of a single moment $A^{(2,k)}(s_{0})$ ($k=0,1,2$), either with FOPT or CIPT. The fitted values are in good agreement with the ones obtained before in Tables~\ref{tablaant} and \ref{tablaanto10}, for the same channel. The sensitivity to the power corrections turns out to be very bad, being all fitted results compatible with zero. On the other side, one obtains very stable values for the strong coupling with moderate errors. The CIPT result is amazingly stable with the three moments giving practically the same value
$\alpha_s(m_\tau^2)^{\mathrm{CIPT}} = 0.335\pm 0.010$, while a weighted average of the three FOPT results (keeping the smallest error) translates into
$\alpha_s(m_\tau^2)^{\mathrm{FOPT}} = 0.323\pm 0.008$. However, all these fits have a very low quality ($\chi^{2}_{\mathrm{min}}/\mathrm{d.o.f.}$), indicating the presence of the neglected duality-violation effects.
Adding quadratically half of the difference between the maximum and minimum values of $\alpha_{s}(m_{\tau}^{2})$ given in Figure~\ref{dualvio} for this channel, as an estimate of duality-violation uncertainties, one gets:
\be
\ba{c}
\alpha_{s}(m_\tau^2)^{\mathrm{CIPT}} \; =\; 0.335 \pm 0.014
\\[5pt]
\alpha_{s}(m_\tau^2)^{\mathrm{FOPT}} \; =\; 0.323 \pm 0.012
\ea
\qquad\longrightarrow\qquad
\alpha_{s}(m_\tau^2) \; =\; 0.329 \pm 0.013 \, .
\label{segundostrong}
\ee

\section{Modeling duality violations}
\label{sec:DV}

In order to study violations of duality, Refs.~\cite{Boito:2011qt,Boito:2012cr,Boito:2014sta} parametrize the differences between the physical spectral functions and their OPE approximations with the following ansatz:
\bel{eq:DVparam}
\Delta\rho^{\mathrm{DV}}_{V/A}(s)\; =\; e^{-(\delta_{V/A}+\gamma_{V/A}s)}\;\sin{(\alpha_{V/A}+\beta_{V/A}s)}\, ,
\qquad\qquad s>
\hat s_0\, .
\ee
Although it is theoretically well motivated, this functional form cannot be derived from first principles, which unavoidably introduces some model dependence in their analyses. This combination of an oscillatory function
with an exponential damping is assumed to describe the fall-off of duality violations at very high energies. However, nobody really knows from which $\hat s_0$
value this could start to be a valid approximation.

Having a model for the spectral function, to be fitted to data, one can then estimate the duality-violation correction to Eq.~\eqn{aomega} through the identity
\cite{Cata:2005zj,Rodriguez-Sanchez:2016jvw,GonzalezAlonso:2010rn,Chibisov:1996wf}
\bel{eq:DVcorr}
\Delta A^{\omega, \mathrm{DV}}_{V/A}(s_{0})\;\equiv\; \frac{i}{2}\;\oint_{|s|=s_{0}}
\frac{ds}{s_{0}}\;\omega(s)\,\left\{\Pi^{\phantom{\mathrm{OPE}}}_{V/A}(s) - \Pi^{\mathrm{OPE}}_{V/A}(s)\right\}
\; =\; -\pi\;\int^{\infty}_{s_{0}} \frac{ds}{s_{0}}\;\omega(s)\; \Delta\rho^{\mathrm{DV}}_{V/A}(s)
\, .
\ee

The strategy adopted in Refs.~\cite{Boito:2011qt,Boito:2012cr,Boito:2014sta} consists in making a global fit to the $s_0$ dependence of the moments $A^{\omega}_{V/A}(s_{0})$, in order to fit $\alpha_s(m_\tau^2)$,
the vacuum condensates and the eight spectral function parameters in Eq.~\eqn{eq:DVparam}, assuming the ansatz to be valid above $\hat s_0
\sim 1.55\, \mathrm{GeV}^{2}$.
Since there are far too many parameters to get a reasonable fit to the highly-correlated $\tau$ data sample, Ref. \cite{Boito:2014sta} concentrates in the $A^{(0,0)}_{V/A}(s_{0})$ moment, which does not
receive OPE corrections and is, moreover, very exposed to duality violation effects because it is not protected
by any pinch factor. The problem with this type of strategy was already analyzed in the previous section. A fit with $n$ $s_0$ points of the $A^{(0,0)}_{V/A}(s_{0})$ moment is equivalent to a fit of
\be
\left\{ A^{(0,0)}_{V/A}(s_{0})\, ,\, \rho_{V/A}^{\phantom{()}}(s_{0})\, ,\,\cdots\, ,\, \rho_{V/A}^{\phantom{()}}(s_{0}+(n-2)\Delta s_{0})\right\}\, ,
\ee
so that $n-1$ of the $n$ fitted points are dedicated just to fit the spectral function.
Once $\rho_{V/A}^{\phantom{()}}(s)$ has been fitted, the ensuing $s_0$-stability of the moments in the fitted region is just a direct consequence of the 5-parameters fit, not a test of the model as
incorrectly claimed in Refs.~\cite{Boito:2011qt,Boito:2012cr,Boito:2014sta}.\footnote{The fact that
$\rho_{V\! -\! A}^{\phantom{()}}(s)\equiv \rho^{(1+0)}_{ud,V}(s)-\rho^{(1+0)}_{ud,A}(s)$ satisfies approximately the Weinberg Sum Rules (WSRs) \cite{Weinberg:1967kj}
is only a consequence of the fit, of the fact that duality violations are exponentially suppressed and that they are already satisfied in $s_{0}=2.8$ GeV$^2$, the last point with  large-enough experimental resolution:
\be
\underbrace{\int^{s_{0}}_{0}ds\; \rho_{V\! -\! A}^{\phantom{()}}(s)}_{\text{Experimental data}}
\; +\;\underbrace{\int^{2.8\, \text{GeV}^{2}}_{s_{0}}ds\; \rho_{V\! -\! A}^{\phantom{()}}(s)}_{\rho_{V\! -\! A}^{\phantom{()}}(s)\, \text{fitted with data}}
\; +\;\underbrace{\int^{\infty}_{2.8\, \text{GeV}^{2}}ds \; \rho_{V\! -\! A}^{\phantom{()}}(s)}_{\approx 0}\; =\; 0 \, .
\ee
The same result would be obtained with any sensible model that fits well the spectral function and whose duality violations are small for $s_{0}> 2.8 \,$GeV$^{2}$.}

The $A$ channel is not useful to determine $\alpha_{s}$ with this strategy. As shown in Fig.~\ref{fig:ALEPHsf}, the tail of the $a_1(3\pi)$ resonance extends to quite high values of $s$,
which questions the use of the ansatz~\eqn{eq:DVparam} except at the highest energy bins where the large experimental errors make the whole game useless. Even if one insists on imposing the model from
$\hat s_0= 1.55\;\mathrm{GeV}^2$, the larger experimental errors of the axial spectral function in $\hat s_0 < s_{0} < m_{\tau}^{2}$ do not allow us to obtain new information about $\alpha_{s}(m_{\tau}^{2})$.
Thus, the strong coupling is finally extracted from a fit to the $s_0$ dependence of
$A^{(0, 0)}_{V}(s_0)$, with 5 parameters: $\delta_V$, $\gamma_V$, $\alpha_V$, $\beta_V$ and $\alpha_s(m_\tau^2)$, taking $\hat s_0 = 1.55\;\mathrm{GeV}^2$.

\begin{figure}[t]\centering
\includegraphics[width=0.49\textwidth]{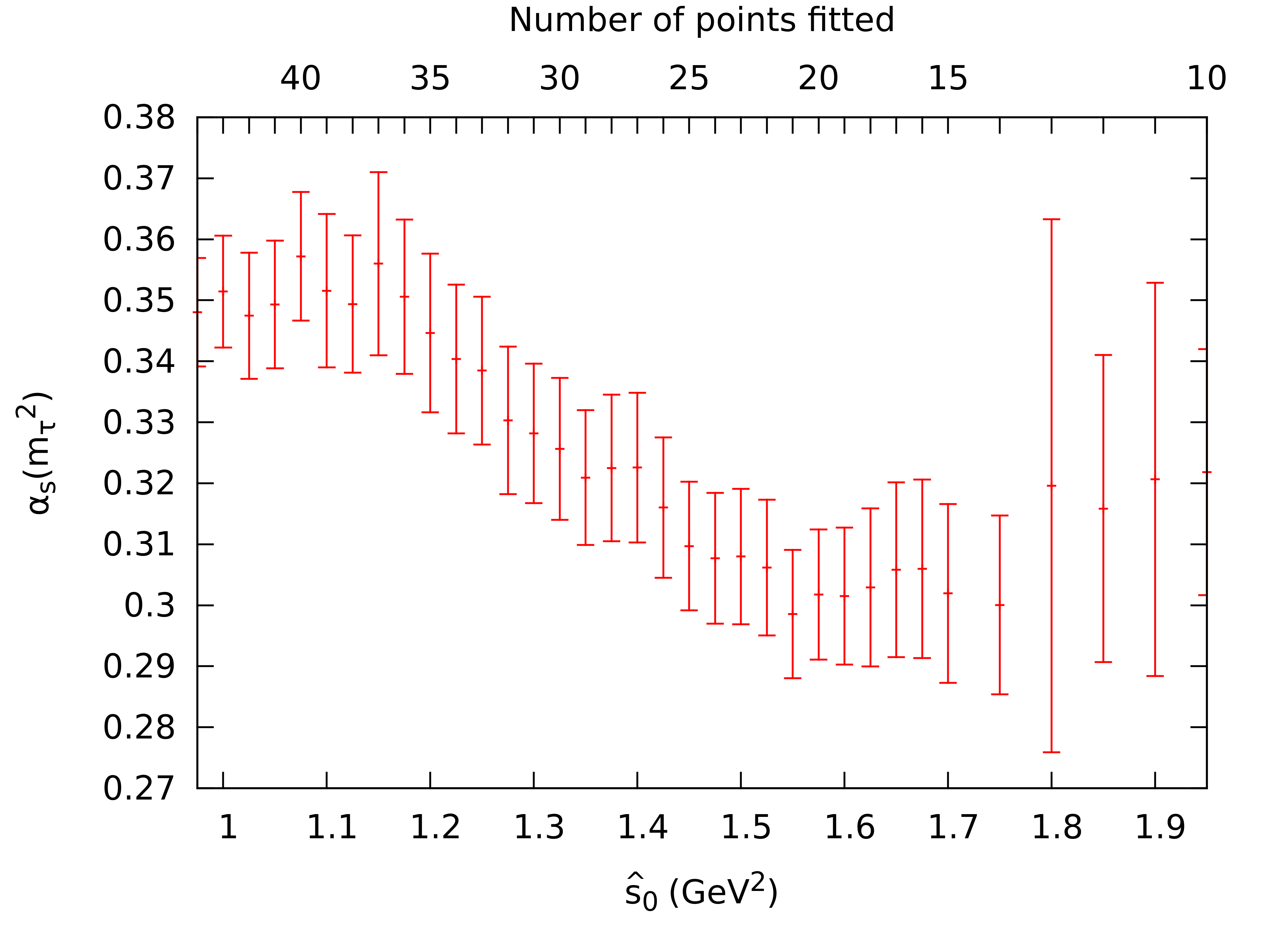}
\includegraphics[width=0.49\textwidth]{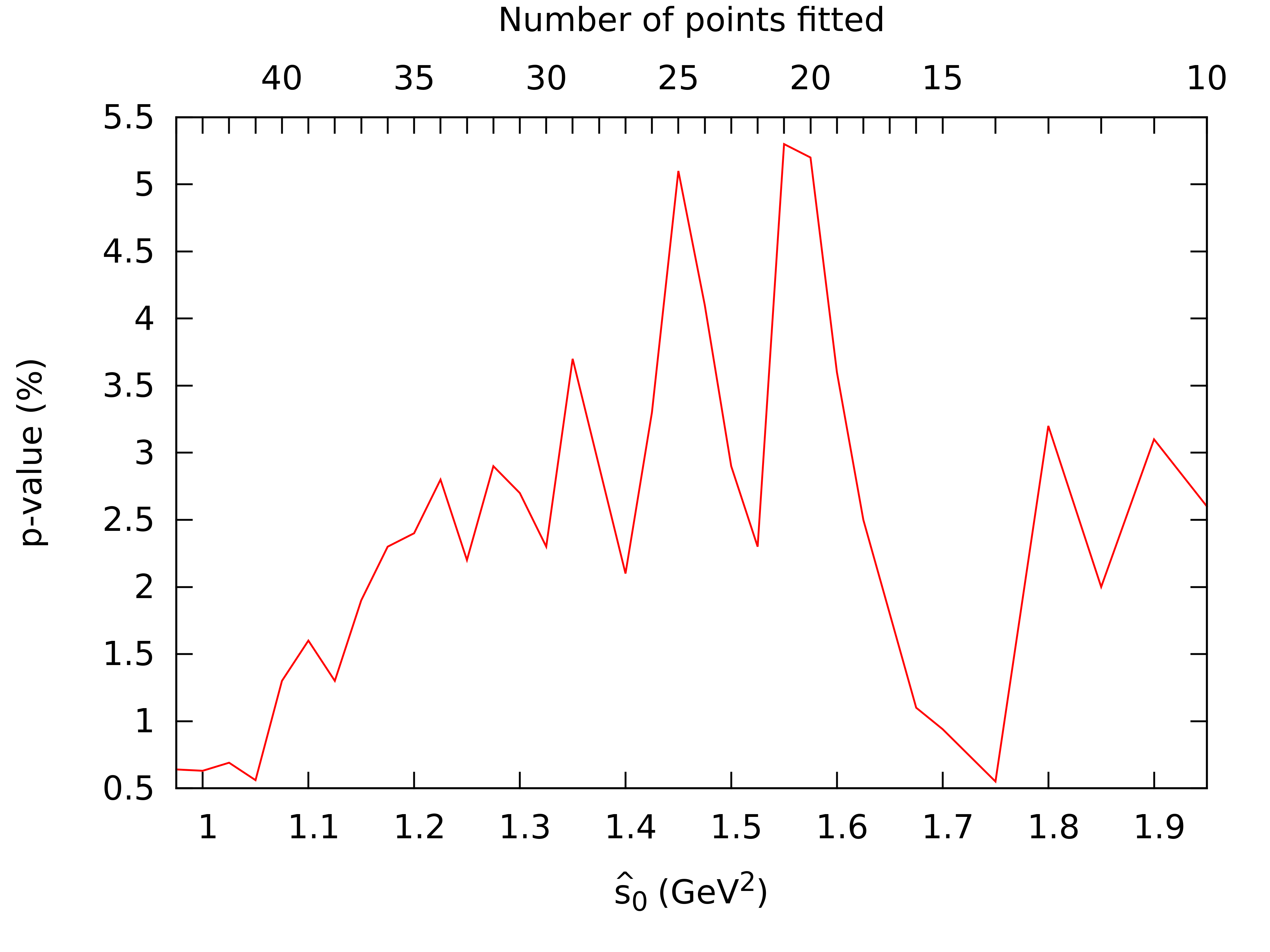}
\caption{\label{boitofig} FOPT determination of $\alpha_s(m_\tau^2)$ from the $s_0$ dependence of
$A^{(0, 0)}_{V}(s_0)$, fitting all $s_0$ bins with $s_{0}>\hat s_0$, as function of $\hat s_0$, using the approach of Ref. \cite{Boito:2014sta}.}
\end{figure}

Following the same strategy as Refs.~\cite{Boito:2011qt,Boito:2012cr,Boito:2014sta}, we have performed a fit to the $s_0$ dependence of $A^{(0, 0)}_{V}(s_0)$, above some minimum value $\hat s_0$.
The results of this exercise are presented in Figure~\ref{boitofig}, using FOPT to handle the perturbative series (the same conclusions, with correspondingly larger values of $\alpha_s(m_\tau^2)$,
are obtained in the CIPT case). The left panel shows, as a function of $\hat s_0$, the value of $\alpha_s(m_\tau^2)$ extracted from a fit to all $s_0$ bins with $s_{0}>\hat s_0$, while the right panel
gives the associated p-values of the different fits. Our results are very similar to the ones obtained in Ref.~\cite{Boito:2014sta} (small differences, much lower than the uncertainties, could have arisen
from the different handling).

%%%%%%%%%%%%%%%%%%%%%%%%%%%%%%%%%%%%%%%%%%%%%%%%%%%%%%%%%%%%%%%%%%%%%%%%%%%%%%%%%%
%%%%%%%%%%%%%%%%%%%%%%%%%%%%%%%%%%%%%%%%%%%%%%%%%%%%%%%%%%%%%%%%%%%%%%%%%%%%%%%%%%

One immediately notices the very poor statistical quality of these fits, with very low p-values in all cases. In ref.~\cite{Boito:2014sta}, the value $\hat s_0 = 1.55\;\mathrm{GeV}^2$ is chosen to perform the $\alpha_s$ determination because it has the larger (although still small) p-value, but this is a completely ad-hoc assumption which is difficult to justify. If the model were reliable, it should work better at higher hadronic invariant masses. However, the p-value becomes worse when we go to higher values of $\hat s_0$ and significant deviations from the model are observed.
Additionally, the fitted values of $\alpha_{s}(m_\tau^2)$ do not present the stability one should expect. Fluctuations of the order of $1\, \sigma$ are observed, just removing 1 of the $\sim 20$ points included in the fit. On the other hand, if the model is valid from $s_{0} \sim 1.55\;\mathrm{GeV}^{2}$, one should expect a soft convergence to it in the left $s_{0}$-side, especially taking into account that we are still relatively far from the $\rho$ resonance. However, looking at Fig. \ref{boitofig} or at Fig. 5 of Ref. \cite{Boito:2014sta}, it is evident that the data deviate from the model dramatically fast as soon as one moves away from the fitted area.

Ref. \cite{Boito:2014sta} performs another fit, including the $s_0$ dependence of the moments
$A^{(1,1)}_V(s_{0})$ and $A^{(2,1)}_V(s_{0})$, in addition to $A^{(0,0)}_V(s_{0})$.
$A^{(1,1)}_V(s_{0})$ adds an additional unknown theoretical parameter to the fit, $\mathcal{O}_{6V}$,
while $A^{(2,1)}_V(s_0)$ adds  $\mathcal{O}_{8V}$ to the previous two moments. However, since one has the relation
\be
A^{(1,1)}_V(s_{0}+\Delta s_{0})\; =\; 2\, \frac{\Delta s_{0}}{s_{0}}\; A^{(0,0)}_V(s_{0})+ \left( 1-3\, \frac{\Delta s_{0}}{s_{0}} \right)\, A^{(1,1)}_V(s_{0})+ \mathcal{O}\left(\frac{\Delta s_{0}}{s_{0}}\right)^{2} \, ,
\ee
adding $\{A^{(1,1)}_V(s_{0})\,|\;  s_{0}>\hat s_0\}$ to the $\{A^{(0,0)}_V(s_{0})\,|\; s_{0}>\hat s_0\}$ fit only gives a new independent point, which is useful to fit $\mathcal{O}_{6V}$, but gives no new information about the other parameters.
The same is true when adding $\{A^{(2,1)}_V(s_{0})\,|\;  s_{0}>\hat s_0\}$ to a fit with $\{A^{(1,1)}_V(s_{0}),A^{(0,0)}_V(s_{0})\,|\;  s_{0}>\hat s_0\}$ (in this case with $\mathcal{O}_{8V}$).
Thus, from the $3N$ points in $\{A^{(1,1)}_V(s_{0}),A^{(0,0)}_V(s_{0}), A^{(2,1)}_V(s_{0})\,|\;  s_{0}>\hat s_0\}$ only $N+2$ bring independent information, leading to a highly correlated fit with all kinds of
numerical problems. In order to accomplish the $\chi^2$ minimization, in Ref. \cite{Boito:2014sta} a deformed fit putting to zero the huge
correlations among the moments is made. However, this procedure is hiding the
real problem. One could just perform a much simpler fit, using the $N+2$ independent points
\be
\left\{ A^{(0,0)}_V(s_{0}),A^{(0,0)}_V(s_{0}+\Delta s_{0}),...,A^{(0,0)}_V(s_{0}+(N-1)\Delta s_{0}),A^{(1,1)}_V(s_{0}),A^{(2,1)}_V(s_{0})\right\}\, ,
\ee
which contain exactly the same experimental information. Nevertheless, since the last two inputs include 2 new unknown parameters, this fit would not bring
additional information on $\alpha_s$.

In conclusion, modeling duality violations does not seem to be the best approach to reduce the possible weaknesses in the $\alpha_{s}$ determinations performed in Ref. \cite{Davier:2013sfa}.
The theoretical basis is much weaker (the OPE cannot be applied on the real axis), the statistical
quality is poor, and a dramatic model dependency is observed with significant deviations from the assumed model arising as soon as one moves slightly away from the fitted area. If nevertheless,
one insists in getting a numerical value from our findings in Figure~\ref{boitofig}, we could apply a pragmatic recipe analogous to the one used in the previous section to derive the determination in
Eq.~\eqn{segundostrong}. Throwing away the last three points in the figure, which have too large experimental uncertainties to provide any useful information, we take an interval of $0.6\;\mathrm{GeV}^2$
in the variable $\hat s_0$, {\it i.e.}, $\hat s_0\in [1.15 , 1.75]~\mathrm{GeV}^2$, to measure the fluctuations in the fitted value of the strong coupling, and add half the difference between the minimum
and maximum values as a theoretical uncertainty coming from the model dependence. Quoting as central value (without any strong justification) the determination at $\hat s_{0} \sim 1.55\;\mathrm{GeV}^{2}$,
as done in Ref.~\cite{Boito:2014sta}, one would get in this way
\be
\ba{c}
\alpha_{s}(m_\tau^2)^{\mathrm{CIPT}} \; =\; 0.312 \pm 0.047
\\[5pt]
\alpha_{s}(m_\tau^2)^{\mathrm{FOPT}} \; =\; 0.298 \pm 0.031
\ea
\qquad\longrightarrow\qquad
\alpha_{s}(m_\tau^2) \; =\; 0.302 \pm 0.032 \, ,
\label{DVstrong}
\ee
which has much larger uncertainties than claimed in Ref.~\cite{Boito:2014sta} and is clearly not competitive with the values derived previously with more solid methods.

One could play the same game assuming slightly different ansätze. For instance one could multiply the functional form \eqn{eq:DVparam} with a polynomial. We have repeated the exercise multiplying the ansatz with a simple power $s^n$,
to avoid increasing the number of parameters. Taking still the ``optimal'' point at $\hat s_0 = 1.55~\mathrm{GeV}^2$
(and ignoring the existing instabilities away from it), one finds significant fluctuations in the fitted value of $\alpha_s(m_\tau^2)$ when varying the power $n$,
reinforcing our error estimate in \eqn{DVstrong}. Worth mentioning, we have found ``better models'' (higher p-values) of the spectral function than the default ``$n=0$'', and they
provide significantly higher values of the strong coupling.
In Table~\ref{tab:models} we illustrate a few examples of this simple exercise, varying $n$ between $0$ and $8$. One immediately appreciates the strong correlation between $\alpha_s(m_\tau^2)$ and the power $n$.
The statistical quality of the fit improves with growing values of $n$, while the exponential parameters $\delta_V$ and $\gamma_V$ adapt themselves to compensate the growing of the ansatz spectral function at high values
of $s$ with the net result of a smaller duality-violation correction. As the fit quality improves, the central value of the fitted $\alpha_s(m_\tau^2)$ approaches the result of the ALEPH-like fit in Eq.~\eqn{strongdav}.
The slightly lower central value in Eq.~\eqn{DVstrong} is then meaningless, as the fit result is clearly model dependent.

%%%%%%%%%%%%%%%%%%%%%%%%%%%%%%% Table n %%%%%%%%%%%%%%%%%%%%%%%%%%%%%%%%%%%%%%%%%%%%
\begin{table}[t]
\centering
\renewcommand\arraystretch{1.2}
\begin{tabular}{|c|c|c|c|c|c|c|}
\hline
n  & $\alpha_{s}(m_{\tau}^{2})$ & $\delta$      & $\gamma$      & $\alpha$       & $\beta$
& p-value (\% )
\\ \hline

0  & $0.298 \pm 0.010$          & $3.6 \pm 0.5$ & $0.6 \pm 0.3$ & $-2.3 \pm 0.9$ & $4.3 \pm 0.5$
& 5.3
\\ \hline
1  & $0.300 \pm 0.012$          & $3.3 \pm 0.5$ & $1.1 \pm 0.3$ & $-2.2 \pm 1.0$ & $4.2 \pm 0.5$
& 5.7
\\ \hline
2  & $0.302 \pm 0.011$          & $2.9 \pm 0.5$ & $1.6 \pm 0.3$ & $-2.2 \pm 0.9$ & $4.2 \pm 0.5$
& 6.0
\\ \hline
4  & $0.306 \pm 0.013$          & $2.3 \pm 0.5$ & $2.6 \pm 0.3$ & $-1.9 \pm 0.9$ & $4.1 \pm 0.5$
& 6.6
\\ \hline
8  & $0.314 \pm 0.015$          & $1.0 \pm 0.5$ & $4.6 \pm 0.3$ & $-1.5 \pm 1.1$ & $3.9 \pm 0.6$
& 7.7
\\ \hline
\end{tabular}
\caption{\label{tab:models} Fitted values of $\alpha_s(m_\tau^2)$, in FOPT, and the spectral function parameters, modifying the ansatz \eqn{eq:DVparam} with a power $s^n$ (GeV units).}
\end{table}
%%%%%%%%%%%%%%%%%%%%%%%%%%%%%%%%%%%%%%%%%%%%%%%%%%%%%%%%%%%%%%%%%%%%%%%%%%%%%%%%%%%%

%%%%%%%%%%%%%%%%%%%%%%%%%%%%%%%%%%%%%%%%%%%%%%%%%%%%%%%%%%%%%%%%%%%%%
\begin{figure}[t]\centering
 \includegraphics[width=0.5\textwidth]{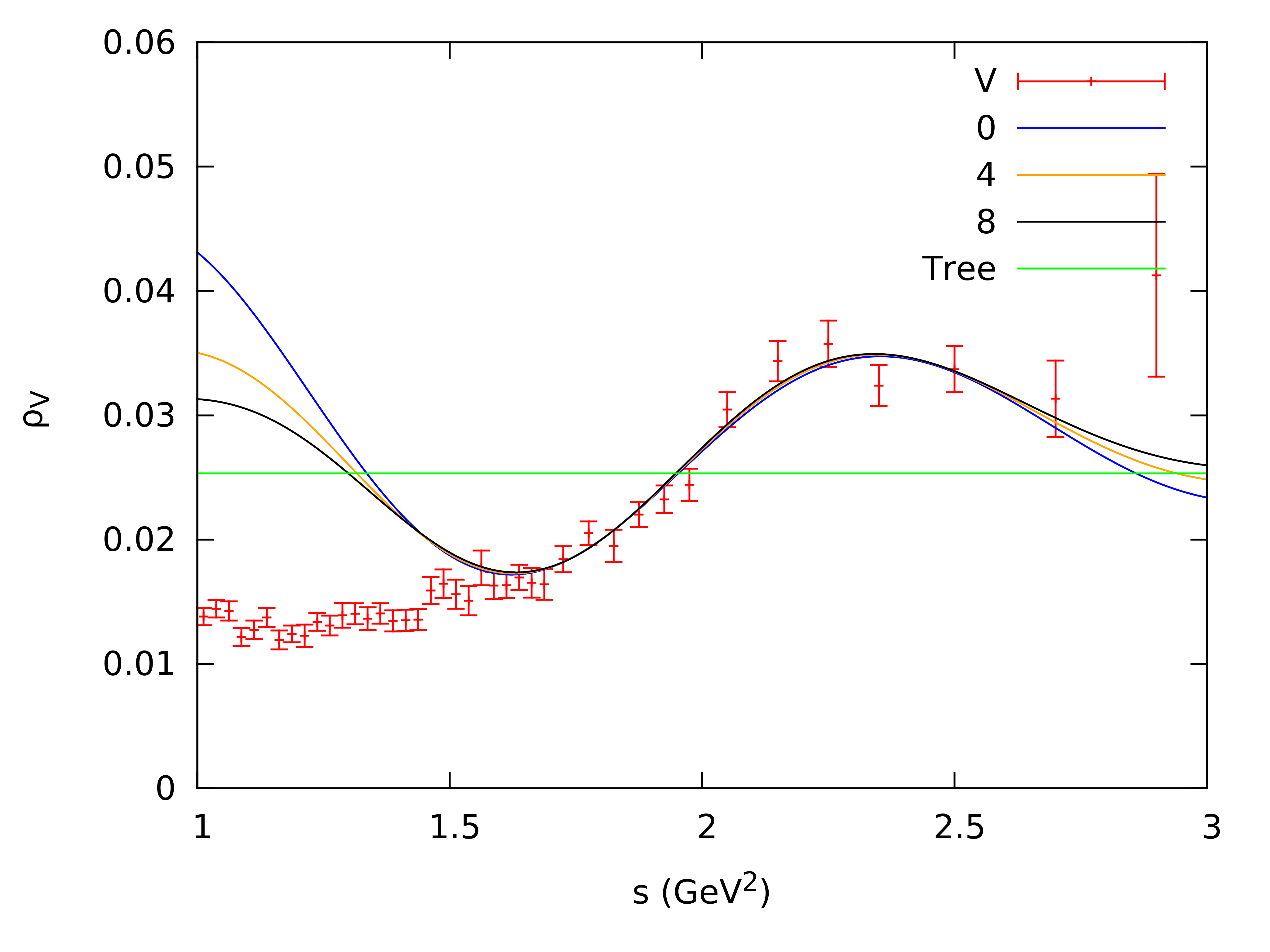}
\caption{Vector spectral function $\rho_V^{\protect\phantom{()}}(s)$, fitted
with the ansatz \eqn{eq:DVparam} multiplied by a power $s^n$, for different values of $n= 0, 4, 8$, compared with the data points.
}
\label{fig:spectral-n}
\end{figure}
%%%%%%%%%%%%%%%%%%%%%%%%%%%%%%%%%%%%%%%%%%%%%%%%%%%%%%%%%%%%%%%%%%%%%

Figure~\ref{fig:spectral-n} compares the measured vector spectral function with the fitted ansatz for $n=0$, $4$ and $8$. Although all models reproduce well the spectral function in the fitted region, they deviate very fast from the data below $1.55\;\mathrm{GeV}^2$, exhibiting a clear failure of the assumed ansatz.
As the power $n$ increases, the fit quality slightly improves and the ansatz slowly approaches the data at values of the invariant mass below the fitted range.

\section{An alternative approach}
\label{sec:Borel}

So far, we have been exploring different strategies adopted in previous works, analyzing their advantages and weaknesses. The standard approach followed in section~\ref{sec:ALEPH} appears to be on solid ground, once systematic uncertainties are properly estimated. Higher-order condensates and violations of duality are neglected, but the numerical impact of these effects can be shown to be small enough when appropriate pinched weight functions are used. In particular, the more inclusive $V+A$ channel provides a very reliable determination of the strong coupling, given in Eq.~\eqn{strongdav}. The stability of this result has been carefully studied in sections~\ref{sec:ALEPH} and \ref{sec:optimal}, using different weights. In all cases, the fits provided consistent determinations of $\alpha_s(m_\tau^2)$, in excellent agreement with \eqn{strongdav}.

The possibility to extract additional information on the higher-dimensional vacuum condensates from the $s_0$ dependence of the moments was investigated in section~\ref{sec:improvements}. It was shown there that varying $s_0$ turns out to be equivalent to a fit of the measured hadronic distribution on the physical region (the positive real axis), where the OPE cannot be applied. Nevertheless, the fits performed with the $V+A$ spectral function exhibit a quite surprising stability, suggesting that higher-order condensates and duality-violation uncertainties are not large in this channel. Taking the fluctuations with $s_0$ into account to conservatively estimate the theoretical uncertainties, we finally obtained a determination of $\alpha_s(m_\tau^2)$ from the $s_0$ dependence, given in Eq.~\eqn{segundostrong}. The amazing agreement with \eqn{strongdav} suggests a much better behaviour of perturbative QCD at low invariant masses than naively expected. This had been already noticed long time
ago in the pioneering analyses of the $s_0$ dependence performed in Refs.~\cite{Schael:2005am,Barate:1998uf,Buskulic:1993sv,Narison:1993sx,Girone:1995xb}.

\begin{figure}[t]\centering
\includegraphics[width=0.49\textwidth]{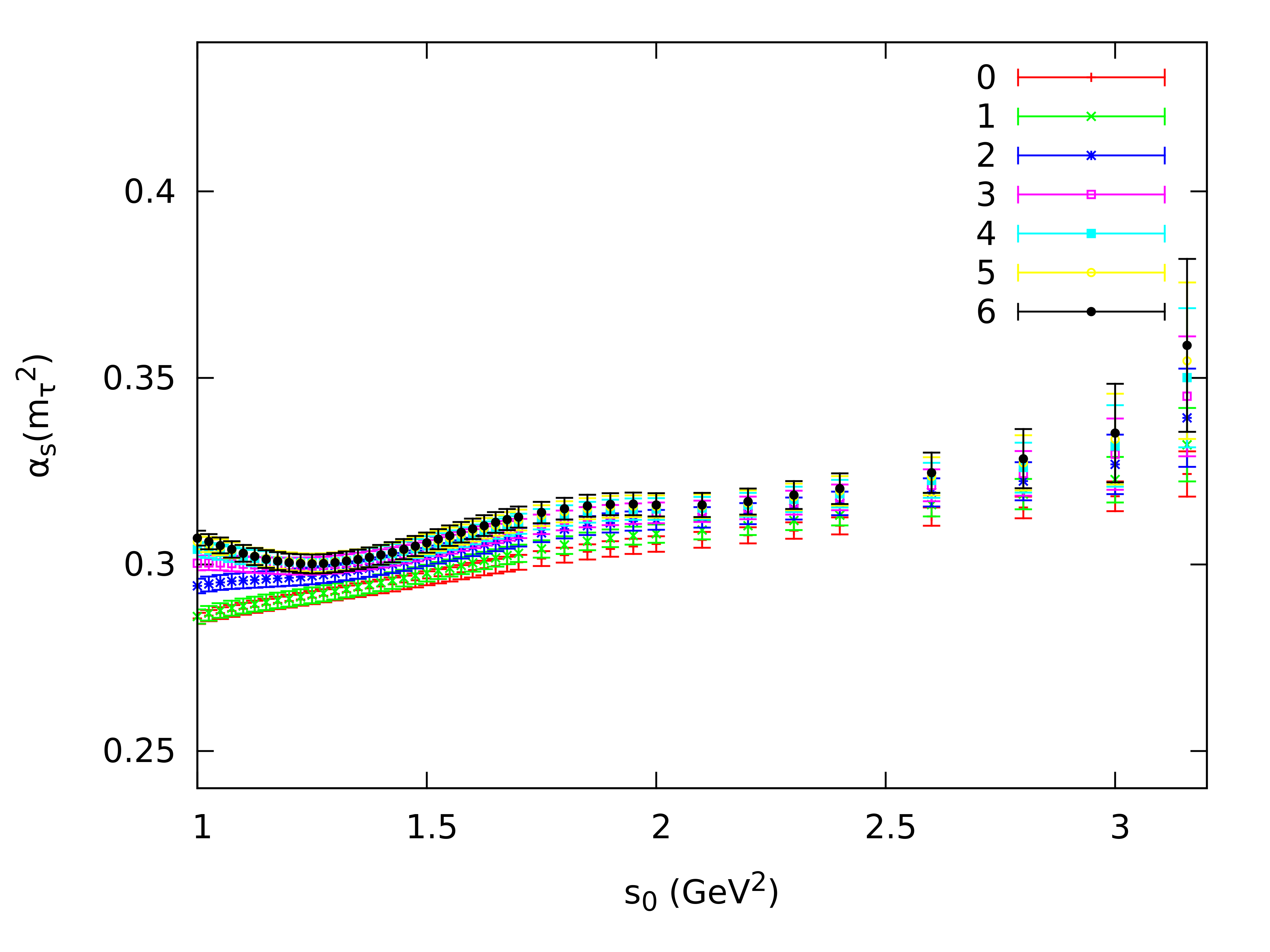}
\includegraphics[width=0.49\textwidth]{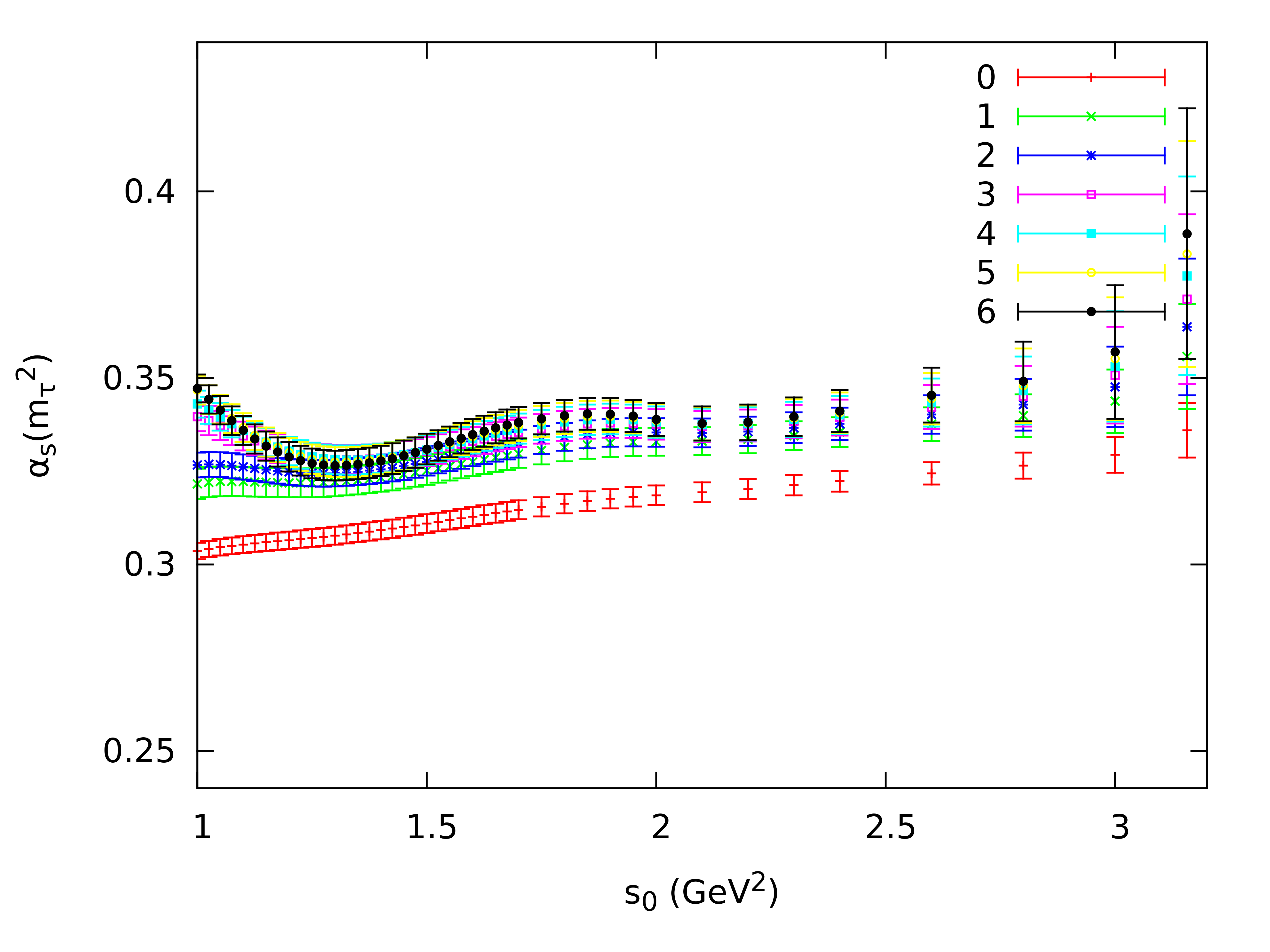}\\
\includegraphics[width=0.49\textwidth]{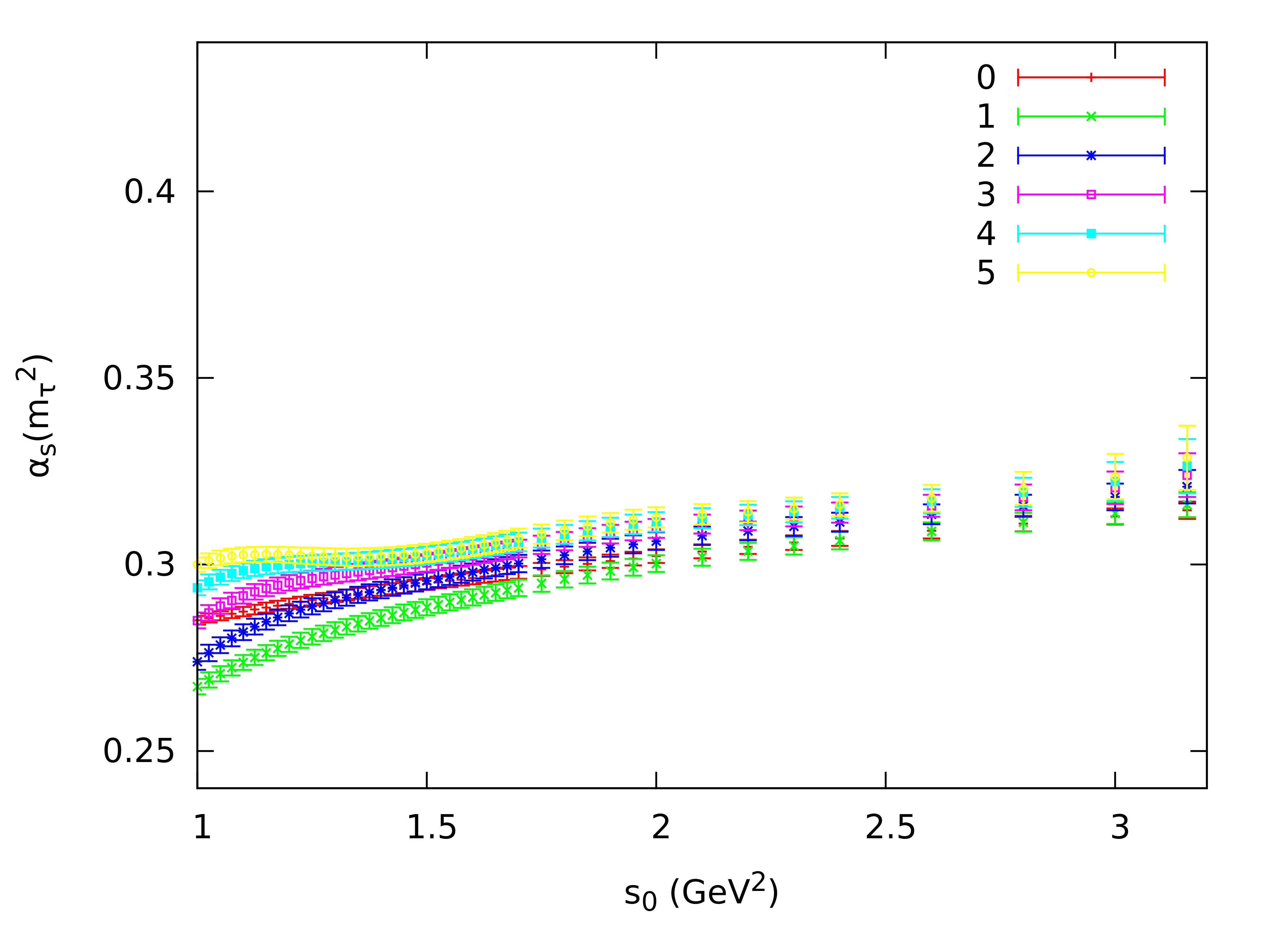}
\includegraphics[width=0.49\textwidth]{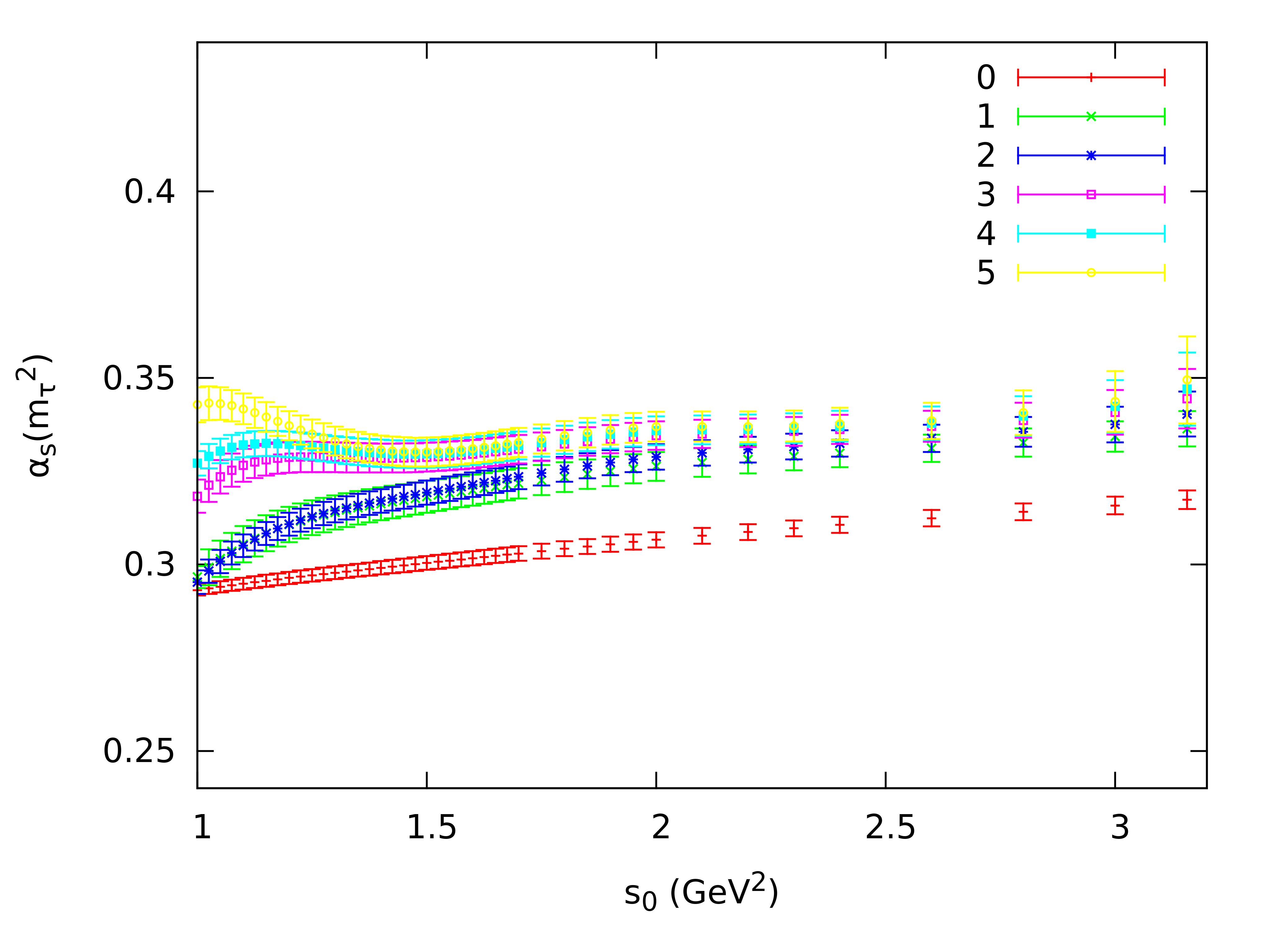}
\caption{\label{todoa0}
$V+A$ determinations of $\alpha_{s}(m_{\tau}^{2})$ from different moments, as function of $s_0$, fitted ignoring all non-perturbative corrections. The top panels show the results extracted from $A^{(1,n)}(s_{0})$ with $\{n=0,...,6\}$, while the bottom panels correspond to $A^{(2,n)}(s_{0})$ with $\{ n=0,...,5\}$. FOPT fits are on the left and CIPT on the right. Only experimental uncertainties have been included.}
\end{figure}

To better appreciate this fact, we plot in Figure~\ref{todoa0}, as function of $s_0$, the results of fits to different $A^{\omega}_{V+A}(s_0)$ moments, ignoring all non-perturbative effects.
%
%%%%%%%%%%%%%%%%%%%%%%%%%%%%%%%%%%%%%%%%%%%%%%%%%%%%%%%%%%%%%%%%%%%%%%%%%%%%%%%%%%%%%%%%
%%%%%%%%%%%%%%%%%%%%%%%%%%%%%%%%%%%%%%%%%%%%%%%%%%%%%%%%%%%%%%%%%%%%%%%%%%%%%%%%%%%%%%%%
%
The different curves correspond to the weight functions $\omega^{(1,n)}(x)$ (top panels) and $\omega^{(2,n)}(x)$ (bottom panels), defined in Eqs. (\ref{eq:omega-1n}) and (\ref{eq:omega-2n}), for $\{n=0,...,6\}$ and $\{n=0,...,5\}$, respectively. These pure perturbative determinations are shown with
the two alternative prescriptions for the $\alpha_s$ expansion, FOPT (left) and CIPT (right). The non-perturbative corrections to these 13 different moments are completely different, carrying a broad variety of inverse
powers of $s_0$:
\beqn\label{nonpertuconc}
A_{V+A}^{(1,n),\mathrm{NP}}(s_0) &=& (-1)^{n}\,\pi\;\frac{\mathcal{O}_{2n+4,V+A}}{s_{0}^{n+2}} \, ,
\\[5pt]
\label{nonpertuconc2}
A_{V+A}^{(2,n),\mathrm{NP}}(s_0) &=&
(-1)^{n}\,\pi\,\left\{ (n+2)\;\frac{\mathcal{O}_{2n+4,V+A}}{s_{0}^{n+2}}+
(n+1)\;\frac{\mathcal{O}_{2n+6,V+A}}{s_{0}^{n+3}}\right\} \, .
\eeqn
Therefore, one would expect a splitting among the different moments for a given value of $s_0$ that should increase at lower energies. This is however not seen in the figure, which exhibits a quite surprising clustering
of the different curves with a very similar dependence on $s_0$. Clearly, the OPE contributions are not the dominant feature behind the slight $s_0$ dependence observed, which should be probably ascribed to a duality-violation
effect. The small difference in normalization observed for the $A^{(k,0)}(s_0)$ CIPT case ($k=0,1$) seems more related to perturbative uncertainties. Notice that only the experimental errors have been shown in the plots.

From this perturbative exercise one could perhaps conclude that we have been too conservative when worrying about possible uncertainties from higher-dimensional condensate corrections, because their effects are not manifest
in the $V+A$ analyses, with the current experimental accuracy.

The situation seems to be different for the separate vector and axial-vector channels, with more resonance structure in their spectral functions which only flatten at higher values of $s$, specially in the $A$ case.
The fitted results are less stable and we have already seen in Figure~\ref{pinchs} a more clear indication of a sizeable power correction with $D=6$, in agreement with theoretical expectations.
Nevertheless, in the higher energy bins the power corrections seem to decrease very fast, and the fitted results from both channels tend to converge towards the more stable $V+A$ values.

Duality violations appear to be more important in the semi-inclusive $V$ and $A$ channels, except for the higher energy bins. We can try to reduce these effects by adding an exponencial term to the weight functions.
The same kind of weights were used long time ago in Refs.~\cite{Shifman:1978bx} to extract the so-called SVZ sum rules. We will pay the prize of enhancing the unknown high-energy condensate contributions. Since they are
smaller for the $A^{(1,n)}$ weights, we will take
\be
\omega_{a}^{(1,n)}(x)\; =\; \left(1- x^{n+1}\right)\; \mathrm{e}^{-ax} \, ,
\ee
which give an OPE correction
\be\label{nonpertucomp}
A_{V/A}^{\omega^{(1,n)}_a\! ,\mathrm{NP}}(s_0)\; =\;
\pi\;\sum_{D}\;\frac{\cO_{D,V/A}}{s_0^{D/2}}\;\frac{a^{\frac{D}{2}-1}}{(\frac{D}{2}-1)!}
\;\left\{ 1 \, +\, \theta (D-4-2 n)\;
\frac{(-1)^n}{a^{n+1}}\;\frac{(\frac{D}{2}-1)!}{(\frac{D}{2}-n-2)!}\right\}\, ,
\ee
with $\theta(z) = 1$ for $z\ge 0$ and zero otherwise. When $a=0$ we recover \eqn{nonpertuconc}. Notice that the OPE corrections become independent of $n$ when $a \gg 1$, since all moments are equal in that limit.

Owing to the exponential weight factor, all vacuum condensates contribute to the moments. Therefore, if the non-perturbative uncertainties are dominated by power corrections, one should expect from
Eq.~(\ref{nonpertucomp}) that a determination of the strong coupling neglecting those terms would become immediately more unstable under variations of $s_{0}$ than in the $a=0$ case, and that the splitting among moments at a given value of $s_0$ would increase, before they converge in the limit $a\to\infty$.

\begin{figure}[t]\centering
\includegraphics[width=0.323\textwidth]{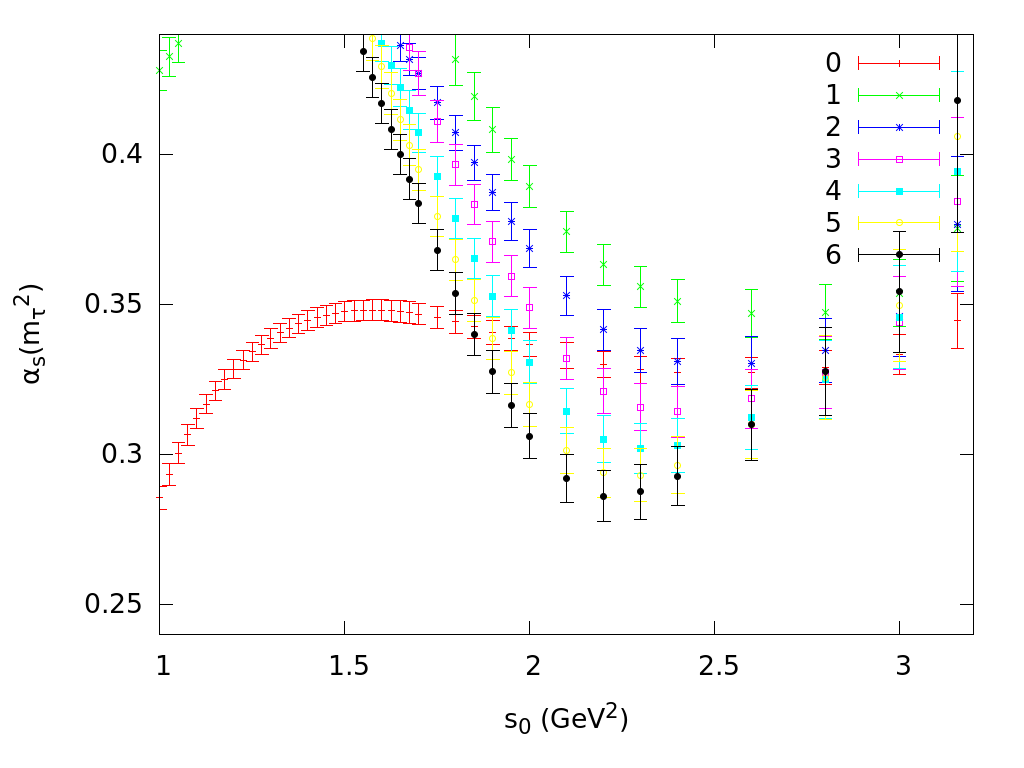}
\includegraphics[width=0.323\textwidth]{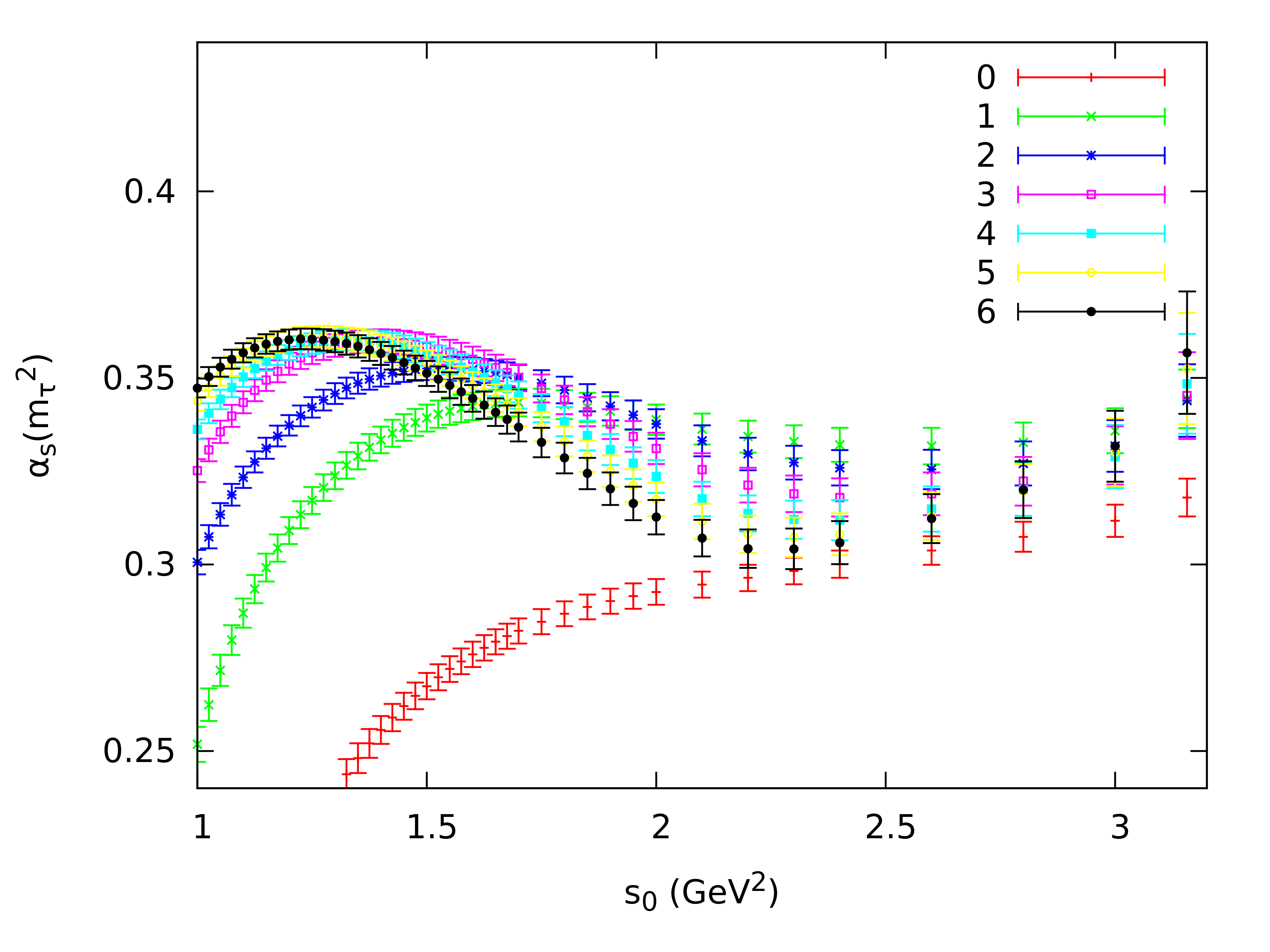}
\includegraphics[width=0.323\textwidth]{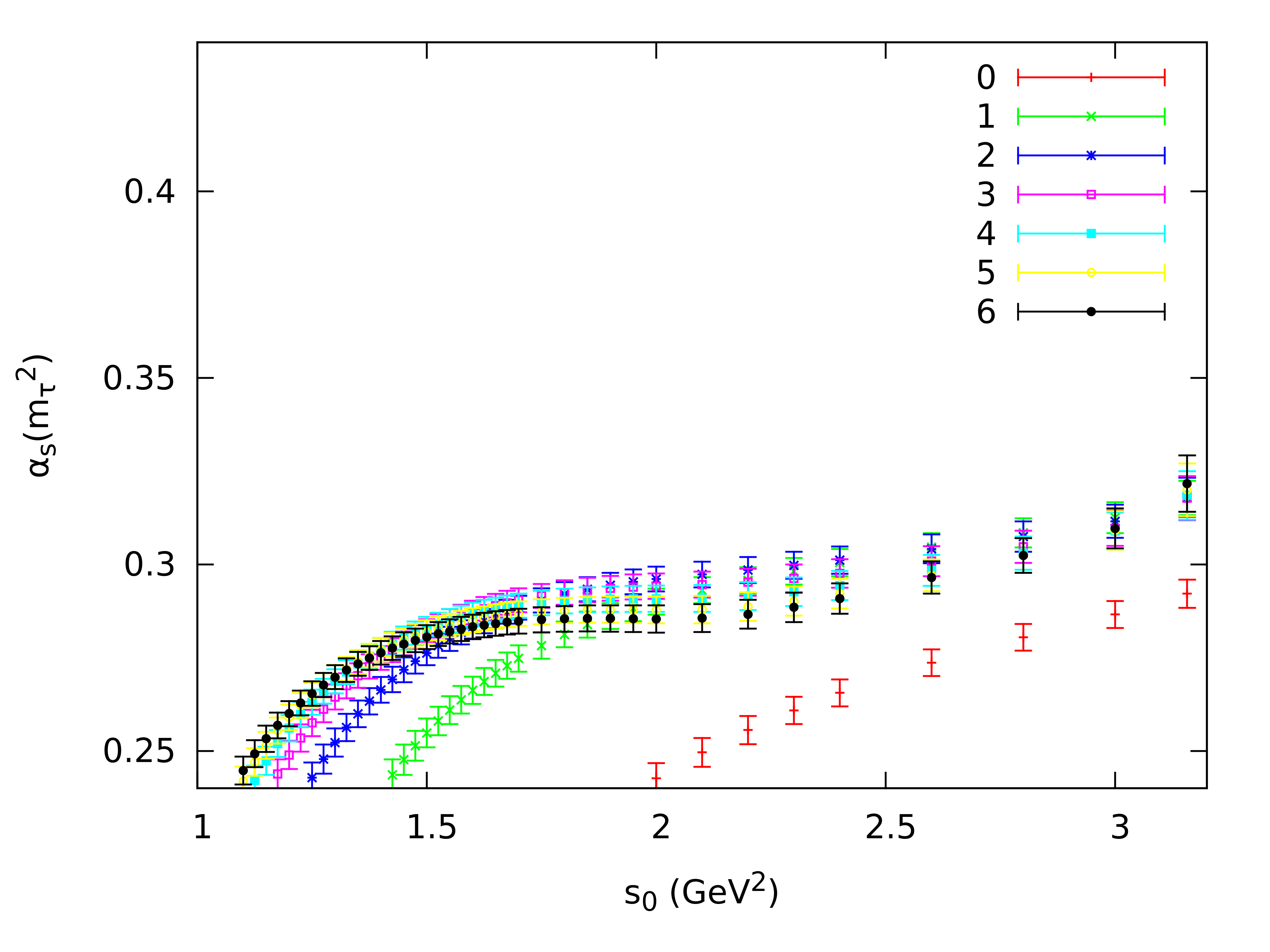}\\
\includegraphics[width=0.323\textwidth]{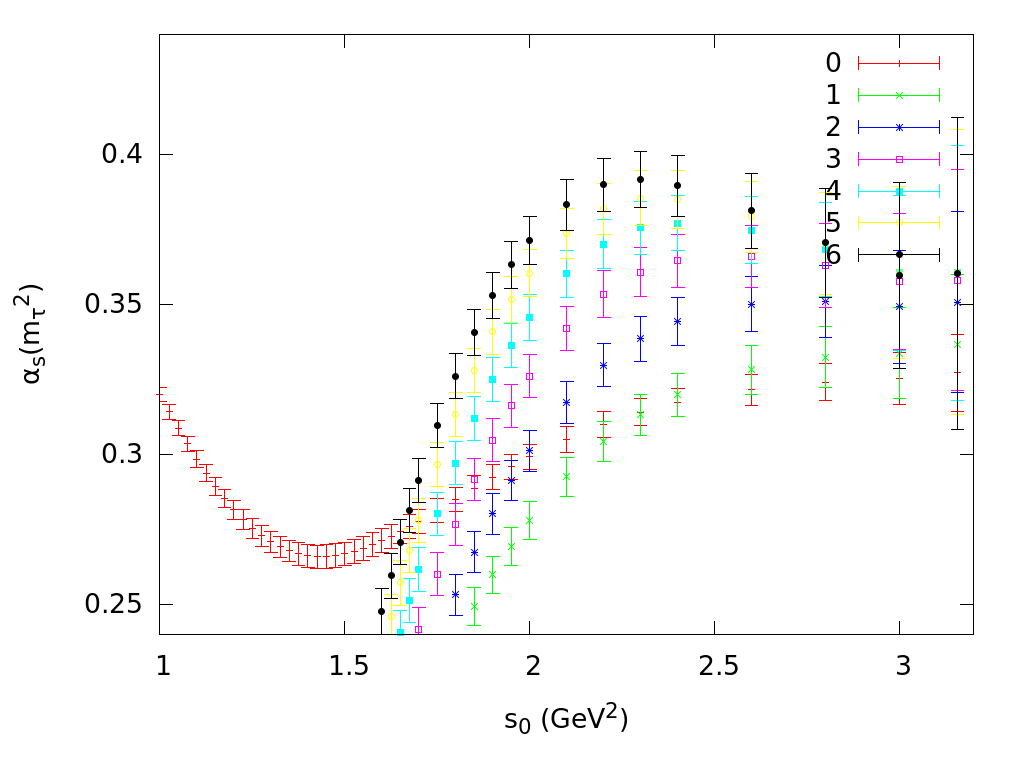}
\includegraphics[width=0.323\textwidth]{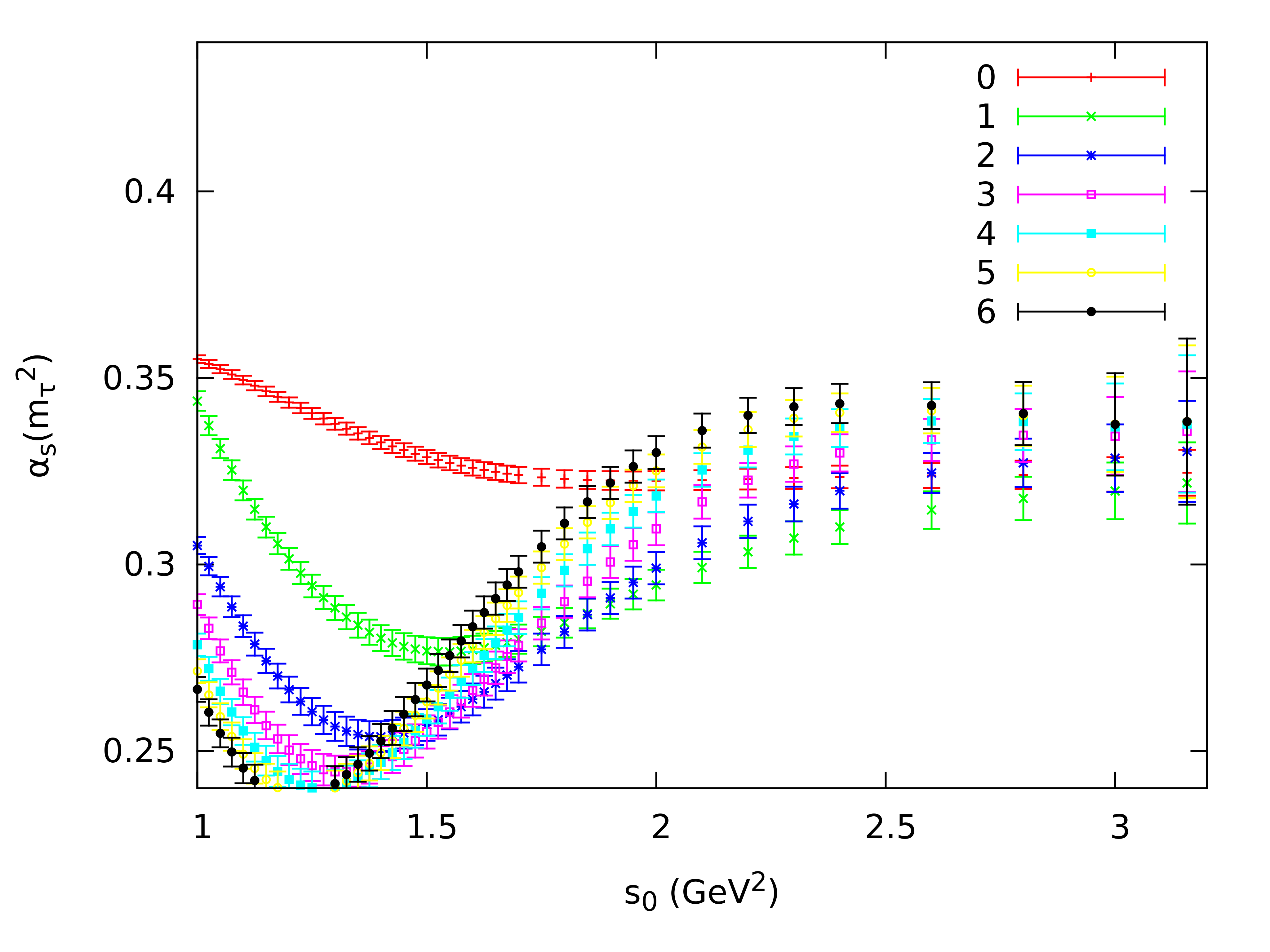}
\includegraphics[width=0.323\textwidth]{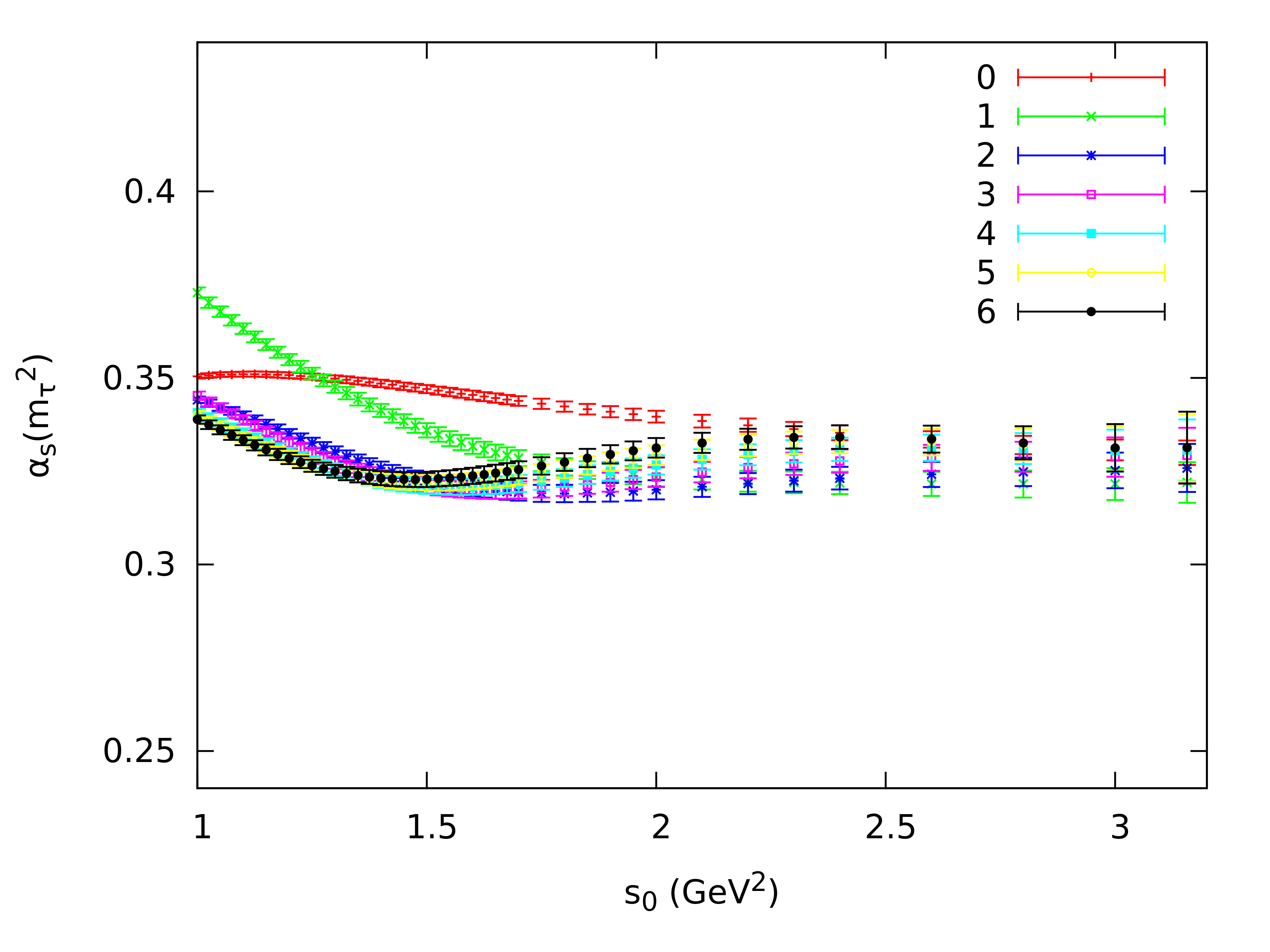}
\caption{\label{transf} CIPT determinations of $\alpha_{s}(m_{\tau}^{2})$ from the moments
$A_{V/A}^{\omega^{(1,n)}_a}(s_0)$, as function of $s_0$, ignoring all non-perturbative corrections and evaluated at $a=0$ (left), 1 (center) and 2 (right). The top (bottom) panels correspond to the vector (axial-vector) distribution. Only experimental uncertainties have been included.}
\end{figure}

In Figure \ref{transf} we show, as function of $s_0$, the determinations of $\alpha_s(m_\tau^2)$ extracted from 7 different $A_{V/A}^{\omega^{(1,n)}_a}(s_0)$ moments ($n=0,\ldots,6$), neglecting all non-perturbative corrections. We show the results obtained with CIPT and three different choices of $a=0,1,2$. Again, only experimental uncertainties have been included in the plots.

%%%%%%%%%%%%%%%%%%%%%%%%%%%%%%%%%%%%%%%%%%%%%%%%%%%%%%%%%%%%%%%%%%%%%%%%%%%%%%%%%%%
%%%%%%%%%%%%%%%%%%%%%%%%%%%%%%%%%%%%%%%%%%%%%%%%%%%%%%%%%%%%%%%%%%%%%%%%%%%%%%%%%%%

It is evident from the panels that with a non-zero Borel parameter $a$ one gets more stable results,
and the different moments converge very soon when $a$ is increased. This indicates that, for these weight functions and for the plotted ranges of $s_0$ and $a$, non-perturbative uncertainties are probably more affected by duality-violation effects than by power corrections. Of course, if one takes $a$ too large, higher-dimensional condensate corrections will become dominant, and the extracted values of $\alpha_{s}(m_{\tau}^{2})$ will depend strongly on $s_{0}$.
We observe in Figure \ref{transf} that this is actually starting to happen in the $V$ channel, at $a\sim 2$.

\begin{figure}[t]\centering
\includegraphics[width=0.49\textwidth]{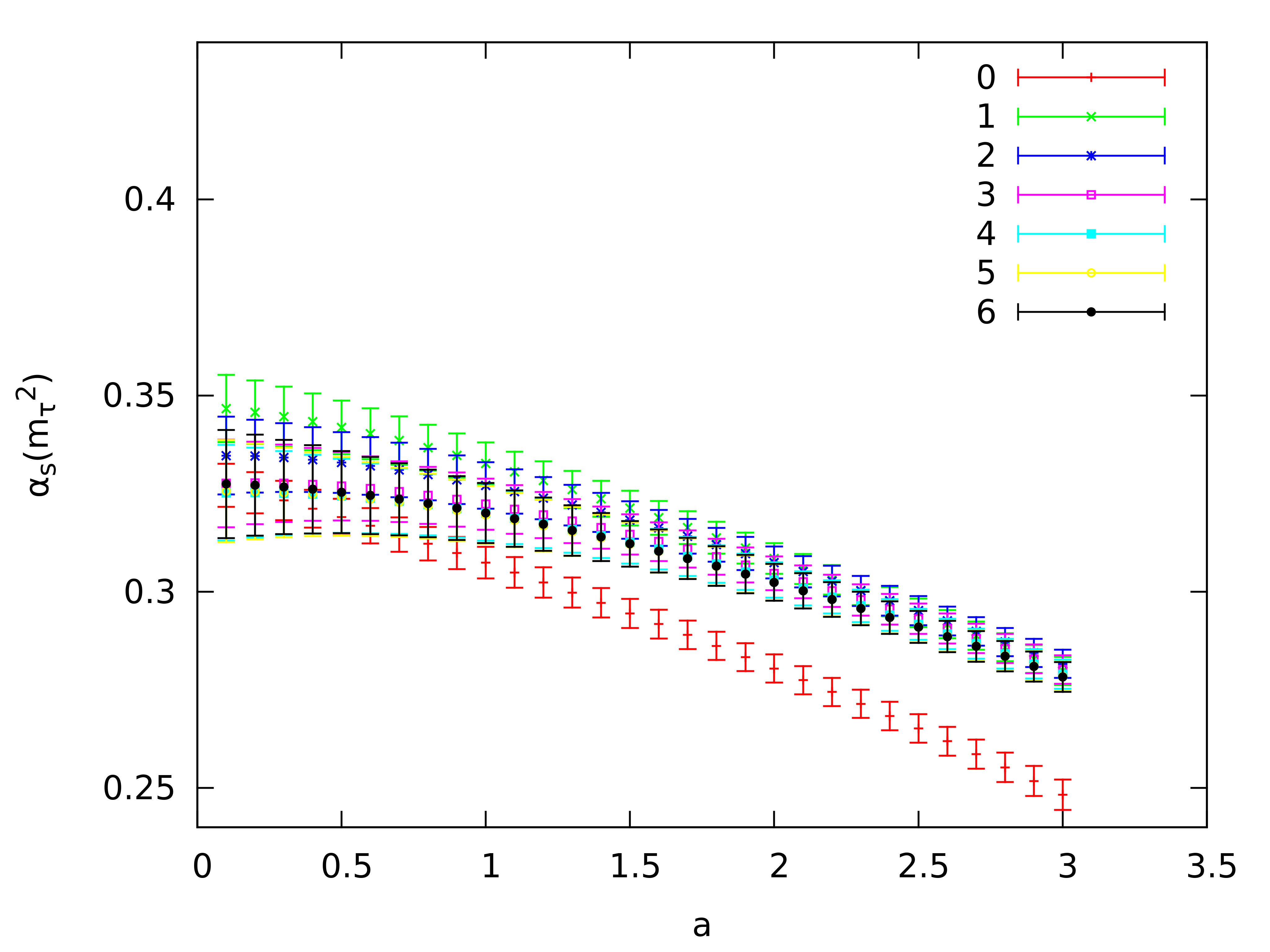}
\includegraphics[width=0.49\textwidth]{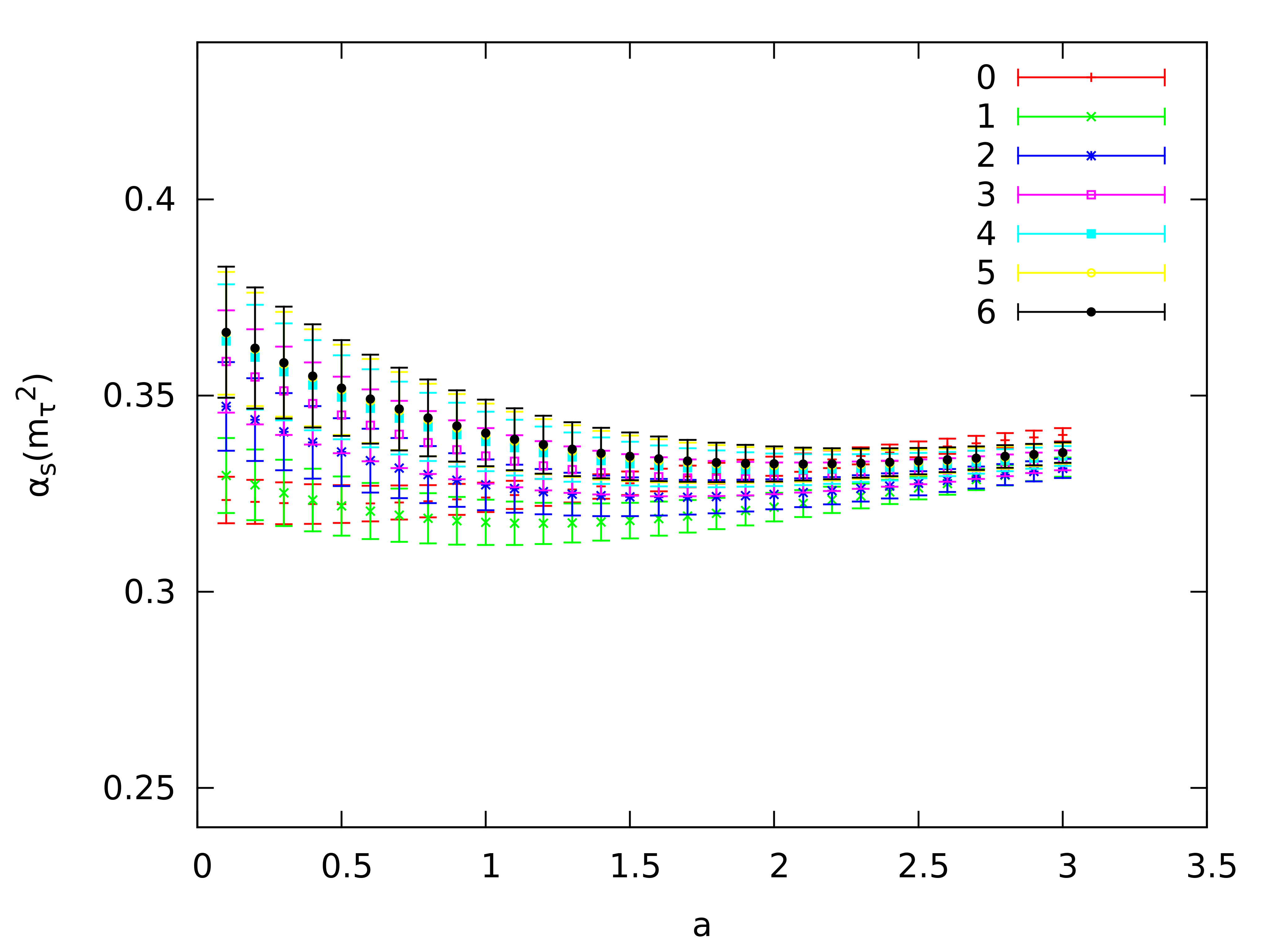}
\caption{\label{bor} CIPT seterminations of $\alpha_{s}(m_{\tau}^{2})$ from the moments
$A_{V/A}^{\omega^{(1,n)}_a}(s_0)$, at $s_{0}=2.8\;\mathrm{GeV}^{2}$ and ignoring all non-perturbative corrections, for different values of the Borel parameter $a$ (V on the left and A on the right). Only experimental uncertainties are included.}
\end{figure}

In Figure \ref{bor}, we plot the determinations of $\alpha_{s}(m_\tau^2)$  at a fixed\footnote{We take this value as a reference point because it is the largest invariant-mass bin with enough experimental resolution.}
value of $s_{0}=2.8\;\mathrm{GeV}^{2}$, as a function of the Borel parameter $a$.
We observe how in the region where the strong coupling is stable, {\it i.e.}, $\frac{d\alpha_{s}}{da}\sim 0$, there is a similar stability range under variations of $s_{0}$, for every moment.
This reinforces the idea that there exist a range of values of $a$, large enough to minimize duality-violation effects and not so large to get dominant condensate corrections, so that it is the best region
for determining $\alpha_{s}(m_{\tau}^{2})$.

%%%%%%%%%%%%%%%%%%%%%%%%%%%%%%%%%%%%%%%%%%%%%%%%%%%%%%%%%%%%%%%%%%%%%%%%%%%%%%%%%%%%
%%%%%%%%%%%%%%%%%%%%%%%%%%%%%%%%%%%%%%%%%%%%%%%%%%%%%%%%%%%%%%%%%%%%%%%%%%%%%%%%%%%%

In order to extract a reliable value for the strong coupling, using what we know, we start by taking as reference the point $s_{0}=2.8$ GeV$^{2}$. For every moment and channel ($V$ or $A$), we only
accept values of $\alpha_{s}$ in the stability region of the Borel transform,\footnote{We have checked that results are not really different if we remove this condition. This happens because the Borel-stable region
is very similar to the $s_{0}$-stable one, for all moments at every experimental channel, as one would expect if all non-perturbative effects are indeed small in that Borel region.}
{\it i.e.}, those whose central values are within the experimental errors of the derivative-zero point $\frac{d\alpha_{s}}{da}=0$.
For every value of $a$ in that region, we have an $\alpha_{s}(m_{\tau}^{2})$ value. Its error is calculated adding quadratically to the experimental error the perturbative uncertainty, estimated varying $K_{5}$ and the scale $\xi$ with the same criteria as above, and the non-perturbative one, calculated conservatively as the maximum value minus the minimum one in the region $s_0\in [2,2.8]\;\mathrm{GeV}^{2}$. We choose as the optimal value for every moment the one that gives the minimum total error.
Finally, we take as central value of the $7$ moments the one closest to the average and its error summed quadratically to half the difference between the maximum and minimum $\alpha_{s}(m_{\tau}^{2})$ value (as a second non-perturbative uncertainty, more related with the neglected vacuum condensates) to get a conservative estimate of the total uncertainty. We obtain in this way:
\beqn
\alpha_{s}(m_\tau^2)^{V, \mathrm{CIPT}} &=& 0.326 \, {}^{+\, 0.021}_{-\, 0.019}\, ,
\no\\[5pt]
\alpha_{s}(m_\tau^2)^{A, \mathrm{CIPT}} &=& 0.325 \, {}^{+\, 0.018}_{-\, 0.014}\, ,
\no\\[5pt]
\alpha_{s}(m_\tau^2)^{V, \mathrm{FOPT}} &=& 0.314 \, {}^{+\, 0.015}_{-\, 0.011}\, ,
\no\\[5pt]
\alpha_{s}(m_\tau^2)^{A, \mathrm{FOPT}} &=& 0.320 \, {}^{+\, 0.019}_{-\, 0.016}\, .
\eeqn
Thus, we find a very good consistency between the determinations performed in the vector and axial-vector channels. Taking the average of both experimental channels and keeping the minimum error, we get finally
\be
\ba{c}
\alpha_{s}(m_\tau^2)^{\mathrm{CIPT}} \; =\; 0.325 \, {}^{+\, 0.018}_{-\, 0.014}
\\[5pt]
\alpha_{s}(m_\tau^2)^{\mathrm{FOPT}} \; =\; 0.317 \, {}^{+\, 0.015}_{-\, 0.011}
\ea
\qquad\longrightarrow\qquad
\alpha_{s}(m_\tau^2) \; =\; 0.321 \, {}^{+\, 0.016}_{-\, 0.012} \, .
\label{alfafin}
\ee

\begin{figure}[p]\centering
\includegraphics[width=0.326\textwidth]{pinch1v+afopt.png}
\includegraphics[width=0.326\textwidth]{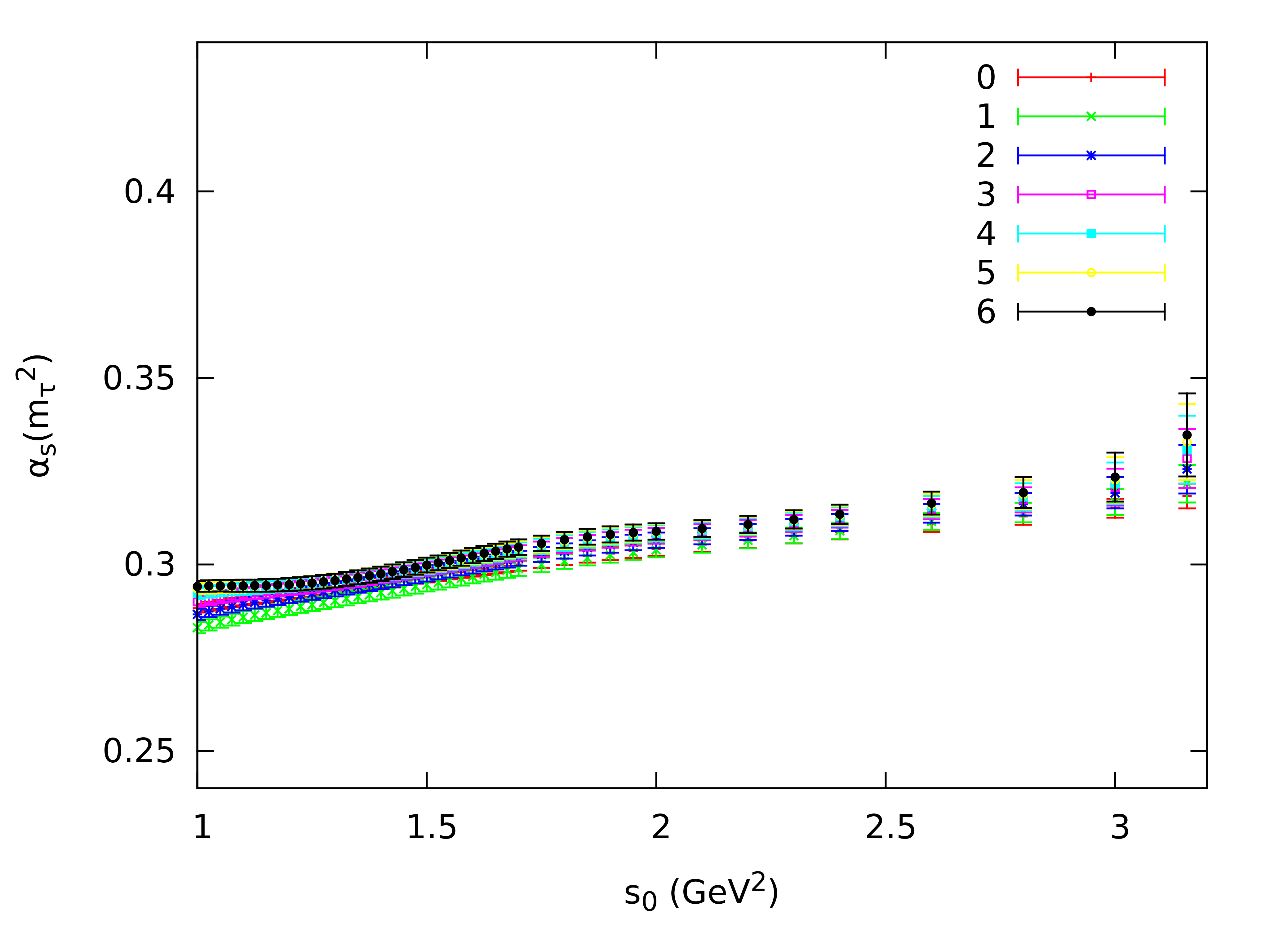}
\includegraphics[width=0.326\textwidth]{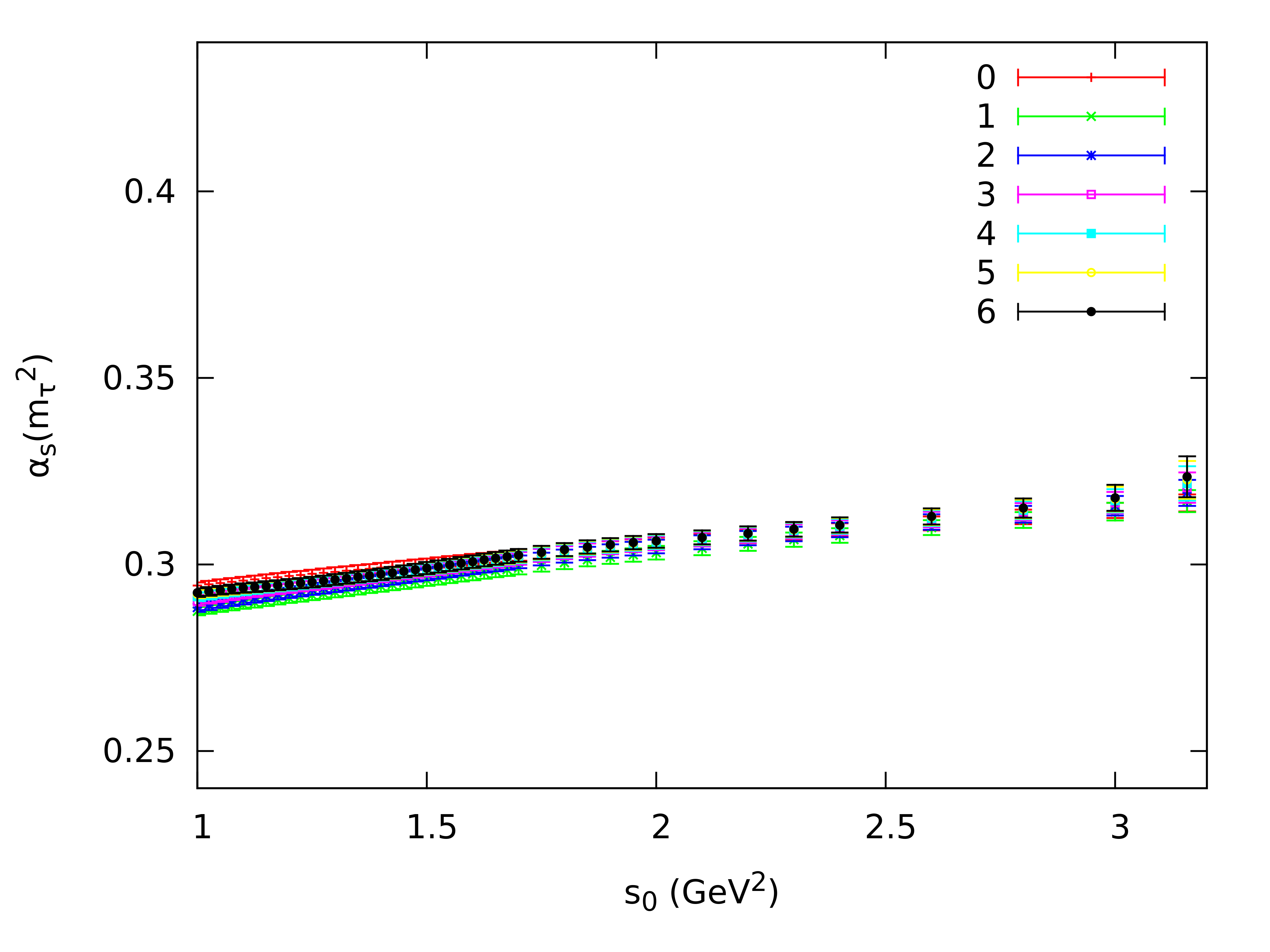}\\
\includegraphics[width=0.326\textwidth]{pinch1v+acipt.png}
\includegraphics[width=0.326\textwidth]{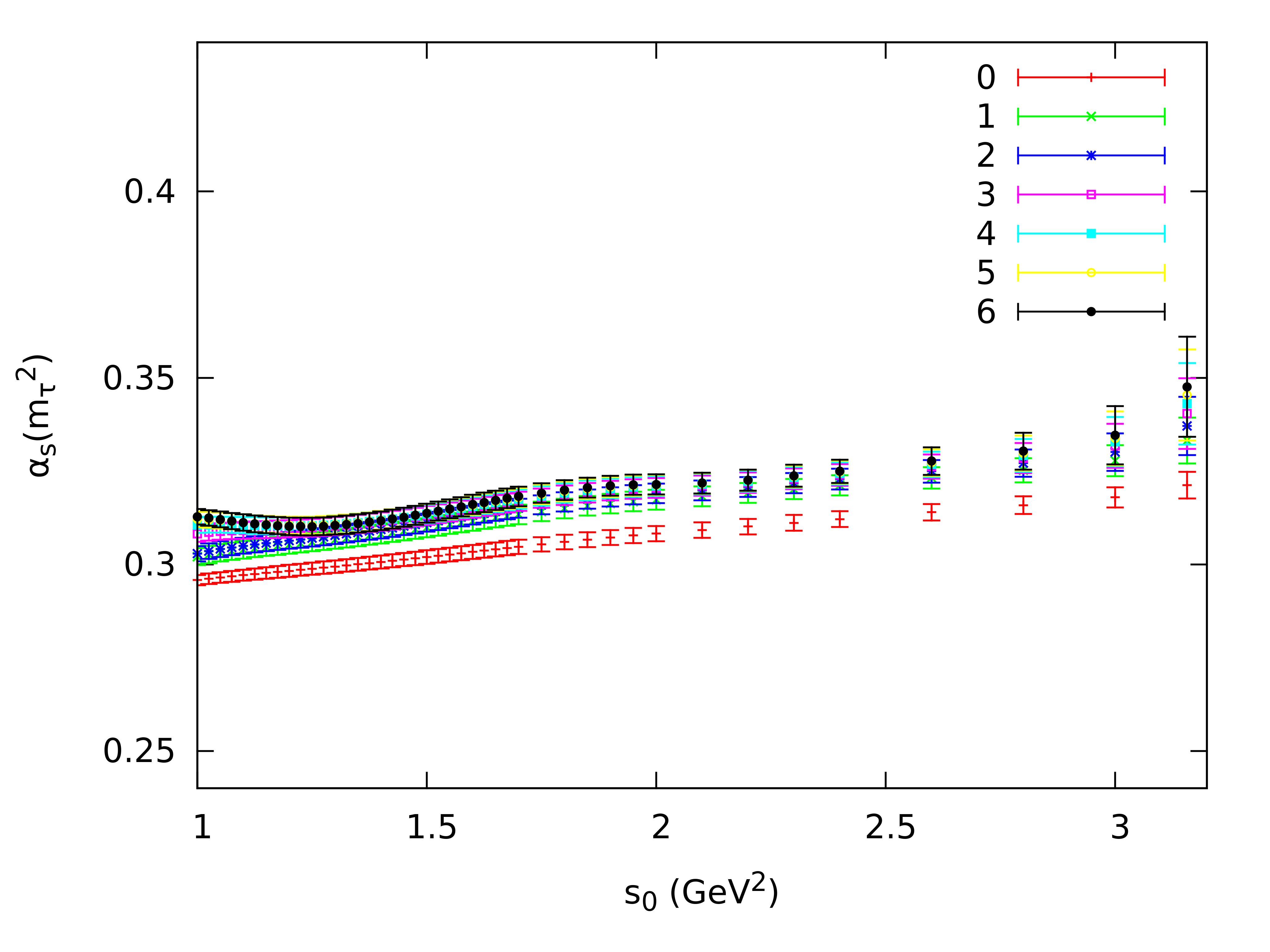}
\includegraphics[width=0.326\textwidth]{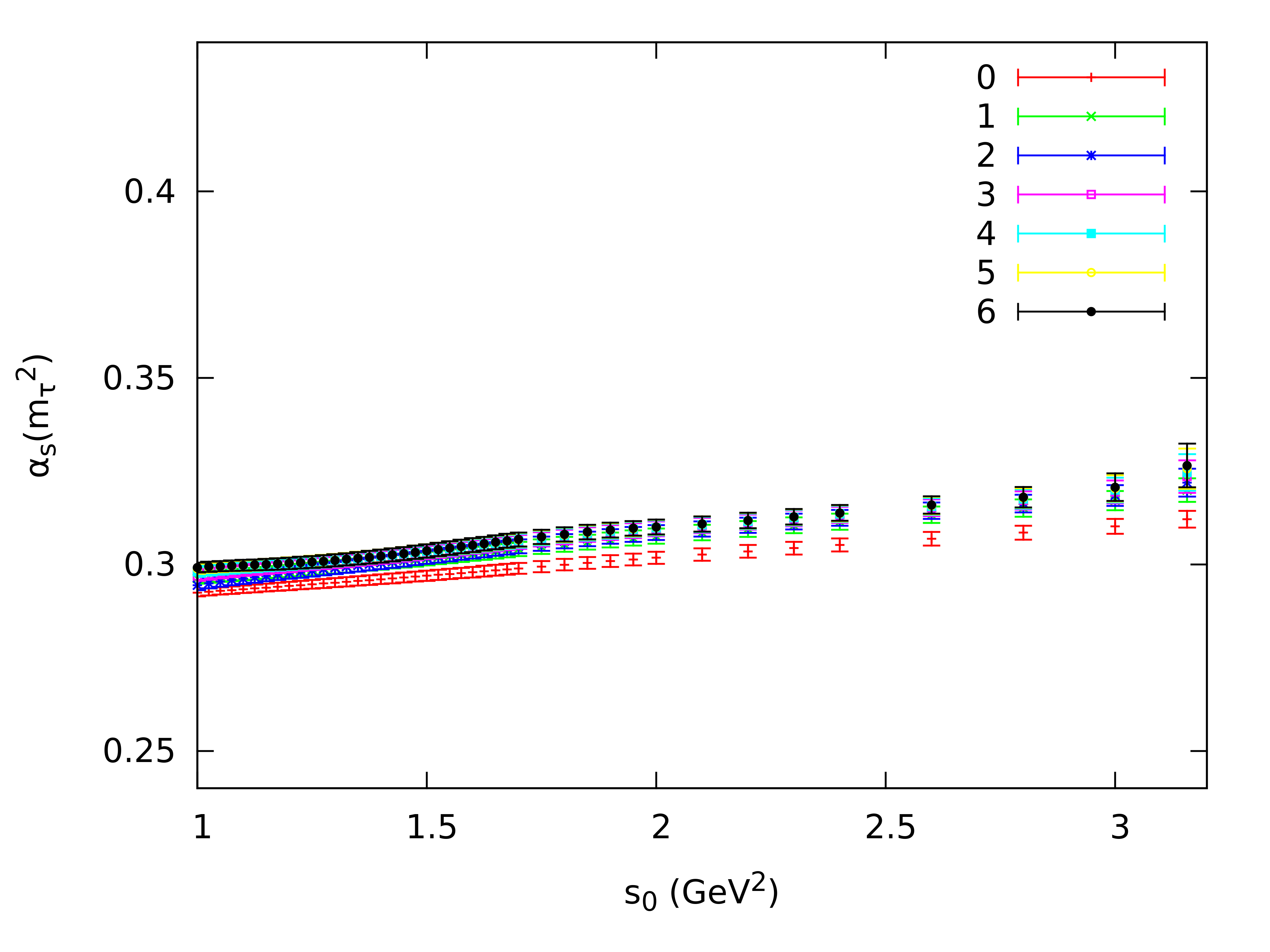}
\caption{\label{transf2} Determinations of $\alpha_{s}(m_{\tau}^{2})$ from the moments
$A_{V+A}^{\omega^{(1,n)}_a}(s_0)$, as function of $s_0$, ignoring all non-perturbative corrections and evaluated at $a=0$ (left), 1 (center) and 2 (right). The top (bottom) panels correspond to FOPT (CIPT). Only experimental uncertainties have been included.}
%\end{figure}
%%%%%%%%%%%%%%%%%%%%%%%%%%%%%%%%%%%%%%%%%%%%%%%%%%%%%%%%%%%%%%%%%%%%%%%%%%%%%%%%%%%
\vskip .5cm
%%%%%%%%%%%%%%%%%%%%%%%%%%%%%%%%%%%%%%%%%%%%%%%%%%%%%%%%%%%%%%%%%%%%%%%%%%%%%%%%%%%%
%\begin{figure}[t]\centering
\includegraphics[width=0.49\textwidth]{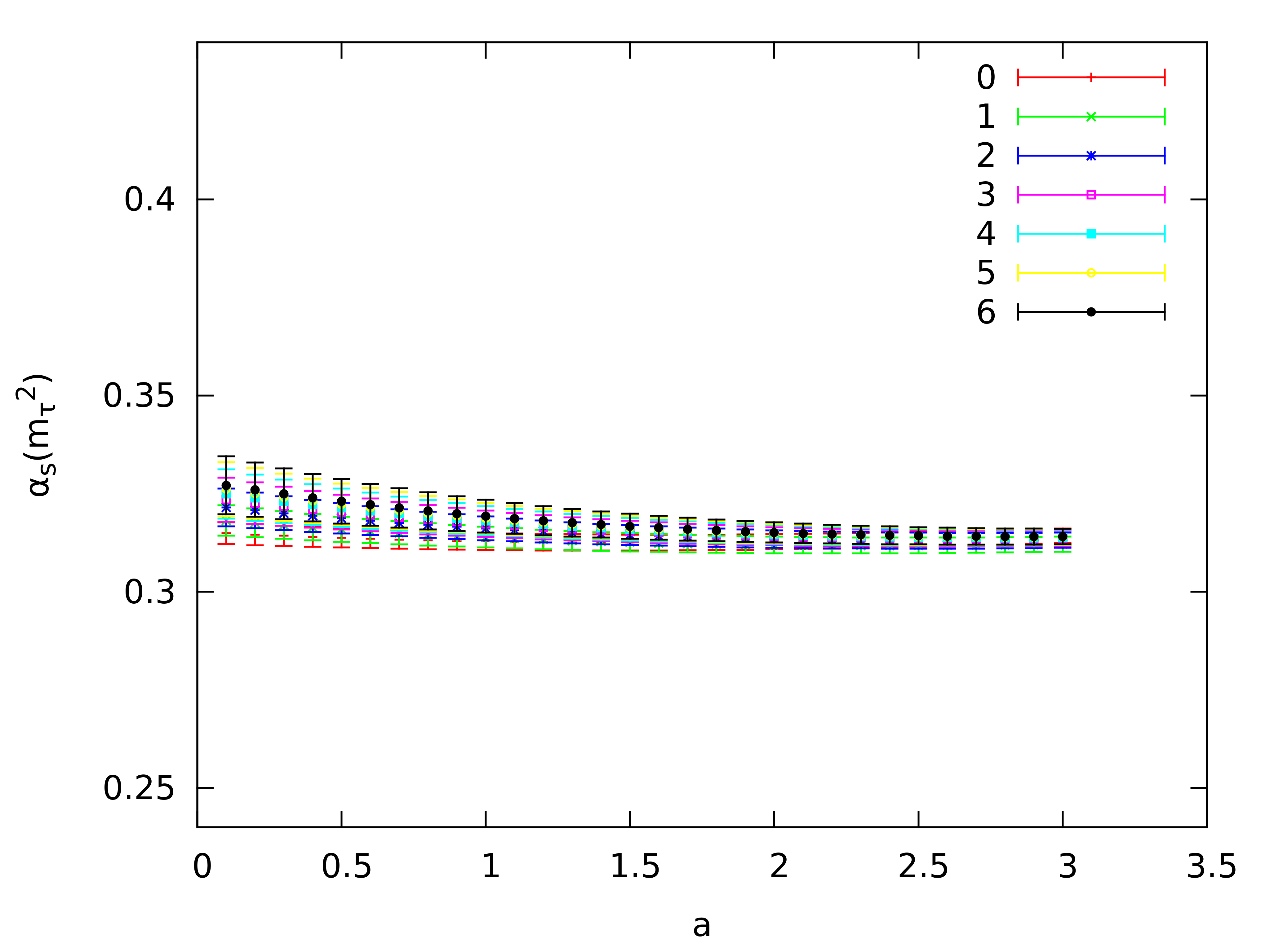}
\includegraphics[width=0.49\textwidth]{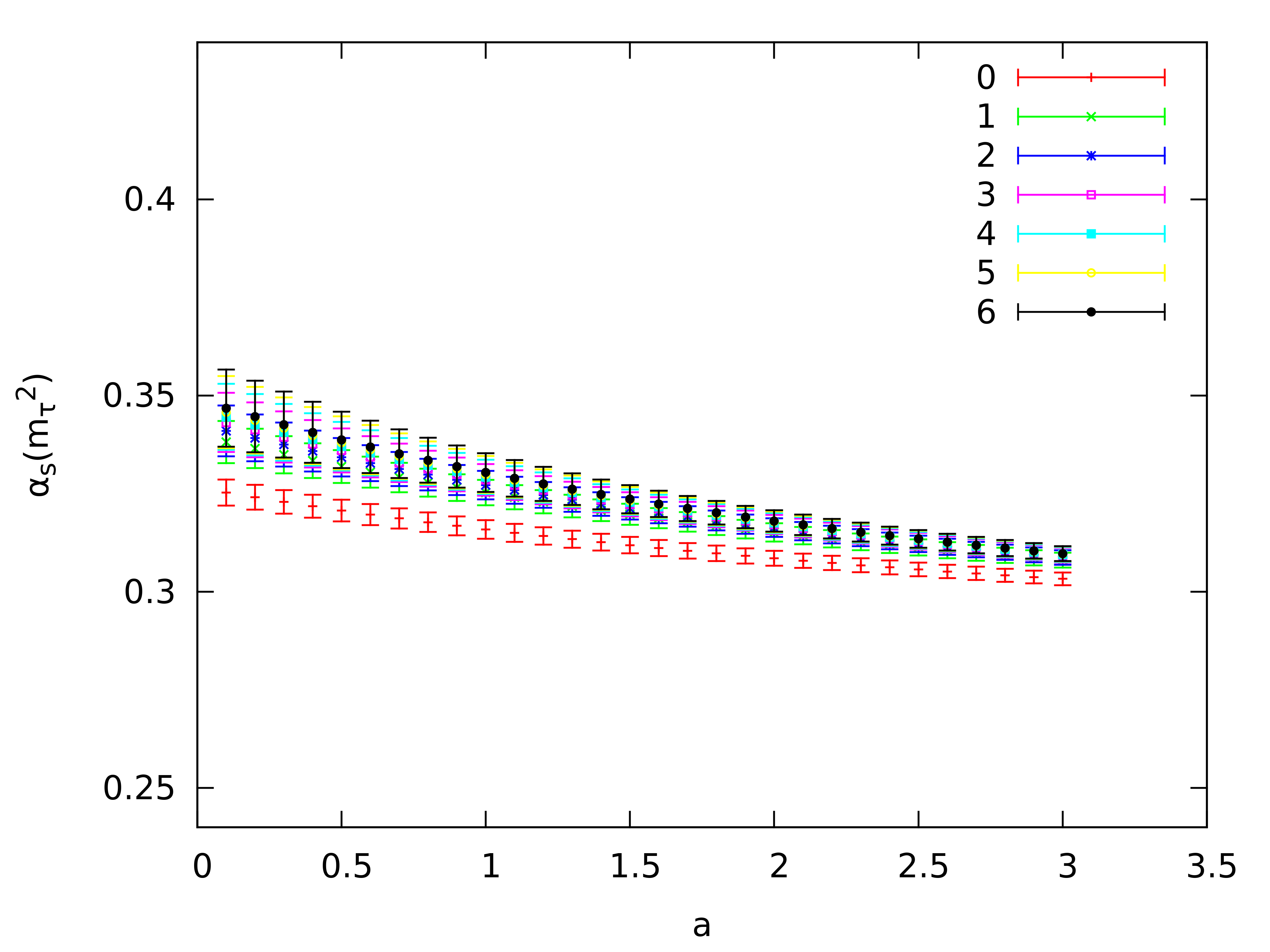}
\caption{\label{bor2} Determinations of $\alpha_{s}(m_{\tau}^{2})$ from the moments
$A_{V+A}^{\omega^{(1,n)}_a}(s_0)$, at $s_{0}=2.8\;\mathrm{GeV}^{2}$ and ignoring all non-perturbative corrections, for different values of $a$ (FOPT on the left and CIPT on the right). Only experimental uncertainties are included.}
\end{figure}

One can play a similar game with the $V+A$ channel. The resulting $\alpha_s(m_\tau^2)$ determinations are plotted in Figures~\ref{transf2} and \ref{bor2}, as function of $s_0$ and $a$, respectively.
Since for $V+A$ one observes a slightly different behaviour in FOPT and CIPT, the results of both perturbative approaches are shown in the figures.
Increasing the Borel parameter $a$ does not bring in this case any clear improvement in the stability under $s_{0}$ (Figure~\ref{transf2}), because the duality-violation effects are smaller for $V+A$.
When we reduce the tiny duality-violation effects, the condensate corrections could become dominant. The different qualitative behaviour observed in Figure~\ref{bor2} for FOPT and CIPT reflects the difficulties
in extracting conclusions with this method about the tiny non-perturbative corrections in the $V+A$ channel, within the much larger perturbative uncertainties.

%%%%%%%%%%%%%%%%%%%%%%%%%%%%%%%%%%%%%%%%%%%%%%%%%%%%%%%%%%%%%%%%%%%%%%%%%%%%%%%%%%%
%%%%%%%%%%%%%%%%%%%%%%%%%%%%%%%%%%%%%%%%%%%%%%%%%%%%%%%%%%%%%%%%%%%%%%%%%%%%%%%%%%%%

Applying the same method as in the separate vector and axial-vector channels, in FOPT we can derive from these plots a combined determination of the strong coupling in a completely straightforward way. A little bit more care has to be taken in CIPT because of the absence of a derivative-zero point ($\frac{d\alpha_{s}(m_\tau^2)}{da}=0$) in Figure~\ref{bor2}. We can try two different possibilities: either accept only the small stability region in the separate V and A channels, or apply the method without imposing that constraint.
We find the same result with both procedures. Our final results from the $V+A$ channel are:
\be
\ba{c}
\alpha_{s}(m_\tau^2)^{\mathrm{CIPT}} \; =\; 0.328 \, {}^{+\, 0.014}_{-\, 0.013}
\\[5pt]
\alpha_{s}(m_\tau^2)^{\mathrm{FOPT}} \; =\; 0.318 \, {}^{+\, 0.015}_{-\, 0.012}
\ea
\qquad\longrightarrow\qquad
\alpha_{s}(m_\tau^2) \; =\; 0.323 \, {}^{+\, 0.015}_{-\, 0.013} \, ,
\label{BorelVpAstrong}
\ee
in good agreement with \eqn{alfafin}.

\section{Summary}
\label{sec:summary}

We have presented a thorough numerical reanalysis of the $\alpha_s$ determination from $\tau$ decay data, using the most recent release of the experimental ALEPH data \cite{Davier:2013sfa}. Our main goal has been to
achieve a quantitative assessment of the role of non-perturbative effects, either from inverse-power corrections or violations of duality. While these corrections are known to be
small \cite{Pich:2013lsa,Braaten:1991qm,Pich:2015ivv,Pich:2011bb}, the current level of $\cO(\alpha_s^4)$ perturbative precision requires a careful study of this type of contributions.

In order to be sensitive to non-perturbative effects, one needs to go beyond the very clean $R_\tau$ ratio \cite{Braaten:1991qm} and investigate moments of the hadronic invariant-mass distribution in $\tau$
decays~\cite{LeDiberder:1992zhd}. Several strategies have been advocated in previous works, with different advantages and disadvantages. We have investigated all of them, trying to uncover their potential
hidden weaknesses and test the stability of the obtained results under slight variations of the assumed inputs. Moreover, we have put forward various novel approaches which allow to study complementary aspects of the problem.

Perturbative uncertainties from unknown higher-order corrections dominate the final error of the $\alpha_{s}(m_{\tau}^{2})$ determination, being at present the main limitation on the achievable
accuracy \cite{Pich:2013lsa}. In particular, two different prescriptions to handle the renormalization-group-improved perturbative series, CIPT and FOPT, lead to systematic differences on the extracted value of the strong coupling, with $\alpha_{s}(m_{\tau}^{2})$ slightly smaller in the FOPT case. While CIPT resums very efficiently the known sources of large logarithms \cite{LeDiberder:1992jjr,Pivovarov:1991rh},
the more naive FOPT procedure has been advocated to approach better the Borel-summed result if the series is already asymptotic at $\cO(\alpha_s^4)$ \cite{Beneke:2008ad,Caprini:2009vf}. In the absence
of a better understanding of the perturbative behaviour at higher orders, we have performed all our analyses with the two prescriptions.

In Table~\ref{tab:summary} we summarize our determinations of $\alpha_{s}(m_{\tau}^{2})$, obtained with different methods from the $V+A$ spectral distribution. The numbers in the table are representative of
the various strategies that we have investigated, and all of them have been corroborated with additional tests and stability studies of the final numerical results. Overall, our results exhibit a very consistent
pattern, being the agreement among them much better than what one should expect from the quoted uncertainties.

%%%%%%%%%%%%%%%%%%%%%%%%%%%%%%% Table n %%%%%%%%%%%%%%%%%%%%%%%%%%%%%%%%%%%%%%%%%%%%
\begin{table}[t]
\centering\
\renewcommand\arraystretch{1.2}
\begin{tabular}{|c|c|c|c|}
\hline
Method \ \& \ Eq. (\#) & \multicolumn{3}{|c|}{$\alpha_{s}(m_{\tau}^{2})$}
\\ \cline{2-4}
& CIPT & FOPT & Average
\\ \hline
ALEPH moments \eqn{strongdav} & $0.339 \,{}^{+\, 0.019}_{-\, 0.017}$ &
$0.319 \,{}^{+\, 0.017}_{-\, 0.015}$ & $0.329 \, {}^{+\, 0.020}_{-\, 0.018}$
\\
Modified ALEPH moments \eqn{strongdavbis} & $0.338 \,{}^{+\, 0.014}_{-\, 0.012}$ &
$0.319 \,{}^{+\, 0.013}_{-\, 0.010}$ & $0.329 \, {}^{+\, 0.016}_{-\, 0.014}$
\\
$A^{(2,m)}$ moments \eqn{strong2mMom} & $0.336 \,{}^{+\, 0.018}_{-\, 0.016}$ &
$0.317 \,{}^{+\, 0.015}_{-\, 0.013}$ & $0.326 \, {}^{+\, 0.018}_{-\, 0.016}$
\\
$s_0$ dependence \eqn{segundostrong} & $0.335 \pm 0.014$ &
$0.323 \pm 0.012$ & $0.329 \pm 0.013$
\\
Borel transform \eqn{BorelVpAstrong} & $0.328 \, {}^{+\, 0.014}_{-\, 0.013}$ &
$0.318 \, {}^{+\, 0.015}_{-\, 0.012}$ & $0.323 \, {}^{+\, 0.015}_{-\, 0.013}$
\\ \hline
\end{tabular}
\caption{Summary of the most reliable determinations of $\alpha_{s}(m_{\tau}^{2})$, performed in the $V+A$ channel.}
\label{tab:summary}
\end{table}
%%%%%%%%%%%%%%%%%%%%%%%%%%%%%%%%%%%%%%%%%%%%%%%%%%%%%%%%%%%%%%%%%%%%%%%%%%%%%%%%%%%%%

Our first determination in Eq.~\eqn{strongdav}, using the same moments as in the standard ALEPH analysis, is in very good agreement with the results of Ref.~\cite{Davier:2013sfa}. We have increased the uncertainties to account for the potential sensitivity to higher-order inverse-power corrections. However, taking away the $(1 + 2 s/m_\tau^2)$ factor from the ALEPH weights \eqn{kl}, we found basically the same results with smaller errors, as shown in~\eqn{strongdavbis}. This suggests that our error enlargement was too pessimistic. In any case, it provides a very strong consistency check.  Taking away the $(1 + 2 s/m_\tau^2)$ factor, one eliminates the highest-dimensional contribution to each moment.

In section~\ref{sec:optimal} we have analyzed alternative families of weights to better understand the potential role of different types of non-perturbative corrections. The study of optimal moments, which are only sensitive to particular condensate dimensions, brings more light on the numerical size of these effects. From a combined fit of five different $A^{(2,m)}$ moments ($1\le m\le 5$), we have obtained the results in Eq.~\eqn{strong2mMom}, in perfect agreement with the previous determinations. Similar values are obtained from the global fit of $A^{(n,0)}$ moments ($0\le n\le 3$) in Table~\ref{n0Fit}.

Neglecting all non-perturbative effects, one can determine the strong coupling with a single moment. The comparison among results extracted from different moments provides then a direct assessment on the missing contributions. While the moment $A^{(2,m)}$ is sensitive to $\cO_{2 (m+2)}$ and $\cO_{2 (m+3)}$, $A^{(1,m)}$ only gets corrections from $\cO_{2 (m+2)}$. In Table~\ref{noOPE} we show the values of $\alpha_s(m_\tau^2)$ extracted from 12 different moments with completely different sensitivity to the neglected inverse power corrections. The good agreement among them clearly indicates that vacuum condensate corrections are very small in the $V+A$ case. Moreover, for all moments the fitted value of the strong coupling agrees with the more solid determinations in Table~\ref{tab:summary} which do take non-perturbative effects properly into account.

A different handle to uncover signals of non-perturbative dynamics is provided by the $s_0$ dependence of the moments. This has been carefully studied in section~\ref{sec:improvements}. Comparing the $s_0$ dependence of a few experimental $A^{(n,0)}(s_0)$ moments ($0\le n\le 3$) with their values predicted with perturbative QCD, one finds the results shown in Figure~\ref{pinchs}, where $\alpha_s(m_\tau^2)$ has been fixed to the value in Eq.~\eqn{strongdav}. In spite of the fact that all non-perturbative contributions have been neglected, the theoretical curves reproduce well the data at large values of $s_0\sim m_\tau^2$. In the $V+A$ distribution the agreement extends to surprisingly low values of $s_0$, specially for $n=0$ and 1. In particular, the data appear to closely follow the CIPT predictions for
$A^{(0,0)}(s_0)$ and $A^{(1,0)}(s_0)$, the moments most exposed to violations of duality. These effects (and $\cO_4$ in the $n=1$ case) appear to be too small to become visible within the much larger perturbative uncertainties. The higher moments seem to indicate a more sizeable $D=6$ contribution, with opposite signs for the $V$ and $A$ distributions, which cancels to a large extent in $V+A$ as expected theoretically. The different $V$, $A$ and $V+A$ curves merge in the higher $s_0$ range, suggesting a tiny numerical effect at $s_0\sim m_\tau^2$.

In Figure~\ref{todoa0} we show, as function of $s_0$, independent determinations of $\alpha_s(m_\tau^2)$ extracted from 13 different moments of the $V+A$ distribution, ignoring all non-perturbative effects. The clear clustering of the different curves is another strong indication that inverse power corrections are small for $V+A$.

One can try to fit the strong coupling, together with the appropriate power corrections, from the $s_0$ dependence of a given moment. However, this is not really justified because it turns out to be equivalent to a direct fit of the spectral function, and the OPE is not valid in the physical real axis. The functional dependence of the moments with $s_0$ should necessarily manifest the violations of quark-hadron duality which are present in the hadronic spectrum. This is seen in Figure~\ref{dualvio} which shows the fitted parameters from the moment $A^{(2,0)}(s_0)$, as function of $\hat s_0$, the starting $s_0$ value of the fit. There is a clear dependence on $\hat s_0$ for the $V$ and $A$ distributions, in the lower $\hat s_0$ range, which however converges to the more stable $V+A$ results at higher values of $\hat s_0$. The stability of the extracted $V+A$ values is surprising, but it can be understood looking to the experimental spectral function in Figure~\ref{fig:ALEPHsf} and observing the rapid flattening of the $V+A$ curve with increasing values of the hadronic invariant mass, which manifests an evident compensation of the vector and axial-vector departures from local duality.

Ignoring duality-violation effects, but including in the uncertainties the variations with $\hat s_0$, one gets from the $V+A$ data in Figure~\ref{fig:ALEPHsf} the values of $\alpha_s(m_\tau^2)$ given in Eq.~\eqn{segundostrong}. The agreement with the other determinations looks amazing. Obviously, the extraction from the $s_0$ dependence has a much lower theoretical basis than the previous ones, since local duality is needed. Nevertheless, it provides a good consistency test of the negligible role of duality-violation effects in the results quoted in Eqs.~\eqn{strongdav}, \eqn{strongdavbis} and \eqn{strong2mMom}.

In section~\ref{sec:DV} we have followed the strategy advocated in Refs.~\cite{Boito:2011qt,Boito:2012cr,Boito:2014sta}, modeling duality-violations through a functional ansatz with several parameters which are directly fitted to the physical spectral functions. Thus, one is heavily relying on local duality which is a dangerous assumption. A short-distance description in terms of quarks and gluons cannot be applied on the physical cut where, due to confinement, only colour-singlet particles can be produced \cite{Poggio:1975af}. While we are able to reproduce the numerical results of Ref.~\cite{Boito:2014sta}, they turn out to be quite unstable and have a very bad statistical quality. In spite of the many caveats of this approach, one gets reasonable values of the strong coupling, although the uncertainties on the fitted parameters are much larger than the very optimistic estimates claimed in Ref.~\cite{Boito:2014sta}. Making small changes in the assumed functional form of the ansatz one finds very significant fluctuations in the fitted value of $\alpha_s(m_\tau^2)$, which is highly correlated with the model parameters. One easily finds models of the spectral function giving the central values for the strong coupling shown in Table~\ref{tab:summary}, and with much better statistical quality ($\chi^2$, p-value) than the model assumed in Ref.~\cite{Boito:2014sta}. Therefore, this determination is model dependent.

An alternative approach, based on Borel weights, has been explored in section~\ref{sec:Borel}.
It has been shown there that the exponential suppression of the weights allows to find stability regions in both $s_0$ and the Borel parameter $a$, where it is possible to extract clean determinations of the strong coupling from the separate vector and axial-vector distributions, in very good agreement with the $V+A$ results shown in Table~\ref{tab:summary}. Applying the same method in the combined $V+A$ channel
one gets the results in Eq.~\eqn{BorelVpAstrong}.

Our final conclusion is that the results quoted in Table~\ref{tab:summary} are very solid (except perhaps the one from the $s_0$ dependence). The overall agreement among determinations extracted under very different assumptions clearly shows their reliability and even indicates that our uncertainties are probably too conservative. In order to quote combined values, we can make a naive average, but taking into account that the uncertainties are fully correlated.
We find:
\be
\ba{c}
\alpha_{s}(m_\tau^2)^{\mathrm{CIPT}} \; =\; 0.335 \pm 0.013\, ,
\\[5pt]
\alpha_{s}(m_\tau^2)^{\mathrm{FOPT}} \; =\; 0.320 \pm 0.012\, .
\ea
\label{FinalValues}
\ee
The same results are obtained irrespective or whether one includes or not in the average the determination from the $s_0$ dependence of the moments in Eq.~\eqn{segundostrong}, exhibiting a very good numerical stability. Averaging the CIPT and FOPT ``averages'' in Table~\ref{tab:summary}, we quote as our final determination of the strong coupling
\be
\alpha_{s}(m_\tau^2) \; =\; 0.328 \pm 0.013 \, .
\ee
These results nicely agree with the value of the strong coupling extracted from $R_\tau$ in Ref.~\cite{Pich:2013lsa}.

After evolution up to the scale $M_Z$, the strong coupling decreases to
\be
\alpha_{s}^{(n_f=5)}(M_Z^{2})\; =\; 0.1197\pm 0.0015 \, ,
\ee
in excellent agreement with the direct measurement at the $Z$ peak from the $Z$ hadronic width,
$\alpha_{s}(M_Z^{2})\; =\; 0.1197\pm 0.0028$ \cite{Agashe:2014kda}.
The comparison of these two determinations provides a beautiful test of the predicted QCD running; {\it i.e.} a very significant experimental verification of asymptotic freedom:
\be
\left.\alpha_{s}^{(n_f=5)}(M_Z^{2})\right|_\tau - \left.\alpha_{s}^{(n_f=5)}(M_Z^{2})\right|_Z
\; =\; 0.0000\pm 0.0015_\tau\pm 0.0028_Z\, .
\ee

Improvements on the determination of $\alpha_{s}(m_\tau^2)$ from $\tau$ decay data would require high-precision measurements of the spectral functions, specially in the higher kinematically-allowed energy bins.
Both higher statistics and a good control of experimental systematics are needed, which could be possible at the forthcoming Belle-II experiment. On the theoretical side, one needs an improved understanding of
higher-order perturbative corrections.

\section*{Acknowledgements}

We want to thank Michel Davier, Andreas Hoecker, Bogdan Malaescu, Changzheng Yuan and
Zhiqing Zhang for making publicly available the updated ALEPH spectral functions, with all
the necessary details about error correlations. Our analysis would have not been possible without
all this precious information.
This work has been supported in part by the Spanish Government and ERDF funds from
the EU Commission [Grants No. FPA2014-53631-C2-1-P and FPU14/02990], by the Spanish
Centro de Excelencia Severo Ochoa Programme [Grant SEV-2014-0398] and by the Generalitat Valenciana [PrometeoII/2013/007].

\clearpage
\bibliography{mybibfile}

\begin{thebibliography}{10}

\bibitem{Pich:2013lsa}
Antonio Pich.
\newblock {Precision Tau Physics}.
\newblock {\em Prog. Part. Nucl. Phys.}, 75:41--85, 2014.

\bibitem{Pich:2015ivv}
Antonio Pich.
\newblock {$\alpha_s$ from hadronic $\tau$ decays}.
\newblock In {\em {High-precision $\alpha_s$ measurements from LHC to FCC-ee}},
  pages 37--40, 2015.

\bibitem{d'Enterria:2015toz}
David d'Enterria and Peter~Z. Skands, editors.
\newblock {\em {High-Precision $\alpha_s$ Measurements from LHC to FCC-ee}},
  2015.
\newblock Eprint: arXiv:1512.05194.

\bibitem{Agashe:2014kda}
K.~A. Olive et~al.
\newblock {Review of Particle Physics}.
\newblock {\em Chin. Phys.}, C38:090001, 2014.

\bibitem{Pich:2013sqa}
Antonio Pich.
\newblock {Review of $\alpha_s$ determinations}.
\newblock 2013.
\newblock PoSConfinementX,022(2012).

\bibitem{Deur:2016tte}
Alexandre Deur, Stanley~J. Brodsky, and Guy~F. de~Teramond.
\newblock {The QCD Running Coupling}.
\newblock 2016.
\newblock Eprint: arXiv:1604.08082.

\bibitem{Narison:1988ni}
Stephan Narison and A.~Pich.
\newblock {QCD Formulation of the tau Decay and Determination of Lambda (MS)}.
\newblock {\em Phys. Lett.}, B211:183, 1988.

\bibitem{Braaten:1988hc}
E.~Braaten.
\newblock {QCD Predictions for the Decay of the tau Lepton}.
\newblock {\em Phys. Rev. Lett.}, 60:1606--1609, 1988.

\bibitem{Braaten:1988ea}
Eric Braaten.
\newblock {The Perturbative QCD Corrections to the Ratio R for tau Decay}.
\newblock {\em Phys. Rev.}, D39:1458, 1989.

\bibitem{Braaten:1991qm}
E.~Braaten, Stephan Narison, and A.~Pich.
\newblock {QCD analysis of the tau hadronic width}.
\newblock {\em Nucl. Phys.}, B373:581--612, 1992.

\bibitem{LeDiberder:1992zhd}
F.~Le~Diberder and A.~Pich.
\newblock {Testing QCD with tau decays}.
\newblock {\em Phys. Lett.}, B289:165--175, 1992.

\bibitem{Baikov:2008jh}
P.~A. Baikov, K.~G. Chetyrkin, and Johann~H. Kuhn.
\newblock {Order $\alpha_s^4$ QCD Corrections to Z and $\tau$ Decays}.
\newblock {\em Phys. Rev. Lett.}, 101:012002, 2008.

\bibitem{LeDiberder:1992jjr}
F.~Le~Diberder and A.~Pich.
\newblock {The perturbative QCD prediction to $R_\tau$ revisited}.
\newblock {\em Phys. Lett.}, B286:147--152, 1992.

\bibitem{Pivovarov:1991rh}
A.~A. Pivovarov.
\newblock {Renormalization group analysis of the tau lepton decay within QCD}.
\newblock {\em Z. Phys.}, C53:461--464, 1992.
\newblock [Yad. Fiz.54,1114(1991)].

\bibitem{Gross:1973id}
David~J. Gross and Frank Wilczek.
\newblock {Ultraviolet Behavior of Nonabelian Gauge Theories}.
\newblock {\em Phys. Rev. Lett.}, 30:1343--1346, 1973.

\bibitem{Politzer:1973fx}
H.~David Politzer.
\newblock {Reliable Perturbative Results for Strong Interactions?}
\newblock {\em Phys. Rev. Lett.}, 30:1346--1349, 1973.

\bibitem{Coleman:1973sx}
Sidney~R. Coleman and David~J. Gross.
\newblock {Price of asymptotic freedom}.
\newblock {\em Phys. Rev. Lett.}, 31:851--854, 1973.

\bibitem{Pich:2011bb}
Antonio Pich.
\newblock {Tau Decay Determination of the QCD Coupling}.
\newblock In {\em {Workshop on Precision Measurements of $\alpha_s$}}, 2011.
\newblock Eprint: arXiv:1107.1123.

\bibitem{Schael:2005am}
S.~Schael et~al.
\newblock {Branching ratios and spectral functions of tau decays: Final ALEPH
  measurements and physics implications}.
\newblock {\em Phys. Rept.}, 421:191--284, 2005.

\bibitem{Davier:2013sfa}
Michel Davier, Andreas Hoecker, Bogdan Malaescu, Chang-Zheng Yuan, and Zhiqing
  Zhang.
\newblock {Update of the ALEPH non-strange spectral functions from hadronic
  $\tau$ decays}.
\newblock {\em Eur. Phys. J.}, C74(3):2803, 2014.

\bibitem{Davier:2008sk}
M.~Davier, S.~Descotes-Genon, Andreas Hocker, B.~Malaescu, and Z.~Zhang.
\newblock {The Determination of $\alpha_s$ from $\tau$ Decays Revisited}.
\newblock {\em Eur. Phys. J.}, C56:305--322, 2008.

\bibitem{Davier:2005xq}
Michel Davier, Andreas Hoecker, and Zhiqing Zhang.
\newblock {The Physics of hadronic tau decays}.
\newblock {\em Rev. Mod. Phys.}, 78:1043--1109, 2006.

\bibitem{Barate:1998uf}
R.~Barate et~al.
\newblock {Measurement of the spectral functions of axial -- vector hadronic
  tau decays and determination of $\alpha_s(M^2_\tau)$}.
\newblock {\em Eur. Phys. J.}, C4:409--431, 1998.

\bibitem{Buskulic:1993sv}
D.~Buskulic et~al.
\newblock {Measurement of the strong coupling constant using tau decays}.
\newblock {\em Phys. Lett.}, B307:209--220, 1993.

\bibitem{Ackerstaff:1998yj}
K.~Ackerstaff et~al.
\newblock {Measurement of the strong coupling constant $\alpha_s$ and the
  vector and axial vector spectral functions in hadronic tau decays}.
\newblock {\em Eur. Phys. J.}, C7:571--593, 1999.

\bibitem{Coan:1995nk}
T.~Coan et~al.
\newblock {Measurement of $\alpha_s$ from tau decays}.
\newblock {\em Phys. Lett.}, B356:580--588, 1995.

\bibitem{Boito:2011qt}
Diogo Boito, Oscar Cata, Maarten Golterman, Matthias Jamin, Kim Maltman, James
  Osborne, and Santiago Peris.
\newblock {A new determination of $\alpha_s$ from hadronic $\tau$ decays}.
\newblock {\em Phys. Rev.}, D84:113006, 2011.

\bibitem{Boito:2012cr}
Diogo Boito, Maarten Golterman, Matthias Jamin, Andisheh Mahdavi, Kim Maltman,
  James Osborne, and Santiago Peris.
\newblock {An Updated determination of $\alpha_s$ from $\tau$ decays}.
\newblock {\em Phys. Rev.}, D85:093015, 2012.

\bibitem{Boito:2014sta}
Diogo Boito, Maarten Golterman, Kim Maltman, James Osborne, and Santiago Peris.
\newblock {Strong coupling from the revised ALEPH data for hadronic $\tau$
  decays}.
\newblock {\em Phys. Rev.}, D91(3):034003, 2015.

\bibitem{Cata:2008ye}
Oscar Cata, Maarten Golterman, and Santi Peris.
\newblock {Unraveling duality violations in hadronic tau decays}.
\newblock {\em Phys. Rev.}, D77:093006, 2008.

\bibitem{Cata:2008ru}
Oscar Cata, Maarten Golterman, and Santiago Peris.
\newblock {Possible duality violations in tau decay and their impact on the
  determination of $\alpha_s$}.
\newblock {\em Phys. Rev.}, D79:053002, 2009.

\bibitem{Marciano:1988vm}
W.~J. Marciano and A.~Sirlin.
\newblock {Electroweak Radiative Corrections to tau Decay}.
\newblock {\em Phys. Rev. Lett.}, 61:1815--1818, 1988.

\bibitem{Braaten:1990ef}
Eric Braaten and Chong-Sheng Li.
\newblock {Electroweak radiative corrections to the semihadronic decay rate of
  the tau lepton}.
\newblock {\em Phys. Rev.}, D42:3888--3891, 1990.

\bibitem{Erler:2002mv}
Jens Erler.
\newblock {Electroweak radiative corrections to semileptonic tau decays}.
\newblock {\em Rev. Mex. Fis.}, 50:200--202, 2004.

\bibitem{Shifman:1978bx}
Mikhail~A. Shifman, A.I. Vainshtein, and Valentin~I. Zakharov.
\newblock {QCD and Resonance Physics. Theoretical Foundations}.
\newblock {\em Nucl. Phys.}, B147:385--447, 1979.

\bibitem{Chibisov:1996wf}
Boris Chibisov, R.~David Dikeman, Mikhail~A. Shifman, and N.~Uraltsev.
\newblock {Operator product expansion, heavy quarks, QCD duality and its
  violations}.
\newblock {\em Int. J. Mod. Phys.}, A12:2075--2133, 1997.

\bibitem{Shifman:2000jv}
Mikhail~A. Shifman.
\newblock {Quark hadron duality}.
\newblock In {\em {Proceedings, 8th International Symposium on Heavy Flavor
  Physics (Heavy Flavors 8)}}, page hf8/013, 2000.

\bibitem{Cirigliano:2002jy}
Vincenzo Cirigliano, John~F. Donoghue, Eugene Golowich, and Kim Maltman.
\newblock {Improved determination of the electroweak penguin contribution to
  $\varepsilon' / \varepsilon$ in the chiral limit}.
\newblock {\em Phys. Lett.}, B555:71--82, 2003.

\bibitem{Cirigliano:2003kc}
Vincenzo Cirigliano, Eugene Golowich, and Kim Maltman.
\newblock {QCD condensates for the light quark $V-A$ correlator}.
\newblock {\em Phys. Rev.}, D68:054013, 2003.

\bibitem{Cata:2005zj}
O.~Cata, M.~Golterman, and S.~Peris.
\newblock {Duality violations and spectral sum rules}.
\newblock {\em JHEP}, 0508:076, 2005.

\bibitem{GonzalezAlonso:2010xf}
Mart\'{\i}n Gonz\'alez-Alonso, Antonio Pich, and Joaquim Prades.
\newblock {Pinched weights and Duality Violation in QCD Sum Rules: a critical
  analysis}.
\newblock {\em Phys. Rev.}, D82:014019, 2010.

\bibitem{Dominguez:2016jsq}
C.~A. Dom\'{\i}nguez, L.~A. Hern\'andez, K.~Schilcher, and H.~Spiesberger.
\newblock {Tau-decay hadronic spectral functions: probing quark-hadron
  duality}.
\newblock 2016.
\newblock Eprint: arXiv:1602.00502.

\bibitem{Rodriguez-Sanchez:2016jvw}
A.~Rodr\'{\i}guez-S\'anchez, M.~Gonz\'alez-Alonso, and A.~Pich.
\newblock {Updated determination of chiral couplings and vacuum condensates
  from hadronic tau decay data}.
\newblock 2016.
\newblock Eprint: arXiv:1602.06112.

\bibitem{Adler:1974gd}
Stephen~L. Adler.
\newblock {Some Simple Vacuum Polarization Phenomenology: $e^+ e^-\to$ Hadrons:
  The $\mu$-Mesic Atom X-Ray Discrepancy and $(g-2)$ of the Muon}.
\newblock {\em Phys. Rev.}, D10:3714, 1974.

\bibitem{Gorishnii:1990vf}
S.~G. Gorishnii, A.~L. Kataev, and S.~A. Larin.
\newblock {The $O(\alpha^{3}_{s})$-corrections to
  $\sigma_{tot}(e^{+}e^{-}\rightarrow \mathrm{hadrons})$ and $\Gamma(\tau^{-}
  \rightarrow \nu_{\tau} + \mathrm{hadrons})$ in QCD}.
\newblock {\em Phys. Lett.}, B259:144--150, 1991.

\bibitem{Surguladze:1990tg}
Levan~R. Surguladze and Mark~A. Samuel.
\newblock {Total hadronic cross-section in $e^+ e^-$ annihilation at the four
  loop level of perturbative QCD}.
\newblock {\em Phys. Rev. Lett.}, 66:560--563, 1991.
\newblock [Erratum: Phys. Rev. Lett.66,2416(1991)].

\bibitem{Chetyrkin:1979bj}
K.~G. Chetyrkin, A.~L. Kataev, and F.~V. Tkachov.
\newblock {Higher Order Corrections to $\sigma_t (e^+ e^- \to\mathrm{Hadrons})$
  in Quantum Chromodynamics}.
\newblock {\em Phys. Lett.}, B85:277, 1979.

\bibitem{Dine:1979qh}
Michael Dine and J.~R. Sapirstein.
\newblock {Higher Order QCD Corrections in $e^+ e^-$ Annihilation}.
\newblock {\em Phys. Rev. Lett.}, 43:668, 1979.

\bibitem{Celmaster:1979xr}
William Celmaster and Richard~J. Gonsalves.
\newblock {An Analytic Calculation of Higher Order Quantum Chromodynamic
  Corrections in $e^+ e^-$ Annihilation}.
\newblock {\em Phys. Rev. Lett.}, 44:560, 1980.

\bibitem{Pich:1999hc}
Antonio Pich and Joaquim Prades.
\newblock {Strange quark mass determination from Cabibbo suppressed tau
  decays}.
\newblock {\em JHEP}, 10:004, 1999.

\bibitem{Pich:2010xb}
Antonio Pich.
\newblock {$\alpha_s$ Determination from $\tau$ Decays: Theoretical Status}.
\newblock {\em Acta Phys. Polon. Supp.}, 3:165--170, 2010.

\bibitem{Ball:1995ni}
Patricia Ball, M.~Beneke, and Vladimir~M. Braun.
\newblock {Resummation of $(\beta_0 \alpha_s)^n$ corrections in QCD: Techniques
  and applications to the tau hadronic width and the heavy quark pole mass}.
\newblock {\em Nucl. Phys.}, B452:563--625, 1995.

\bibitem{Neubert:1995gd}
Matthias Neubert.
\newblock {QCD analysis of hadronic tau decays revisited}.
\newblock {\em Nucl. Phys.}, B463:511--546, 1996.

\bibitem{Altarelli:1994vz}
G~Altarelli, P.~Nason, and G.~Ridolfi.
\newblock {A Study of ultraviolet renormalon ambiguities in the determination
  of $\alpha_s$ from tau decay}.
\newblock {\em Z. Phys.}, C68:257--268, 1995.

\bibitem{Beneke:2008ad}
Martin Beneke and Matthias Jamin.
\newblock {$\alpha_s$ and the tau hadronic width: fixed-order, contour-improved
  and higher-order perturbation theory}.
\newblock {\em JHEP}, 09:044, 2008.

\bibitem{Beneke:2012vb}
Martin Beneke, Diogo Boito, and Matthias Jamin.
\newblock {Perturbative expansion of tau hadronic spectral function moments and
  $\alpha_s$ extractions}.
\newblock {\em JHEP}, 01:125, 2013.

\bibitem{DescotesGenon:2010cr}
S.~Descotes-Genon and B.~Malaescu.
\newblock {A Note on Renormalon Models for the Determination of
  $\alpha_s(m_\tau)$}.
\newblock 2010.
\newblock Eprint: arXiv:1002.2968.

\bibitem{Jamin:2005ip}
Matthias Jamin.
\newblock {Contour-improved versus fixed-order perturbation theory in hadronic
  tau decays}.
\newblock {\em JHEP}, 09:058, 2005.

\bibitem{Caprini:2009vf}
Irinel Caprini and Jan Fischer.
\newblock {$\alpha_s$ from tau decays: Contour-improved versus fixed-order
  summation in a new QCD perturbation expansion}.
\newblock {\em Eur. Phys. J.}, C64:35--45, 2009.

\bibitem{Caprini:2011ya}
Irinel Caprini and Jan Fischer.
\newblock {Expansion functions in perturbative QCD and the determination of
  $\alpha_s(M_\tau^2)$}.
\newblock {\em Phys. Rev.}, D84:054019, 2011.

\bibitem{Abbas:2012fi}
Gauhar Abbas, B.~Ananthanarayan, Irinel Caprini, and Jan Fischer.
\newblock {Perturbative expansion of the QCD Adler function improved by
  renormalization-group summation and analytic continuation in the Borel
  plane}.
\newblock {\em Phys. Rev.}, D87(1):014008, 2013.

\bibitem{Abbas:2012py}
Gauhar Abbas, B.~Ananthanarayan, and Irinel Caprini.
\newblock {Determination of $\alpha_s(M_{\tau}^2)$ from Improved Fixed Order
  Perturbation Theory}.
\newblock {\em Phys. Rev.}, D85:094018, 2012.

\bibitem{Abbas:2013usa}
Gauhar Abbas, B.~Ananthanarayan, Irinel Caprini, and Jan Fischer.
\newblock {Expansions of $\tau$ hadronic spectral function moments in a
  nonpower QCD perturbation theory with tamed large order behavior}.
\newblock {\em Phys. Rev.}, D88(3):034026, 2013.

\bibitem{Pich:1998yn}
Antonio Pich and Joaquim Prades.
\newblock {Perturbative quark mass corrections to the tau hadronic width}.
\newblock {\em JHEP}, 06:013, 1998.

\bibitem{GellMann:1968rz}
Murray Gell-Mann, R.~J. Oakes, and B.~Renner.
\newblock {Behavior of current divergences under SU(3) $\times$ SU(3)}.
\newblock {\em Phys. Rev.}, 175:2195--2199, 1968.

\bibitem{Pich:1995bw}
A.~Pich.
\newblock {Chiral perturbation theory}.
\newblock {\em Rept. Prog. Phys.}, 58:563--610, 1995.

\bibitem{Ecker:1994gg}
G.~Ecker.
\newblock {Chiral perturbation theory}.
\newblock {\em Prog. Part. Nucl. Phys.}, 35:1--80, 1995.

\bibitem{NarisonBook}
Stephan Narison.
\newblock {\em {QCD as a Theory of Hadrons: From Partons to Confinement}}.
\newblock Cambridge Monographs on Particle Physics, Nuclear Physics and
  Cosmology, Vol.~17. Cambridge University Press, New York, 2004.

\bibitem{Narison:2009vy}
Stephan Narison.
\newblock {Power corrections to alpha(s)(M(tau)),|V(us)| and anti-m(s)}.
\newblock {\em Phys. Lett.}, B673:30--36, 2009.

\bibitem{Boito:2010fb}
D.~R. Boito, O.~Cata, M.~Golterman, M.~Jamin, K.~Maltman, J.~Osborne, and
  S.~Peris.
\newblock {Duality violations in tau hadronic spectral moments}.
\newblock {\em Nucl. Phys. Proc. Suppl.}, 218:104--109, 2011.

\bibitem{Poggio:1975af}
E.~C. Poggio, Helen~R. Quinn, and Steven Weinberg.
\newblock {Smearing the Quark Model}.
\newblock {\em Phys. Rev.}, D13:1958, 1976.

\bibitem{GonzalezAlonso:2010rn}
Martin Gonzalez-Alonso, Antonio Pich, and Joaquim Prades.
\newblock {Violation of Quark-Hadron Duality and Spectral Chiral Moments in
  QCD}.
\newblock {\em Phys. Rev.}, D81:074007, 2010.

\bibitem{Weinberg:1967kj}
Steven Weinberg.
\newblock {Precise relations between the spectra of vector and axial vector
  mesons}.
\newblock {\em Phys. Rev. Lett.}, 18:507--509, 1967.

\bibitem{Narison:1993sx}
Stephan Narison and A.~Pich.
\newblock {Semi-inclusive tau decays involving the vector or axial-vector
  hadronic currents}.
\newblock {\em Phys. Lett.}, B304:359--365, 1993.

\bibitem{Girone:1995xb}
Maria Girone and Matthias Neubert.
\newblock {Test of the running of $\alpha_s$ in $\tau$ decays}.
\newblock {\em Phys. Rev. Lett.}, 76:3061--3064, 1996.

\end{thebibliography}
%\begin{thebibliography}{99}
%\end{thebibliography}
\end{document}